%% file: sample-acmsmall.tex
\documentclass[acmsmall]{acmart}

\usepackage{enumitem}
\usepackage{array}
\usepackage{multirow}
\usepackage{tikz}
\usepackage{caption}
\usepackage{subfigure}
\usepackage{colortbl}
\usepackage{graphicx}
\usepackage{makecell}
\usepackage{subcaption}
\theoremstyle{definition}
\newtheorem{definition}{Definition}[section]

\theoremstyle{plain}

\usepackage[linesnumbered,ruled,vlined]{algorithm2e} 
\usepackage{algorithm2e,setspace}

\makeatletter
\renewcommand\subsubsection{\@startsection{subsubsection}{3}{\z@}%
  {-3.25ex\@plus -1ex \@minus -.2ex}%
  {1.5ex \@plus .2ex}%
  {\normalfont\normalsize\bfseries}}
\makeatother

\AtBeginDocument{%
  \providecommand\BibTeX{{%
    \normalfont B\kern-0.5em{\scshape i\kern-0.25em b}\kern-0.8em\TeX}}}

\setcopyright{acmlicensed}
\copyrightyear{2018}
\acmYear{2018}
\acmDOI{XXXXXXX.XXXXXXX}

\acmJournal{JACM}
\acmVolume{37}
\acmNumber{4}
\acmArticle{111}
\acmMonth{8}

\AtBeginDocument{%
  \providecommand\BibTeX{{%
    Bib\TeX}}}
\setlist[itemize]{leftmargin=0.3cm,itemsep=0.1cm,topsep=0.1cm}

\begin{document}

\title{Motif Counting in Complex Networks: A Comprehensive Survey}


\author{Haozhe Yin}
\email{unswyhz@gmail.com}
\affiliation{%
  \institution{The University of New South Wales}
  \city{Sydney}
  \country{Australia}
}

\author{Kai Wang}
\affiliation{%
  \institution{Antai College of Economics and Management, Shanghai Jiao Tong University}
  \city{Shanghai}
  \country{China}
}
\email{w.kai@sjtu.edu.cn}

\author{Wenjie Zhang}
\affiliation{%
  \institution{The University of New South Wales}
  \city{Sydney}
  \country{Australia}
}
\email{wenjie.zhang@unsw.edu.au}

\author{Yizhang He}
\affiliation{%
  \institution{The University of New South Wales}
  \city{Sydney}
  \country{Australia}
}
\email{yizhang.he@unsw.edu.au}

\author{Ying Zhang}
\affiliation{%
  \institution{Zhejiang Gongshang University}
  \city{Hangzhou}
  \country{China}
}
\email{ying.zhang@zjgsu.edu.cn}

\author{Xuemin Lin}
\affiliation{%
  \institution{Antai College of Economics and Management, Shanghai Jiao Tong University}
  \city{Shanghai}
  \country{China}
}
\email{xuemin.lin@sjtu.edu.cn}

\renewcommand{\shortauthors}{Haozhe Yin et al.}

\begin{abstract}
Motif counting plays a crucial role in understanding the structural properties of networks. By computing motif frequencies, researchers can draw key insights into the structural properties of the underlying network. As networks become increasingly complex, different graph models have been proposed, giving rise to diverse motif patterns. These variations introduce unique computational challenges that require specialized algorithms tailored to specific motifs within different graph structures.

This survey provides a comprehensive and structured overview of motif counting techniques across general graphs, heterogeneous graphs, and hypergraphs. We categorize existing algorithms according to their underlying computational strategies, emphasizing key similarities and distinctions. In addition to reviewing current methodologies, we examine their strengths, limitations, and computational trade-offs. Furthermore, we explore future directions in motif counting, including scalable implementations to improve efficiency in large-scale networks, algorithmic adaptations for dynamic, temporal, and attributed graphs, and deeper integration with large language models (LLMs) and graph-based retrieval-augmented generation (GraphRAG). By offering a detailed analysis of these approaches, this survey aims to support researchers and practitioners in advancing motif counting for increasingly complex network data.

\end{abstract}



\received{20 February 2007}
\received[revised]{12 March 2009}
\received[accepted]{5 June 2009}

\maketitle

\input{Content/1_introduction}
\input{Content/2_preliminaries}
\input{Content/3_general_graph}

\input{Content/4_bipartite_graph}

\input{Content/5_heterogeneous_graph}
\input{Content/6_hypergraph}

\input{Content/7_conclusion}

%
\bibliographystyle{ACM-Reference-Format}
\bibliography{sample-base}

\appendix

\end{document}

%% file: Content/1_introduction.tex
\section{\textbf{INTRODUCTION}}

\label{sec:introduction}

In recent years, graphs have become a fundamental tool for modeling interactions among different types of entities across various domains such as social networks \cite{coyle2008social,hampton2011social}, biological networks \cite{wong2012biological,masoudi2012building}, citation networks \cite{choobdar2012comparison,boekhout2021investigating}, etc. While modeling these networks is a critical step in network analytics, a key challenge lies in understanding their structural properties. In networks, motifs (i.e., small repeated sub-graphs) are regarded as basic building blocks, and identifying these motifs can gain valuable insights into underlying interactions among entities within networks  \cite{wang2020efficient, ma2019linc, gao2022scalable, marcus2010efficient, cai2023efficient, liu2020truss}.

Currently, the field of motif counting has garnered substantial interest from the research community, focusing on the development of motif models and diverse counting algorithms within general graphs \cite{danisch2017large, lee2010survey, angel2012dense, letsios2016finding}. As data becomes increasingly complex in modern applications, different graph models are proposed including bipartite graphs \cite{ye2023efficient, yang2021p, qiu2024accelerating, wang2022efficient, wang2023efficient} and heterogeneous graphs \cite{fang2020effective, luo2021detecting} to achieve effective data modeling. For example, bipartite graphs naturally represent relationships between two distinct sets of entities, making them particularly useful for applications such as user-item networks in e-commerce \cite{li2020hierarchical} and people-location networks in contact tracing \cite{chen2021efficiently}. Additionally, researchers have developed heterogeneous graph models, characterized by multiple types of nodes and relations, to better model complex relational structures, as in graph-based retrieval-augmented generation (GraphRAG) \cite{chen2025pathrag,cai2024simgrag,li2024simple}, where entity and relation diversity is critical for accurate retrieval and reasoning. However, the diverse structural characteristics of different types of graphs mean that the motif patterns and counting algorithms applicable to one may not be transferable to another. Consequently, specialized motif patterns and corresponding counting techniques have been developed for specific graph types. For example, in general graphs, extensive studies have focused on counting and listing triangles \cite{itai1977finding, burkhardt2017graphing, schank2005finding, cohen2009graph, alon1997finding, low2017first}. In contrast, such triangles cannot exist in bipartite graphs due to their structural constraints. Instead, researchers have proposed the concept of the butterfly \cite{wang2014rectangle, sanei2018butterfly, wang2019vertex, zhou2021butterfly, zhou2023butterfly}, which serves as an analogue to the triangle in general graphs, capturing similar local connectivity patterns within bipartite graphs. 

In this paper, we systematically survey motif patterns and their corresponding algorithms in complex networks, enabling researchers to uncover valuable insights into network structure, application-specific motif patterns, and computational techniques associated with different graph types. Our main contributions are outlined as follows:

\begin{itemize}
    \item \textbf{Review of Motif Counting Algorithms Across Different Graph Types.} Our survey covers motif counting algorithms across three graph types: general graphs, heterogeneous graphs, and hypergraphs. Furthermore, since bipartite graphs are a special case of heterogeneous graphs and have been extensively studied, we further categorize heterogeneous graphs into two main types: bipartite graphs and other heterogeneous graphs. We present the distinct motif models introduced in these graph structures, along with the corresponding counting algorithms.
    \item \textbf{Investigation and Classification of Motif-Specific Counting Algorithms.} We focus on algorithms specifically designed for counting particular motif patterns. These algorithms encompass a range of approaches, including exact counting methods, estimation techniques, and their extended versions, such as parallel implementations and GPU-based approaches. To provide a clear understanding of their fundamental principles, we systematically classify algorithms based on their computational logic and provide complete overview tables. 
    \item \textbf{Comparative Analysis and Future Directions.} We conduct a comprehensive comparison of motif counting algorithms and identify key areas for future research. Our analysis highlights the potential of integrating motif counting into LLMs via motif-aware retrieval-augmented generation, which enables more efficient and structured reasoning over graph-structured data. Additionally, we examine the scalability and efficiency challenges in large-scale motif discovery and identify opportunities for further optimization. Moreover, we emphasize the need for more effective motif counting in complex graphs by proposing new motif models and developing techniques that adapt to evolving network structures, including dynamic, temporal, and attributed graphs.

\end{itemize}

The remainder of this paper is structured as follows. Section~\ref{preliminaries} introduces the preliminaries and essential concepts. Section~\ref{general graph} focuses on motifs in general graphs and the corresponding counting algorithms. In Section~\ref{bipartite graph}, we examine motifs in bipartite graphs and their counting methods. Section~\ref{heterogeneous graph} is dedicated to motifs in heterogeneous graphs, while Section~\ref{hypergraph} explores motifs in hypergraphs, both along with their respective counting algorithms. Finally, Section~\ref{conclusion} provides a summary of the survey and discusses future research directions.

\begin{figure}[htb]
	\begin{center}
            \label{process1}	
            \includegraphics[scale=0.12]{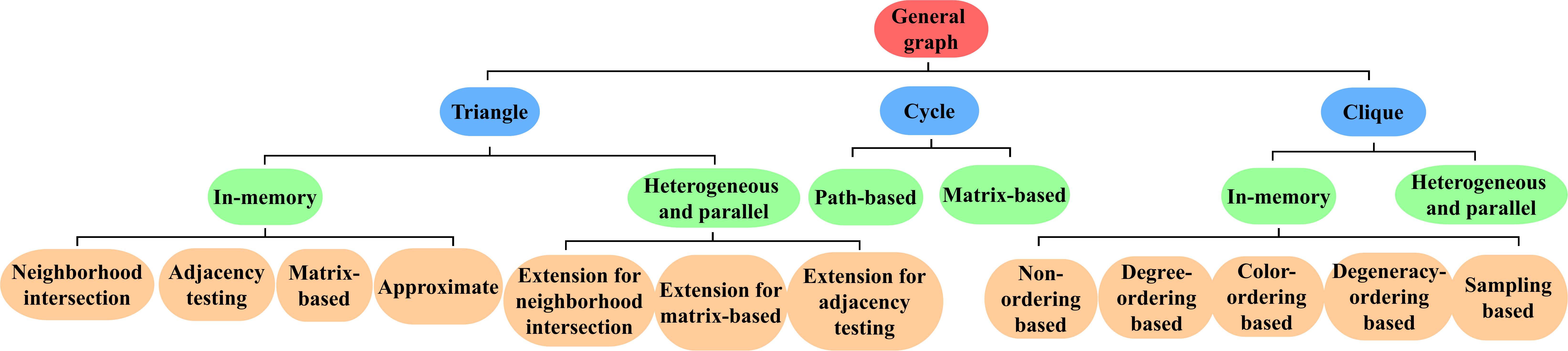}  
            \label{process2}	
            \includegraphics[scale=0.12]{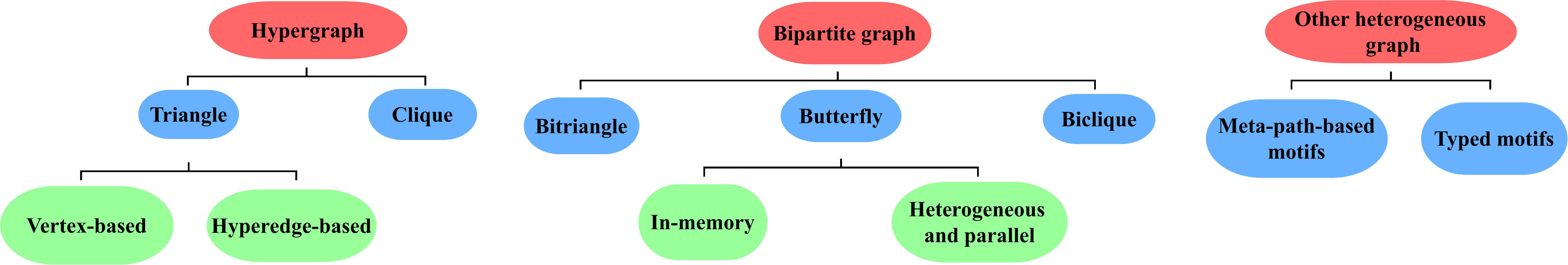} 
	\end{center}
        \vspace{-3mm}
	\caption{Outline of the Survey.}
    
    \label{Organization}
    \vspace{-3mm}
\end{figure}

%% file: Content/2_preliminaries.tex
\section{\textbf{PRELIMINARIES}}
\label{preliminaries}

In this paper, we focus on three fundamental types of graphs: general graphs, heterogeneous graphs, and hypergraphs. In addition, we explore bipartite graphs, a special form of heterogeneous graphs. In this section, we introduce the definitions of these graph models, laying the groundwork for the motif counting techniques discussed later.

An unweighted undirected graph $G$ is comprised of a set \( V \) of vertices and a set \( E \) of edges, where vertices represent entities and edges denote relationships between them. In this paper, we refer to this essential graph type as a general graph. 

\begin{definition}
[{\bf General Graph}] A general graph is a simple graph $G=(V, E)$ where \( V \) is a set of vertices and \( E \) is a set of edges. 
\end{definition}

While general graphs provide a framework for modeling relationships between entities, they assume that all vertices and edges are of the same type. However, real-world datasets often contain entities with diverse attributes and multiple types of relationships \cite{rossi2013multi, kong2013inferring, bassett2006small, bullmore2009complex}. To more precisely describe the different attributes or characteristics of entities in a dataset, researchers propose the concept of heterogeneous graphs.

\begin{definition}
[{\bf Heterogeneous Graph}] A heterogeneous graph is a graph $G=(V, E)$ with a vertex type mapping function $\phi:V \rightarrow \mathcal{T}_{V}$ and an edge type mapping function $\xi:E \rightarrow \mathcal{T}_E$, where $\mathcal{T}_{V}$ and $\mathcal{T}_{E}$ denote the set of vertex object types and edge types, respectively. The type of vertex $i$ is defined as $\phi_i$ whereas the type of edge $e=(i, j)\in E$ is defined as $\xi_{ij}=\xi_{e}$.
\end{definition}

\begin{definition}
[{\bf Heterogeneous Graph Schema}] Given a heterogeneous graph $G=(V, E)$ with the vertex type mapping function $\phi:V \rightarrow \mathcal{T}_{V}$ and the edge type mapping function $\xi:E \rightarrow \mathcal{T}_E$, its schema $S_G$ is a directed graph defined over vertex types $\mathcal{T}_{V}$ and edge types (as relations) from $\mathcal{T}_{E}$, i.e., $S_G=(\mathcal{T}_{V}, \mathcal{T}_{E})$.
\end{definition}


In heterogeneous graphs, there exists a special form known as bipartite graphs, which can be considered a specific case containing only two types of vertices. Due to their ability to effectively model various real-world scenarios \cite{borgatti1997network,latapy2008basic}, they have been widely studied by researchers. Their formal definition is given below.

\begin{definition}
[{\bf Bipartite Graph}] A bipartite graph is a graph $G=(V=(U, L), E)$ where $U(G)$ denotes the set of vertices in the upper layer, $L(G)$ denotes the set of vertices in the lower layer, $U(G)\cap L(G)=\emptyset$, $V(G)=U(G)\cup L(G)$ denotes the vertex set, and $E(G)\subseteq U(G)\times L(G)$ denotes the edge set.
\end{definition}

Besides heterogeneous graphs, hypergraphs have also attracted increasing attention in recent years because of their flexibility in modeling a broad range of systems where high-order relationships exist among interacting components. In this paper, we will also introduce motif counting algorithms in hypergraphs. Their formal definition is as follows:

\begin{definition}
[{\bf Hypergraph}] A hypergraph is a graph \( G = (V, E) \) where \( V \) is a set of vertices and $E=\{e_1,...,e_{|E|} \}$ is a set of hyperedges. Each hyperedge $e_i\in E$ is a non-empty set of vertices.
\end{definition}

\noindent
{{\bf Remark.} Note that in this paper, we focus on surveying practical algorithms specifically designed for counting certain motif patterns. We exclude algorithms tailored to specialized settings (such as dynamic \cite{buriol2006counting, stefani2017triest}, streaming \cite{sanei2019fleet, sheshbolouki2022sgrapp}, or temporal graphs \cite{kovanen2011temporal, liu2021temporal}). While there is growing interest in motif counting under these scenarios, many of these algorithms build on existing counting techniques. Therefore, we restrict our survey to algorithms that focus solely on the motif counting problem.}


%% file: Content/3_general_graph.tex
\section{\textbf{GENERAL GRAPHS}}
\label{general graph}

Motif counting on general graphs has received considerable attention, with various types of motifs being explored, including $k$-cliques, cycles, and their special form triangles. For example, clique counting is crucial for identifying dense subgraphs and plays a key role in applications such as community detection \cite{dourisboure2009extraction}. Triangle counting serves as a foundation for various graph-related tasks, including $k$-truss decomposition \cite{wang2012truss} and clustering coefficient computation \cite{watts1998collective}. Additionally, other motifs, such as cycles \cite{alon1997finding, abboud2022listing}, have garnered significant attention for their ability to reveal deeper insights into network connectivity and structural properties. In this section, we provide an overview of key algorithms designed for counting different types of motifs in general graphs.



\subsection{\textbf{Triangle Counting}}

Triangles are a fundamental type of small subgraph widely used in the analysis of complex graphs. In a general graph, a triangle consists of three vertices, each pair of which is connected by an edge. Notably, a triangle is the shortest non-trivial cycle (i.e., a cycle of length 3) and the smallest non-trivial clique (i.e., a clique of size 3). Triangles play a crucial role in various aspects of network analysis, including clustering coefficient computation \cite{suri2011counting, xu2007scan, yin2017local}, community search \cite{berry2011tolerating, radicchi2004defining}, web spam detection \cite{park2016pte, arifuzzaman2013patric, becchetti2010efficient}, and role discovery \cite{chou2010discovering}. Triangle counting in general graphs has been a central research topic, leading to numerous studies on efficient counting and listing techniques \cite{itai1977finding, burkhardt2017graphing, schank2005finding, cohen2009graph, alon1997finding, low2017first, azad2015parallel, latapy2008main, chiba1985arboricity, danisch2018listing, yu2020aot, lecuyer2023tailored, becchetti2010efficient, chu2011triangle, ahmed2021triangle, arifuzzaman2019fast, polak2016counting, arifuzzaman2013patric, pashanasangi2021faster}. Based on the computational environment and optimization objectives, we categorize existing methods into two groups: in-memory algorithms, which optimize performance for single-machine computations; heterogeneous and parallel algorithms, which leverage GPUs, distributed systems, or specialized hardware to accelerate processing.

\subsubsection{\textbf{In-memory triangle counting algorithms}}

\begin{table}[t]
    \centering
    \resizebox{1\textwidth}{!}{%
    \begin{tabular}{|c|c|c|c|}
        \hline
        \multicolumn{1}{|c|}{\cellcolor{gray!25}\textbf{Category}} & 
        \multicolumn{1}{c|}{\cellcolor{gray!25}\textbf{Strategy}} & 
        \multicolumn{1}{c|}{\cellcolor{gray!25}\textbf{In-memory}}  & 
        \multicolumn{1}{c|}{\cellcolor{gray!25}\textbf{Heterogeneous and parallel}} \\
        \hline
        \multirow{8}{*}{Exact}  & \multirow{3}{*}{Neighborhood intersection} & Chiba-Nishizeki \cite{chiba1985arboricity}, Edge-iterator \cite{batagelj2001subquadratic, schank2005finding, schank2007algorithmic} & Polak \cite{polak2016counting}\\
         & & Forward \cite{schank2007algorithmic}, Compact forward \cite{latapy2006theory, latapy2008main} & A-direction, A-order \cite{hu2021accelerating} \\
         & & AOT \cite{yu2020aot} & TRUST \cite{pandey2021trust}\\
         \cline{2-4}
        & \multirow{4}{*}{Adjacency testing} & Tree-listing \cite{schank2005finding, schank2007algorithmic, latapy2008main, latapy2006theory}  & Chu \cite{chu2011triangle}  \\
        &  & Node-iterator \cite{becchetti2010efficient, latapy2006theory, latapy2008main}, Node-counting-ayz \cite{alon1997finding} & Suri-Vassilvitskii \cite{suri2011counting}\\
        & & Listing-ayz \cite{schank2005finding, schank2007algorithmic}, Node-iterator-core \cite{schank2005finding, schank2007algorithmic} & Patric  \cite{arifuzzaman2013patric}\\
        & & NodeIterator++ \cite{suri2011counting}, NodeIteratorN \cite{arifuzzaman2019fast} & Arifuzzaman  \cite{arifuzzaman2019fast}\\
         \cline{2-4}
         & \multirow{2}{*}{Matrix-based} & MatrixProduct \cite{itai1977finding}  & Cohen \cite{cohen2009graph}, Azad \cite{azad2015parallel} \\
        & & Low \cite{low2017first} & Wolf \cite{wolf2017fast} \\
        \cline{3-4}
        \hline
         \multirow{4}{*}{Approx} & Lanczos \cite{golub2013matrix, demmel1997applied} & EigenTriangle, EigenTriangleLocal \cite{tsourakakis2008fast} & \multirow{4}{*}{\textemdash{}}\\
        \cline{2-3}
         & Probabilistic edge-sampling & Doulion \cite{tsourakakis2009doulion} &  \\
        \cline{2-3}
         & Single-pass sampling & gSHT \cite{ahmed2014graph} & \\
        \cline{2-3}
         & Matrix partitioning & Kollias \cite{kollias2024counting} &  \\
        \hline
    \end{tabular}%
    }
    \vspace{1mm}
    \caption{Summary of in-memory, heterogeneous and parallel triangle counting algorithms}
    \label{DatasetT} 
    \vspace{-8mm}
\end{table}

Existing in-memory triangle counting algorithms can be broadly classified into exact and approximate methods. Among the exact methods, three major categories emerge: neighborhood-intersection algorithms, adjacency testing algorithms, and matrix-based algorithms.

The first category, exemplified by the {\em compact forward} algorithm \cite{latapy2006theory, latapy2008main}, identifies triangles by computing the intersection of the neighborhoods of connected vertices. This method is particularly efficient when vertex neighborhoods are stored in a structured manner, such as in sorted adjacency lists, allowing for fast set intersections. On the other hand, adjacency testing algorithms, such as the {\em node-iterator} algorithm \cite{becchetti2010efficient}, determine triangles by iterating through a vertex’s neighbors and verifying the existence of edges between them. While neighborhood-intersection methods typically benefit from reduced memory access costs and structured vertex ordering, adjacency testing methods can be more efficient in sparse graphs, where direct edge lookups are computationally cheaper than set intersections. Finally, matrix-based algorithms \cite{low2017first} count triangles using the adjacency matrix representation of the graph. This approach leverages linear algebra techniques to efficiently compute triangle counts. Although computationally expensive for large graphs, matrix-based methods are well-suited for parallel processing and have been extended to frameworks like MapReduce and GPU-based computing. In this part, we will first introduce exact triangle counting algorithms, discussing their key principles and computational trade-offs. We will then explore approximate algorithms, highlighting their advantages and applications in large-scale graph analysis.


\noindent
\textbf{\underline{Neighborhood-intersection-based triangle counting.}} In 1985, Chiba and Nishizeki proposed an enhanced neighborhood intersection algorithm \cite{chiba1985arboricity}. This algorithm processes each vertex $v$ in non-increasing order of degree, then scans the subgraph formed by the neighbors of each vertex to detect and list triangles. Once a vertex $v$ has been processed, all edges incident to it are removed from the graph to prevent redundant intersections of the same neighborhood pairs in future scans. 

In \cite{batagelj2001subquadratic, schank2005finding, schank2007algorithmic}, the authors propose {\em edge-iterator} algorithms, which also leverage neighborhood intersections to efficiently count triangles. These algorithms traverse all edges in the graph and compare the neighborhoods of the two endpoint vertices. If a common neighbor exists in both neighborhoods, a triangle is identified. Compared to earlier approaches, these algorithms store each vertex’s neighborhood as a sorted adjacency array rather than a hash set. An improved version of the \texttt{edge-iterator} algorithms, called the {\em forward} algorithm, is introduced in \cite{schank2007algorithmic}. The \texttt{forward} algorithm operates on a directed graph induced by an ordering of the vertices. It maintains a dynamic adjacency structure $A(v)$ for each vertex $v$, which includes only a subset of its neighbors to reduce the search space. The algorithm processes vertices in order, updating $A(v)$ as edges are traversed. For each vertex $s$, it considers adjacent vertices $t$ where $s < t$. Triangles ${v, s, t}$ are counted by identifying common neighbors $v$ shared between $A(s)$ and $A(t)$. 

Matthieu Latapy proposes the {\em compact forward} algorithm in \cite{latapy2006theory, latapy2008main}, which is a more efficient version of the \texttt{forward} algorithm. The core idea of the \texttt{compact forward} algorithm is to use iterators to compare subsets of adjacencies. To achieve this, adjacencies must be sorted, and the comparison stops once a specified index is reached. The algorithm prioritizes low-degree vertices to minimize redundant triangle checks that would otherwise involve high-degree vertices later in the process. It maintains an array to track already processed vertices and iterates over the graph's edges. The \texttt{compact forward} algorithm has the same asymptotic worst-case bounds for running time and space consumption as the \texttt{forward} algorithm. However, unlike the \texttt{forward} algorithm, it does not require storing the complete adjacency array of the graph. Instead, it only stores partial neighbor information for each vertex, which reduces both space requirements and computational overhead in practice.

The \texttt{compact forward} algorithm simplifies triangle listing by assigning directions to graph edges based on a chosen vertex ordering. This technique is known as the orientation technique. Based on this, Yu et al. propose the {\em adaptive oriented triangle-listing (AOT)} algorithm \cite{yu2020aot}, which uses an adaptive orientation approach that dynamically adjusts edge traversal directions based on the out-degrees of adjacent vertices. This reduces redundant lookups and computational complexity. The \texttt{AOT} algorithm categorizes triangles into two types: positive and negative, depending on the direction of their constituent edges. It applies distinct pivot selection strategies for each type to minimize overlap and avoid redundant computations. Additionally, \texttt{AOT} organizes vertices within neighborhoods by degree, enhancing memory access patterns and optimizing cache usage.


\noindent
\textbf{\underline{Adjacency-testing-based triangle counting.}} In 1978, Itai and Rodeh proposed a triangle-finding algorithm with a time complexity of $O(m^{3/2})$ \cite{itai1977finding}. Although the original algorithm stops after finding the first triangle, it can be extended to list all triangles without increasing the asymptotic running time. This extended version, known as the {\em tree-listing} algorithm \cite{schank2005finding, schank2007algorithmic, latapy2008main, latapy2006theory}, computes a covering tree for each connected component, identifies triangles by checking edges not in the trees, removes those edges, and repeats the process until no edges remain. However, since the algorithm repeatedly constructs covering trees and checks for edge existence throughout the search, it is highly time-consuming and requires extra space in practice.

The {\em node-iterator} algorithm, introduced in \cite{becchetti2010efficient, latapy2006theory, latapy2008main}, identifies triangles by iterating over all vertices and checking pairs of their neighbors. Alon et al. \cite{alon1997finding} optimized this approach with the {\em node-counting-ayz} algorithm. This algorithm computes triangle counts by partitioning vertices into low-degree and high-degree sets, applying the \texttt{node-iterator} method to the low-degree set, and leveraging fast matrix multiplication for the high-degree set. Building on the \texttt{node-counting-ayz} algorithm, Thomas Schank and Dorothea Wagner proposed the {\em listing-ayz} algorithm \cite{schank2005finding, schank2007algorithmic}. The \texttt{listing-ayz} algorithm extends this approach to list triangles in $O(m^{3/2})$ time by additionally applying the \texttt{node-iterator} method to the high-degree subgraph. Furthermore, Thomas Schank and Dorothea Wagner refined the \texttt{node-iterator} algorithm with a degree-based approach, called the {\em node-iterator-core} algorithm \cite{schank2005finding, schank2007algorithmic}. This algorithm iteratively selects the vertex $v$ with the lowest degree, processes it using the node-iterator method, and removes $v$ from the graph.

In \cite{suri2011counting}, a degree-based method called {\em NodeIterator++} is introduced. This algorithm begins by ordering vertices by degree and, for each vertex $v$, processes only its neighbors $u$ with higher degrees, further considering neighbors $w$ of $v$ that also have higher degrees than $u$. If an edge exists between $u$ and $w$, the triangle count is incremented. Based on this, Arifuzzaman et al. present the {\em NodeIteratorN} algorithm \cite{arifuzzaman2019fast}, which enhances both runtime and memory efficiency. \texttt{NodeIteratorN} starts with a preprocessing step that sorts nodes by degree and, in cases of ties, by node ID. This sorting allows the algorithm to maintain an effective neighbor set for each node, consisting solely of neighbors with higher degrees. By precomputing the order and using these reduced neighbor sets, \texttt{NodeIteratorN} efficiently identifies triangles through set intersections. Compared to \texttt{NodeIterator++}, this approach eliminates repeated degree comparisons during the counting process and significantly reduces memory usage by storing only effective neighbors.

\noindent
\textbf{\underline{Matrix-based triangle counting.}} For a graph $G$, we can derive an adjacency matrix $A$ based on the connections between its vertices. Then, for any vertex $v$, the diagonal elements of $A^3$ represent twice the number of triangles to which $v$ belongs. Clearly, by calculating $A^3$ , we can determine the number of triangles in $O(n^3)$ time, where $n$ is the number of vertices in the graph. In \cite{itai1977finding}, Itai et al. propose a faster algorithm to count triangles incorporating the fast matrix multiplication method introduced by Coppersmith and Winograd \cite{coppersmith1987matrix}. This algorithm achieves a time complexity of $O(n^{2.376})$, which is significantly faster than $O(n^3)$ and even exceeds Strassen's algorithm \cite{strassen1986asymptotic}, which has a complexity of $O(n^{2.81})$.


Low et al. \cite{low2017first} propose a linear algebra-based algorithm for exact triangle counting that avoids matrix multiplication. The algorithm partitions the graph's vertex set into two disjoint subsets, {\em VTL} and {\em VBR}, and counts triangles by iteratively moving vertices from \texttt{VBR} to \texttt{VTL}. As each vertex moves, the algorithm updates the triangle count by evaluating its interactions with vertices remaining in \texttt{VBR} and those already in \texttt{VTL}. This eliminates redundant calculations typical of other linear algebra-based methods, which often involve multiple matrix multiplications and scaling operations. Instead, the method employs a compact adjacency matrix representation and leverages sparse matrix-vector operations to count triangles precisely once. This approach eliminates the computational overhead of matrix-matrix multiplication, which is both time-consuming and prone to redundant data movements. The implementation employs the {\em compressed sparse row (CSR)} format for adjacency matrix storage, enabling efficient traversal and vertex updates in each iteration. 

\noindent
\textbf{\underline{Approximate algorithms for triangle counting.}} Despite the efficiency of exact algorithms, their computational cost grows significantly for large graphs, making them impractical for massive real-world networks. As a result, researchers develop approximate triangle counting algorithms to trade off some accuracy for improved efficiency. These methods leverage techniques such as sampling, probabilistic data structures, and sketching techniques to estimate triangle counts with high confidence while significantly reducing time and space complexity.

Tsourakakis \cite{tsourakakis2008fast} introduces {\em EigenTriangle} and {\em EigenTriangleLocal}, two fast approximate algorithms for triangle counting. These algorithms rely on the principle that the total triangle count is proportional to the sum of the cubes of the adjacency matrix’s eigenvalues. They exploit the spectral properties of real-world networks, which typically exhibit a few dominant eigenvalues due to their small-world and scale-free nature. By leveraging these dominant eigenvalues, the algorithms approximate triangle counts efficiently, circumventing the high computational cost of exact enumeration. \texttt{EigenTriangle} estimates the global triangle count by computing the leading eigenvalues of the adjacency matrix via the {\em Lanczos} method, a projection technique for symmetric eigenvalue problems \cite{golub2013matrix, demmel1997applied}, and summing their cubes until a predefined threshold is met. \texttt{EigenTriangleLocal} extends this approach to estimate local triangle counts, leveraging eigenvalues and eigenvectors to approximate the number of triangles each node participates in.


The {\em Doulion} algorithm \cite{tsourakakis2009doulion} employs a probabilistic edge-sampling approach to efficiently approximate the number of triangles in large graphs. Each edge $e$ in the original graph is retained with probability $p$, determined by a biased coin toss. If retained, the edge is assigned a weight of $\frac{1}{p}$; otherwise, it is discarded. This sparsification process yields a reduced graph with significantly fewer edges, making triangle counting more efficient. Once the sparsified graph is constructed, an adjacency testing algorithm, such as \texttt{node-iterator}, is applied to identify and count triangles. Each triangle in the sparsified graph corresponds to multiple triangles in the original graph. To compensate, the algorithm scales the count by $\frac{1}{p^3}$, ensuring an unbiased estimate of the original triangle count.

In \cite{ahmed2014graph}, Ahmed et al. propose the {\em graph sample and hold (gSH)} algorithm, which efficiently estimates graph properties without analyzing the entire graph. The \texttt{gSH} algorithm employs a single-pass sampling method for large graphs, using two probabilities, $p$ and $q$, to guide edge selection based on adjacency. Each incoming edge is sampled with probability $p$ if it is not adjacent to any previously sampled edge, enabling the discovery of previously unconnected graph regions. Conversely, if the edge connects to an already-sampled edge, it is sampled with a higher probability $q$, preserving the graph’s local structure. This approach yields a representative sample that balances diversity and connectivity while minimizing memory usage. Ahmed et al. extend \texttt{gSH} for triangle counting by introducing the {\em triangle-focused variant (gSHT)}. This variant prioritizes triangle formations: any edge forming a triangle with two sampled edges is always included. By retaining sufficient structural information, \texttt{gSHT} improves the accuracy of triangle count estimation.

Kollias et al. propose a matrix-based triangle counting approach, combining matrix partitioning with low-rank approximations for improved efficiency \cite{kollias2024counting}. Their method partitions the adjacency matrix $A$ into smaller submatrices, enabling independent processing. Specifically, $A$ is split into two components: $D$, capturing intra-subgraph connections, and $F$, representing inter-subgraph edges. This decomposition allows triangle counting within and across subgraphs via trace operations on submatrices. To further reduce complexity, the algorithm applies truncated {\em singular value decomposition (SVD)} for matrix approximation. By leveraging low-rank approximations, it replaces large matrix multiplications with operations on smaller, dense vectors, significantly reducing computational cost.

\subsubsection{\textbf{Heterogeneous and parallel algorithms for triangle counting}} As graphs grow larger, in-memory triangle counting becomes increasingly costly. To improve efficiency, researchers develop heterogeneous and parallel algorithms by extending existing in-memory methods and leveraging multi-core processors, GPUs, distributed systems, and specialized hardware. Multi-core parallel algorithms \cite{chu2011triangle, suri2011counting} utilize shared-memory architectures to distribute computations across multiple CPU threads, employing techniques such as work-stealing and cache optimization. GPU-based approaches \cite{polak2016counting, bisson2017high} exploit thousands of lightweight threads for fast adjacency list processing and efficient set intersections, often using warp-based or memory-efficient strategies. Distributed computing methods \cite{pandey2021trust, suri2011counting}, implemented in frameworks like MapReduce and Apache Spark, partition graphs across multiple machines, reducing memory constraints but introducing communication overhead. In this section, we categorize these algorithms based on their underlying in-memory foundations, examining how they extend and adapt different in-memory triangle counting strategies to heterogeneous and parallel environments.

\noindent
\textbf{\underline{Extension for neighborhood-intersection-based triangle counting.}} In \cite{polak2016counting}, Polak presents a GPU-optimized parallel version of the \texttt{forward} algorithm for triangle counting using CUDA. The algorithm has two phases: preprocessing and triangle counting. In preprocessing, vertices are sorted by degree, adjacency lists are filtered to retain higher-degree neighbors, and lists are reordered for efficient computation. This phase is parallelized to leverage GPU capabilities. In the triangle counting phase, each GPU thread processes an edge, intersecting adjacency lists via a two-pointer merge technique. Key optimizations include structuring edge data as arrays for better memory access, encoding edges as 64-bit integers, and leveraging texture cache to reduce memory latency. For multi-GPU setups, preprocessing is performed on a single GPU, and data is distributed across devices for concurrent triangle counting.


In \cite{hu2021accelerating}, Hu et al. propose a lightweight graph preprocessing method to enhance the performance of existing GPU-based triangle counting algorithms \cite{bisson2017high, fox2018fast, green2018logarithmic, hu2019triangle} without modifying their core structures or implementations. The method employs two key strategies: edge directing and vertex ordering. Edge directing converts an undirected graph into a directed one, reducing redundant computations and improving workload balance. The authors analyze two common strategies: ID-based and degree-based edge directing. The ID-based approach directs edges from lower to higher vertex IDs, while the degree-based approach directs edges from lower- to higher-degree vertices. Given the NP-hardness of optimal edge directing, the authors propose an approximate solution, {\em A-direction (Analytic Direction)}, to achieve a near-optimal strategy efficiently. Vertex ordering further optimizes GPU resource utilization by balancing memory- and compute-intensive tasks across GPU blocks. To achieve this, the authors introduce the {\em A-order (Analytic Order)} algorithm, which analyzes vertex adjacency lists to group tasks with complementary resource demands. This reordering minimizes idle GPU resources, improving execution efficiency.

By combining the strengths of Hu’s algorithm \cite{hu2021accelerating} with the H-index \cite{pandey2019h}, Pandey et al. propose the {\em TRUST} algorithm \cite{pandey2021trust}. \texttt{TRUST} employs a vertex-centric design integrated with hashing to improve workload distribution and reduce memory usage. The algorithm begins by reordering vertices based on their degrees to minimize hash collisions. This approach reduces collision rates, making hash table construction and lookups more efficient. \texttt{TRUST} also utilizes shared memory for frequently accessed data structures to enhance data locality and reduce latency. To address workload imbalance, \texttt{TRUST} introduces a virtual combination technique that combines two-hop neighbors to evenly distribute workloads across threads. A degree-aware resource allocation mechanism ensures that high-degree vertices are allocated additional computational resources. For scalability, \texttt{TRUST} incorporates a hashing-based 2D partitioning scheme, allowing the graph to be distributed across up to 1,000 GPUs. This partitioning approach enables communication-free and workload-balanced processing, ultimately achieving more than a trillion traversed edges per second.

\noindent
\textbf{\underline{Extension for adjacency-testing-based triangle counting.}} In \cite{chu2011triangle}, Chu et al. introduce an I/O-efficient triangle listing algorithm that partitions graphs into smaller subgraphs to fit into memory, thereby reducing random disk access. For each subgraph, an extended version is created that includes adjacent vertices, and triangles are listed independently within these subgraphs. The partitioning methods include sequential partitioning and dominating-set-based partitioning. During the partitioning process, three types of triangles are identified: triangles where all vertices belong to one subgraph, triangles where two vertices belong to the same subgraph, and triangles where all vertices belong to different subgraphs. The algorithm iteratively simplifies more complex triangle types into simpler ones. This approach efficiently lists all triangles while minimizing I/O costs.

The MapReduce framework provides a powerful approach for parallelizing triangle counting algorithms, as demonstrated in \cite{suri2011counting} by Suri and Vassilvitskii. They present two main algorithms for triangle counting using MapReduce: the {\em node-iterator in MapReduce} and the {\em partition} algorithm. The \texttt{node-iterator in MapReduce} is an adaptation of the \texttt{NodeIterator++} algorithm, designed to efficiently distribute the task of triangle counting across multiple machines. This adaptation generates 2-paths in a graph by pivoting around each node, followed by a verification step to check for edges that complete the triangles. By introducing degree-based restrictions, the \texttt{node-iterator} ensures that each triangle is counted only once, reducing redundancy and computational load. In contrast, the \texttt{partition} algorithm takes a different approach by dividing the graph into overlapping subgraphs, ensuring that each triangle appears in at least one partition. This enables the use of any sequential triangle counting algorithm on the subgraphs, effectively balancing memory and disk usage while maintaining work efficiency.

In \cite{arifuzzaman2013patric}, Arifuzzaman et al. introduce {\em Patric}, an MPI-based distributed memory parallel algorithm to count triangles in massive networks. \texttt{Patric} partitions the graph into subgraphs, enabling each processor to store and process only a portion of the network. Using a modified \texttt{node-iterator} algorithm, each processor counts triangles within its local partition without redundancy. To address computational imbalances caused by skewed degree distributions, \texttt{Patric} incorporates degree-based partitioning for load balancing, ensuring an even workload distribution among processors. Compared to the algorithm in \cite{suri2011counting}, \texttt{Patric} offers several key improvements. It significantly reduces the intermediate data volume by directly counting triangles in subgraphs, while \cite{suri2011counting} generates many 2-paths, increasing overhead. \texttt{Patric} also employs efficient load balancing schemes to distribute tasks evenly, avoiding performance bottlenecks caused by workload imbalances in the previous method. Furthermore, its memory optimization reduces the overall footprint by storing only the relevant parts of the network. These improvements enable \texttt{Patric} to achieve better speed, memory efficiency, and near-linear scalability for massive networks.

Based on the algorithm \texttt{NodeIteratorN} \cite{arifuzzaman2019fast}, Arifuzzaman et al. propose a space-efficient parallel algorithm with non-overlapping partitioning to further optimize memory usage. In this approach, the graph is divided into non-overlapping partitions, with each edge belonging to only one partition. This eliminates redundant storage of edges and reduces memory consumption. However, because no edges are shared between partitions, the algorithm requires communication between processors to count triangles involving nodes in different partitions. To minimize communication overhead, a surrogate approach is used, where processors delegate computations to others holding the required data. This space-efficient algorithm is ideal for scenarios where memory is a primary constraint, and it achieves significant space savings while maintaining efficient parallel computation.

\noindent
\textbf{\underline{Extension for matrix-based triangle counting.}} Cohen et al. present a framework for adapting graph algorithms to the MapReduce paradigm \cite{cohen2009graph}, enabling their deployment in distributed cloud or streaming environments. Specifically, they describe a two-step approach for triangle detection in graphs using MapReduce. The first step enumerates open triads, or pairs of edges sharing a common vertex. Each edge is assigned to its lower-degree vertex to ensure that each potential triangle is counted only once, minimizing redundancy. The reducer pairs all edges for a vertex to generate open triads. The second step checks if these open triads form closed triangles by finding an edge between the two vertices at the ends of each triad. A second MapReduce job is used to combine triads and edges, and the reducer identifies and outputs complete triangles. These steps leverage MapReduce's parallel processing to make triangle enumeration efficient for large-scale distributed systems, breaking down the problem into smaller, independent tasks.

Azad et al. extend Cohen’s algorithm \cite{cohen2009graph} by introducing a matrix-based framework for triangle enumeration with both serial and parallel approaches \cite{azad2015parallel}. The serial algorithm splits the adjacency matrix into lower and upper triangular components. By multiplying these components, it identifies wedge structures (pairs of edges), and an element-wise multiplication with the original matrix verifies if these wedges close to form triangles. To improve efficiency, masked multiplication limits computations to relevant edges, reducing unnecessary operations. The parallel algorithm distributes the computation across multiple processors, using masked multiplication to focus on relevant data and minimize communication overhead. Additional optimizations, such as bloom filters to reduce redundant checks and techniques to compress data exchanges and ensure efficient load balancing, further improve scalability.

In \cite{wolf2017fast}, Wolf combines the algorithm in \cite{azad2015parallel} with the {\em KokkosKernels} technique \cite{edwards2014kokkos, deveci2017performance} to develop a new linear algebra-based approach aimed at achieving optimal parallel performance on multicore architectures. Specifically, the proposed algorithm focuses on using the lower triangular portion of the adjacency matrix to perform triangle counting through matrix operations. The process starts by multiplying the lower triangular matrix with itself, generating a matrix representing all two-paths (or wedges) in the graph. Subsequently, an element-wise multiplication is applied to filter these wedges and identify valid triangles. In contrast to the algorithms in \cite{azad2015parallel}, which use both the lower and upper triangular matrices and perform a separate masking operation involving the full adjacency matrix, the proposed algorithm integrates the masking directly into the multiplication of the sparse matrix. This optimization minimizes memory usage by reducing the need for intermediate storage of wedge counts and avoids redundant triangle counting. Moreover, the algorithm incorporates a reordering strategy based on vertex degrees, which reorders rows to reduce the number of nonzeros, thereby improving the computational efficiency of subsequent matrix operations.

\subsection{\textbf{Clique Counting}}

Identifying dense subgraphs is a significant topic in graph mining research \cite{danisch2017large, lee2010survey}, with diverse applications spanning real-time story tracking \cite{angel2012dense, letsios2016finding}, motif identification in biological systems, social network analysis, and more. Dense subgraphs, often indicative of tightly connected clusters or communities, provide valuable insights into the underlying structure of complex networks. Existing algorithms typically identify these subgraphs by listing all $k$-cliques in a graph, where a $k$-clique is a subgraph with $k$ nodes, each pair of which is connected by an edge (formally defined below). This problem naturally generalizes the well-studied task of listing triangles, as a triangle is simply a 3-clique. In fact, state-of-the-art algorithms can efficiently list all triangles in real-world graphs, making the identification of dense subgraphs in smaller sizes more manageable. However, as the value of $k$ increases, the task of listing all $k$-cliques becomes significantly more complex, requiring more sophisticated computational approaches and optimizations. In this section, we introduce both in-memory, heterogeneous and parallel algorithms for $k$-clique counting.

\begin{definition} 
[{\bf $k$-clique}] Given a general graph $G=(V, E)$, a subgraph $C=(V_c, E_c)$ is said to be a $k$-clique if and only if it has $k$ vertices and has an edge between every pair of vertices, i.e., $|V_c|=k$, $E_c=\{(u,v)|u,v\in V_c, u\neq v \}$. 
\end{definition}

\subsubsection{\textbf{In-memory clique counting algorithms}}


The existing in-memory exact $k$-clique counting algorithms are broadly classified into non-ordering based algorithms and ordering-based algorithms, with the latter further divided into degree-ordering based algorithms, degeneracy-ordering based algorithms, and color-ordering based algorithms. Non-ordering based algorithms \cite{chiba1985arboricity} do not enforce a specific traversal order on vertices and typically rely on recursive expansion or combinatorial enumeration. To improve efficiency, ordering-based algorithms impose a predefined vertex ordering to direct the search space, reducing redundant enumeration. These algorithms first assign an ordering to vertices and then transform the graph into a directed acyclic graph to enforce a structured exploration. Degree-ordering based algorithms \cite{li2020ordering} sort vertices in ascending order of degree and restrict the $k$-clique enumeration process to edges that respect this order. Degeneracy-ordering based algorithms \cite{danisch2018listing} further refine this strategy by leveraging $k$-core decomposition or $k$-truss decomposition, ensuring that each vertex processes only a limited subset of its neighbors, which is especially beneficial for large, sparse graphs. More recently, color-ordering based algorithms \cite{li2020ordering} use greedy graph coloring to assign an ordering that enables search space pruning. By prioritizing vertices based on their assigned colors, these algorithms efficiently eliminate unpromising search paths, significantly improving performance for large $k$ values. Despite the efficiency improvements introduced by ordering-based algorithms, exact $k$-clique counting remains computationally expensive, especially for large $k$ values and dense graphs. To address this challenge, researchers further introduce sampling-based approximate algorithms, which trade off some accuracy for significantly improved efficiency.

In this section, we categorize all in-memory clique counting algorithms based on their computational approach, providing a structured overview of both exact and approximate methods.

\begin{table}[t]
    \centering
    \resizebox{1\textwidth}{!}{%
    \begin{tabular}{|c|c|c|c|c|}
        \hline
        \multicolumn{1}{|c|}{\cellcolor{gray!25}\textbf{Category}} & 
        \multicolumn{1}{c|}{\cellcolor{gray!25}\textbf{Strategy}} & 
        \multicolumn{1}{c|}{\cellcolor{gray!25}\textbf{Algorithms}}  & 
        \multicolumn{1}{c|}{\cellcolor{gray!25}\textbf{Time complexity}} & 
        \multicolumn{1}{c|}{\cellcolor{gray!25}\textbf{Space complexity}} \\
        \hline
        \multirow{13}{*}{Exact}  & \multirow{2}{*}{None ordering} & Chiba-Nishizeki \cite{chiba1985arboricity} & $O(km\alpha^{k-2})$ & $O(m+n)$\\
        \cline{3-5}
         & & Makino-Uno \cite{makino2004new, takeaki2012implementation} & $O(knm\alpha^{k-2})$ & $O(m+n)$\\
         \cline{2-5}
        & \multirow{3}{*}{Degree ordering} & Degree \cite{li2020ordering} & $O(km(\eta/2)^{k-2})$ & $O(m+n)$ \\
        \cline{3-5}
        &  & DDegree \cite{li2020ordering} & $O(km(\delta/2)^{k-2})$ & $O(m+n)$ \\
        \cline{3-5}
        & & SDegree \cite{yuan2022efficient} & $O(km(\delta/2)^{k-2})$ & $O(m+n)$ \\
         \cline{2-5}
         & \multirow{3}{*}{Degeneracy ordering} & kClist\cite{danisch2018listing} & $O(km(c(G)/2)^{k-2}+m)$ & $O(m+n)$ \\
        \cline{3-5}
        & & Pivoter \cite{jain2020power} & $O(\alpha^2|SCT(G)|)$ & $O(m+n)$\\
        \cline{3-5}
        & & EBBkC-T \cite{wang2024efficient} & $O(km(\tau/2)^{k-2}+m\delta)$ & $O(m+n)$ \\
        \cline{2-5}
         & \multirow{5}{*}{Color ordering} & DegCol \cite{li2020ordering} & $O(km(\Delta/2)^{k-2})$ & $O(m+n)$ \\
         \cline{3-5}
         &  & DegenCol \cite{li2020ordering} & $O(km(\Delta/2)^{k-2})$ & $O(m+n)$ \\
         \cline{3-5}
         &  & DDegCol \cite{li2020ordering} & $O(km(\delta/2)^{k-2})$ & $O(m+n)$ \\
         \cline{3-5}
         &  & BitCol \cite{li2020ordering} & $O(km(\delta/2)^{k-2})$ & $O(m+n)$ \\
        \cline{3-5}
        &  & EBBkC-C \cite{wang2024efficient}& $O(km(\Delta/2)^{k-2})$ & $O(m+n)$ \\
        \hline
         \multirow{4}{*}{Approx} & Turán theorem \cite{turn1941extremal} & Turán-shadow \cite{jain2017fast} & $O(n\alpha^{k-1})$ & $O(n\alpha^{k-2}+m)$ \\
        \cline{2-5}
         & Random sampling & Eden \cite{eden2018approximating} & $O(n/C_{k}^{\frac{1}{k}}+m^{\frac{k}{2}}/C_{k})$ & $O(m+n)$ \\
        \cline{2-5}
         & $K$-color set sampling & DPColor \cite{ye2022lightning, ye2023efficient2} & $O((|S|+t)\chi k+k^2t+m+n)$ &  $O(m+n+\chi k)$ \\
        \cline{2-5}
         & $K$-color path sampling & DPColorPath \cite{ye2022lightning, ye2023efficient2} & $O(|S|\delta^2 k+(\delta k+k^2)t+m+n)$ & $O(m+n+\chi k)$ \\
        
        \hline
    \end{tabular}%
    }
    \vspace{1mm}
    \caption{Summary of in-memory $k$-clique counting algorithms ($\alpha$: arboricity; $C_k$: the number of $k$-clique; $\eta$: $h$-index; $\delta$: degeneracy; $\Delta$: the maximum degree; $c(G)$: core value of the graph $G$; $SCT(G)$: succinct clique tree of the graph $G$; $S$: subgraph; $\chi$: number of colors to color all nodes in the graph; $t$: sample size; $\tau$: number related to the maximum truss number of the graph; $\Delta$: maximum degree)}
    \label{Datasetl} 
    \vspace{-8mm}
\end{table}

\noindent
\textbf{\underline{Non-ordering-based clique counting.}} The first practical algorithm for listing $k$-cliques is proposed by Chiba and Nishizeki \cite{chiba1985arboricity}, building on the algorithm introduced by Tsukiyama et al. \cite{tsukiyama1977new}. Specifically, Tsukiyama et al. develop an algorithm for generating all maximal independent sets in a graph using a recursive backtracking technique to ensure that no duplicates are produced. Based on this approach, Chiba and Nishizeki adapt the recursive backtracking method for efficiently listing $k$-cliques. The algorithm begins by ordering the vertices to minimize redundant operations during $k$-clique generation. Then it recursively constructs $k$-cliques by starting with an empty set and incrementally adding vertices while ensuring that each addition preserves the $k$-clique property, that is, all selected vertices remain fully connected. At each step, the algorithm applies two key tests. The maximality test ensures that the generated set forms a maximal $k$-clique by verifying that adding any more vertices would break complete connectivity. Meanwhile, the lexicographical test prevents duplicate enumeration by retaining only the lexicographically largest $k$-clique among potential candidates. This combination of vertex ordering, recursive expansion, and duplicate elimination improves efficiency and guarantees that the algorithm generates unique results.

In \cite{makino2004new, takeaki2012implementation}, Makino and Uno propose two algorithms for enumerating all maximal cliques in a graph, which can be used to count $k$-cliques. For dense graphs, the proposed algorithm leverages matrix multiplication. Using the graph's adjacency matrix, it repeatedly performs matrix multiplications to reveal vertex relationships and identify potential cliques. Maximal cliques are then extracted by analyzing subsets of vertices that form complete subgraphs based on these matrix operations. For sparse graphs, the algorithm follows a different approach optimized for their structure. It begins with preprocessing, computing vertex degrees, and partitioning the graph into two subsets: $F_1$, which contains cliques involving high-degree vertices, and $F_2$, which includes the remaining cliques. The cliques in $F_1$ are precomputed to avoid redundancy, while those in $F_2$ are enumerated during traversal. A reverse search framework systematically constructs a tree rooted at the lexicographically largest maximal clique, generating child cliques by adding or removing vertices based on adjacency constraints. This approach ensures that all maximal cliques are enumerated without duplication. 


\noindent
\textbf{\underline{Degree-ordering-based clique counting.}} In \cite{li2020ordering}, Li et al. propose the {\em Degree} algorithm, which is based on degree ordering. The algorithm first constructs a {\em directed acyclic graph (DAG)} by orienting edges from nodes with lower degrees to those with higher degrees. It then recursively enumerates $k$-cliques by exploring the outgoing neighbors of each node in the \texttt{DAG}, ensuring that no duplicate cliques are listed. This approach leverages the h-index of the graph to bound the out-degree of nodes, effectively pruning the search space. Building on the \texttt{Degree} algorithm, Li et al. introduce an optimized degree-ordering-based algorithm, {\em DDegree}, which improves performance through a subgraph-based pruning strategy. After constructing the \texttt{DAG}, the algorithm induces a subgraph for each node using its outgoing neighbors. Instead of directly enumerating $k$-cliques, \texttt{DDegree} recursively processes these subgraphs to list ($k-1$)-cliques, reducing redundant computations.

Existing algorithms, such as \texttt{DDegree}, rely on hash-based methods for set intersections, which introduce significant overhead due to the need for maintaining induced subgraphs. To address these challenges, Long et al. propose {\em SDegree} \cite{yuan2022efficient}, which uses merge-based joins for set intersections. \texttt{SDegree} begins with a preprocessing phase designed to reduce the graph size and eliminate redundant computations. This phase incorporates two techniques: \texttt{Pre-Core}, which removes nodes that cannot form $k$-cliques based on degree constraints, and \texttt{Pre-List}, which directly identifies and outputs cliques where applicable. After preprocessing, \texttt{SDegree} constructs a \texttt{DAG} using degree ordering. The $k$-clique listing process is then executed recursively. For each node with a sufficient out-degree, \texttt{SDegree} initializes a candidate set with its out-neighbors and iteratively refines it using merge-based joins to identify larger cliques.


\noindent
\textbf{\underline{Degeneracy-ordering-based clique counting.}} In \cite{danisch2018listing}, building on the Chiba-Nishizeki algorithm \cite{chiba1985arboricity}, Danisch et al. propose the {\em kClist} algorithm, which incorporates several critical modifications to improve performance. The core idea of \texttt{kClist} is to use a \texttt{DAG} representation induced by the core-based degeneracy ordering of nodes. This ordering minimizes the graph’s maximum out-degree, aligning with its core value. Unlike the Chiba-Nishizeki algorithm, \texttt{kClist} eliminates the need to reorder nodes within recursive steps and avoids redundant clique enumeration by leveraging the \texttt{DAG} structure. The algorithm processes the subgraph induced by the out-neighbors of each node and recursively explores it until the desired clique size $k$ is reached.

Extending maximal clique enumeration techniques, Jain and Seshadhri propose a novel algorithm called {\em Pivoter} \cite{jain2020power}, which is based on degeneracy ordering. The key innovation lies in the construction of the {\em succinct clique tree (SCT)}, a compressed data structure that uniquely represents all cliques in the graph. The \texttt{SCT} is built using a pivoting technique, where a pivot vertex is selected at each recursive step to partition cliques into disjoint categories. The \texttt{Pivoter} algorithm operates in two main phases. In the first phase, the \texttt{SCT} is constructed using degeneracy ordering, simplifying the graph’s structure and enabling efficient recursive calls. In the second phase, the \texttt{SCT} is traversed to extract both global and local $k$-clique counts. This traversal leverages the unique encoding of cliques within the \texttt{SCT}, ensuring that each clique is processed exactly once.

Most existing $k$-clique listing algorithms rely on a vertex-oriented branching strategy, where a partial $k$-clique is expanded by adding one vertex at a time \cite{li2020ordering, yuan2022efficient}. While effective, these methods often incur high computational costs due to limited pruning capabilities. To address this limitation, Wang et al. propose an {\em edge-oriented branching (EBBkC)} framework for $k$-clique listing \cite{wang2024efficient}. Based on this framework, they introduce the {\em EBBKC-T} algorithm, which employs truss-based edge ordering to optimize branching and improve efficiency. Specifically, the algorithm expands $k$-cliques by selecting edges instead of individual vertices. At each branching step, it reduces the subgraph to the common neighbors of the two vertices in the selected edge. Edges are ordered based on their truss numbers, prioritizing those with fewer common neighbors. This truss-based edge ordering minimizes the size of subgraphs generated during recursion. By combining edge-oriented branching with truss-based ordering, the algorithm significantly reduces computational overhead and achieves better performance compared to traditional vertex-oriented methods.

\begin{figure}[t]
	\begin{center}
            \subfigure[A degree-based graph]{
			\label{clique1}	
			\includegraphics[scale=0.33]{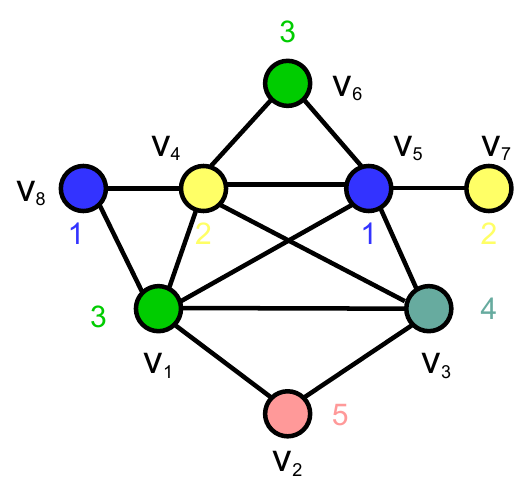}      
		}\hspace{0mm}
		\subfigure[A degree-based DAG]{
			\label{clique2}
			\includegraphics[scale=0.33]{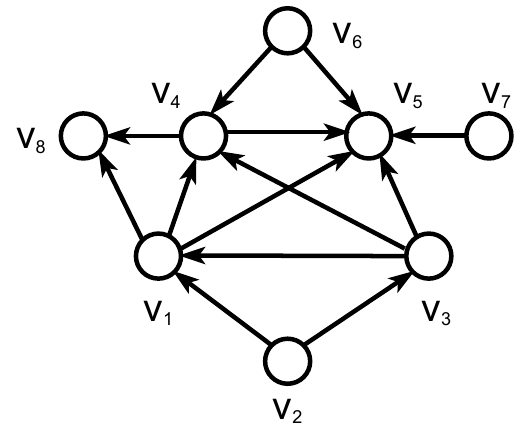}   
		}
            \subfigure[A degeneracy-based graph]{\hspace{3mm}
			\label{clique3}
			\includegraphics[scale=0.33]{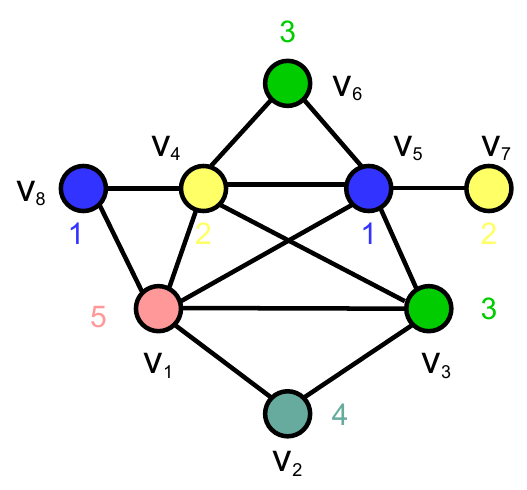}
		}
            \subfigure[A degeneracy-based DAG]{\hspace{2mm}
			\label{fclique4}
			\includegraphics[scale=0.33]{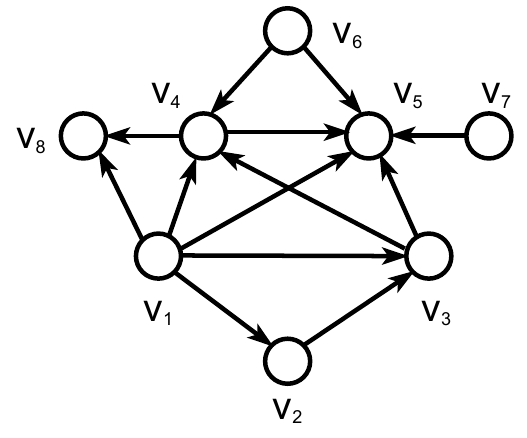}
		}
	\end{center}
        \vspace{-4.5mm}
	\caption{Illustration of the degree-based/degeneracy-based colored graph and its induced DAG.}
    
    \label{CliqueExample}
    \vspace{-6mm}
\end{figure}

\noindent
\textbf{\underline{Color-ordering-based clique counting.}} Building on the greedy graph coloring technique \cite{hasenplaugh2014ordering, yuan2017effective}, Li et al. propose color ordering and introduce three $k$-clique listing algorithms that leverage it for efficient pruning. {\em DegCol} assigns colors based on degree ordering, resolving ties by node ID, and constructs a color-ordered \texttt{DAG} (as shown in \autoref{CliqueExample}(a) and \autoref{CliqueExample}(b)), where each node’s outgoing neighbors have lower color values. A pruning strategy excludes nodes with insufficient color values for $k$-cliques, reducing computation. {\em DegenCol} improves \texttt{DegCol} by using degeneracy ordering from core decomposition, prioritizing denser regions (as shown in \autoref{CliqueExample}(c) and \autoref{CliqueExample}(d)), making it particularly effective for sparse graphs. {\em DDegCol} further optimizes \texttt{DegenCol} by refining the search space. Instead of applying $k$-clique listing to the entire graph, it first constructs a color-ordered \texttt{DAG} and, for each node, extracts the subgraph induced by its outgoing neighbors. Within this reduced search space, it applies \texttt{DegenCol} to enumerate $(k-1)$-cliques efficiently. This localized approach reduces redundant computations and improves scalability for large $k$.

In \cite{yuan2022efficient}, Long et al. extend their proposed \texttt{SDegree} algorithm by introducing {\em BitCol}, which integrates color ordering to enhance pruning effectiveness. \texttt{BitCol} first constructs a \texttt{DAG} using degeneracy ordering and generates induced subgraphs for each node. It then applies color ordering, derived from greedy graph coloring, to these induced subgraphs, assigning color values to vertices. Vertices whose color values prevent them from contributing to a complete $k$-clique are pruned early, reducing unnecessary exploration. To further accelerate set intersections, \texttt{BitCol} encodes adjacency lists into compact bitmap representations, enabling efficient intersection operations through bitwise computations. By leveraging these optimizations, \texttt{BitCol} achieves superior performance, particularly for large values of $k$, significantly reducing memory overhead and computational complexity compared to existing algorithms.

In \cite{wang2024efficient}, Wang et al. propose the {\em EBBkC-C} algorithm, which is based on the edge-oriented branching strategy and employs a color-based edge ordering to further enhance efficiency. Specifically, the algorithm begins by performing vertex coloring on the input graph, assigning each vertex the smallest color value not already assigned to its neighbors. Edges are then ordered based on the color values of their endpoints, prioritizing edges between lower-colored vertices for processing. At each branching step, the algorithm expands the partial $k$-clique by adding an edge and reduces the subgraph to the common neighbors of the edge's two vertices. To prune redundant branches, the algorithm applies two key rules: (1) a branch is pruned if the color value of either endpoint is insufficient to form the remaining $l$-clique, and (2) a branch is pruned if the vertices in the resulting subgraph have fewer than $l-2$ distinct color values. By combining edge-oriented branching with color-based edge ordering and advanced pruning mechanisms, the \texttt{EBBkC-C} algorithm achieves substantial improvements in the efficiency of $k$-clique listing.

\noindent
\textbf{\underline{Sampling-based clique counting.}} Traditional clique counting methods often become computationally infeasible due to combinatorial explosion. To address this challenge, Jain and Seshadhri propose the {\em Turán-shadow} algorithm \cite{jain2017fast}, a randomized approach based on extremal graph theory, specifically a strengthened version of \texttt{Turán}'s theorem \cite{turn1941extremal}. The algorithm introduces the concept of a \texttt{Turán} shadow, a collection of dense subgraphs that cover all possible $k$-cliques in the graph. By leveraging this shadow, it applies a sampling technique to estimate clique counts efficiently. To construct the \texttt{Turán} shadow, the algorithm employs degeneracy ordering to iteratively form subgraphs with high edge density, capturing regions of the original graph most likely to contain $k$-cliques. This process reduces the problem size while preserving essential structural information. Once the \texttt{Turán} shadow is built, the algorithm estimates the number of $k$-cliques by randomly selecting subgraphs and using an unbiased estimator to count cliques within them.

In \cite{eden2018approximating}, Eden et al. develop an efficient sublinear-time algorithm for approximating the number of $k$-cliques. Their algorithm provides a $(1 + \epsilon)$-approximation for $k$-clique counts using standard query types, including degree queries, neighbor queries, and pair queries. The core approach leverages random sampling while addressing the high variance that arises with larger cliques. To manage this variance effectively, vertices are classified as either sociable (highly connected) or shy (less connected). The algorithm focuses its sampling efforts on shy vertices to produce more uniform estimates. The process begins by selecting edges and attempting to extend them into $k$-cliques. For low-degree vertices, the algorithm applies simple rejection sampling to select random neighbors. For high-degree vertices, it manages complexity using an auxiliary set of sampled vertices, facilitating efficient neighbor selection while maintaining uniformity. By employing distinct strategies for different vertex types, the algorithm effectively handles both high- and low-degree vertices, making it scalable and computationally feasible.


To count $k$-cliques in large graphs, especially dense ones, Ye et al. propose a framework that combines exact algorithms with novel sampling-based techniques \cite{ye2022lightning, ye2023efficient2}. The framework adopts distinct methods tailored to different regions of the graph. For sparse regions, it applies the exact \texttt{Pivoter} algorithm \cite{jain2020power}, while for dense regions, where exact enumeration becomes prohibitively expensive, it introduces two sampling techniques: $k$-color set sampling and $k$-color path sampling. The $k$-color set sampling method, referred to as {\em DPColor}, begins by applying a graph coloring procedure that assigns different colors to adjacent nodes. Based on these color classes, the algorithm constructs $k$-color sets, each containing one node from a distinct color class, and estimates the number of $k$-cliques by sampling from these sets. To ensure efficiency and unbiased estimation, it uses a dynamic programming approach to uniformly sample valid color combinations. The second technique, $k$-color path sampling, called {\em DPColorPath}, further refines this strategy by sampling only those $k$-color sets that form connected subgraphs (i.e., $k$-color paths). By focusing on connectivity, \texttt{DPColorPath} increases the likelihood of encountering actual $k$-cliques, thereby improving estimation accuracy and reducing variance.

\subsubsection{\textbf{Heterogeneous and parallel algorithms for clique counting}}


Finocchi et al. propose the first scalable exact algorithm for counting $k$-cliques in large graphs using the MapReduce framework \cite{finocchi2015clique}. To avoid redundant computation, the algorithm imposes a global ordering on nodes and, for each node $u$, focuses only on its high-neighborhood: the neighbors that come later in the order. This ensures that each $k$-clique is counted exactly once. The algorithm runs in three MapReduce rounds: (1) computing high-neighborhoods for all nodes, (2) identifying which edges belong to the high-neighborhood subgraphs, and (3) locally enumerating ($k-1$)-cliques within each subgraph to count the corresponding $k$-cliques. To further enhance scalability, the authors introduce sampling-based estimators that probabilistically sample neighbor pairs during subgraph construction. These approximations maintain unbiased estimates while dramatically reducing computation time and memory usage. 

Danisch et al. propose a parallel version of the \texttt{kClist} algorithm in \cite{danisch2018listing}, introducing two key strategies: node-parallel and edge-parallel. The node-parallel approach distributes the workload by assigning the subgraph induced by each node's out-neighbors to a thread. While effective for a small number of threads, this method often encounters workload imbalance due to the uneven distribution of $k$-cliques across subgraphs. To address this issue, the edge-parallel strategy refines the parallelization granularity by assigning edges to threads. Each thread processes the subgraph induced by the common out-neighbors of the edge's endpoints, ensuring a more balanced distribution of computational tasks. By using the \texttt{DAG} structure of the input graph, the edge-parallel variant prevents redundant clique enumeration and achieves near-linear speedups.

In \cite{shi2021parallel}, Shi et al. design a new parallel algorithm for $k$-clique counting and peeling. The proposed {\em Arb-count} algorithm builds upon the classic Chiba-Nishizeki approach \cite{chiba1985arboricity}. A distinguishing feature of \texttt{Arb-count} is its use of low out-degree graph orientations, which reduce redundant computations by reordering vertices so that only neighbors with higher orientations are considered. This orientation enables efficient recursive pruning of candidate vertices, ensuring each recursive call intersects adjacency lists of bounded size. By adopting parallel primitives like prefix sums and parallel hash tables, \texttt{Arb-count} achieves a low computational span, making it both efficient and highly scalable. Additionally, the algorithm significantly reduces memory usage, requiring far less space per processor than existing methods.


Currently, most state-of-the-art parallel $k$-clique counting algorithms are designed for CPUs \cite{jain2020power, shi2021parallel}. In \cite{almasri2021k, almasri2022parallel}, Almasri et al. introduce a GPU-based algorithm to enhance the efficiency of dense subgraph analysis. 
This algorithm integrates graph orientation and pivoting and adapts them for GPU execution. 
To overcome the inherent inefficiencies of recursion on GPUs, the algorithm replaces recursive search tree traversal with an iterative method, utilizing a shared stack. 
A key innovation is the usage of binary encoding for induced subgraphs, which are critical for efficient adjacency list intersections. 
This encoding compresses adjacency lists into compact bit vectors, enabling high-speed bitwise operations during intersections and reducing memory consumption. The algorithm also incorporates sub-warp partitioning, a GPU-specific optimization that divides threads within a warp into smaller groups, enabling finer-grained parallelism and maximizing resource utilization. The implementation supports both vertex-centric and edge-centric parallelization schemes. The vertex-centric approach assigns entire trees to thread blocks, amortizing the cost of induced subgraph extraction across larger trees. Conversely, the edge-centric approach processes smaller subtrees, offering better load balancing for higher $k$-values.

\subsection{ \textbf{Cycle Counting}}

Counting cycles in graphs is a fundamental problem in network analysis, with applications in computational biology \cite{klamt2009computing} and social networks \cite{giscard2017evaluating}. Cycles are key structures that reveal hidden relationships, connectivity properties, and irregular behaviors in large-scale graphs. Existing cycle counting algorithms can be broadly classified into matrix-based and path-based approaches. Matrix-based algorithms leverage linear algebra techniques to count cycles. In contrast, path-based algorithms decompose cycles into smaller substructures, such as 2-paths or 3-paths, and use efficient enumeration or indexing techniques to count or list cycles directly. While matrix-based methods offer algebraic efficiency, they can be computationally expensive for large graphs. Path-based methods, on the other hand, often scale better for practical applications by focusing on local structures and reducing redundant computations. In this section, we review representative algorithms from both categories.

\noindent
\textbf{\underline{Matrix-based cycle counting algorithms.}} In \cite{alon1997finding}, Alon et al. present an efficient algorithm for counting simple cycles $C_k$ of a given length $k$ in both directed and undirected graphs. The key technique relies on matrix exponentiation: the trace of the $k$-th power of the adjacency matrix provides the number of closed walks of length $k$ in the graph. However, since these walks may revisit vertices, additional refinements are required to extract the count of simple cycles. To address this, the authors introduce the concept of $k$-cyclic graphs, which are homomorphic images of $C_k$-subgraphs that preserve cycle structures while allowing vertex contractions. Using this framework, they systematically subtract contributions from non-simple closed walks, isolating the exact number of $C_k$. By combining matrix exponentiation with combinatorial counting techniques, the algorithm significantly improves upon previous methods, offering an efficient approach for cycle enumeration in both sparse and dense graphs.

\noindent
\textbf{\underline{Path-based cycle counting algorithms.}} Abboud et al. address efficient listing of 4-cycles in a graph \cite{abboud2022listing}. First, they propose a baseline approach that iterates over all pairs of nodes, identifies 2-paths, and then checks for shared neighbors to form 4-cycles. While simple, this approach becomes inefficient for large graphs due to the excessive number of 2-paths that need to be considered. To improve efficiency, they introduce a refined method that partitions nodes into low-degree ($L$) and high-degree ($H$) categories, allowing the algorithm to focus on the most relevant 2-paths for forming 4-cycles. A key insight behind this optimization is that the distribution of these 2-paths follows a structured pattern, which enables a more targeted enumeration strategy. By leveraging this structure, the algorithm efficiently avoids unnecessary computations and significantly reduces runtime, making it scalable to larger graphs.

In \cite{burkhardt2023simple}, Burkhardt and Harris further optimize the algorithm for counting 4-cycles. By replacing randomized hashing with array-based operations, their algorithm ensures predictable performance and reduces memory overhead. The key innovation is the use of the average degeneracy parameter $\delta(G)$, which is often smaller than other sparsity measures like arboricity or core number, leading to improved performance on large-scale graphs. The method efficiently processes each vertex by iterating through its neighbors and using an array-based structure to track 2-paths, avoiding unnecessary computations. This enables efficient 4-cycle counting per graph, per vertex, and per edge, as well as explicit enumeration of 4-cycles.

Jin et al. \cite{jin2024listing} extend prior work on 4-cycle listing, where the algorithm stores 2-paths in tables indexed by their endpoints. Their new algorithm for listing 6-cycles adopts this idea of indexing intermediate structures but instead stores 3-paths. 
A 6-cycle is formed by combining two 3-paths, ($a \rightarrow b_1 \rightarrow c_1 \rightarrow d$) and ($a \rightarrow b_2 \rightarrow c_2 \rightarrow d$), while ensuring that $b_1 \neq b_2$ and $c_1 \neq c_2$ to avoid degenerate cases. The method consists of two key stages: first, it precomputes adjacency relationships and constructs efficient lookup tables; then, it systematically enumerates valid 6-cycles using these precomputed structures. Additionally, the paper introduces a refinement of this approach for listing exactly $t$ 6-cycles. By performing a binary search over subgraphs and applying a filtering step, the algorithm efficiently extracts exactly $t$ cycles without unnecessary computation.

%% file: Content/4_bipartite_graph.tex
\section{\textbf{BIPARTITE GRAPHS}}
\label{bipartite graph}


Bipartite graphs have gained significant attention in recent years due to their ability to effectively model relationships between two distinct sets of objects. As a specialized form of heterogeneous graph, a bipartite graph consists of two types of vertices, with edges existing only between vertices of different types. In many real-world applications \cite{borgatti1997network,latapy2008basic}, bipartite graphs naturally emerge as a data model. For instance, in online shopping platforms like Amazon and Alibaba, purchase interactions between users and products can be represented as a bipartite graph, where users form one layer, products form the other, and connections between them represent purchase records. A key aspect of bipartite graph analysis is motif discovery, which enhances the effectiveness of data mining in such structures. Among the most commonly studied motifs are bitriangles \cite{opsahl2013triadic, yang2021efficient, zhang2023scalable}, bicliques \cite{ye2023efficient, yang2021p, qiu2024accelerating, wang2022efficient, wang2023efficient}, and butterflies \cite{wang2014rectangle, sanei2018butterfly, wang2019vertex, zhou2021butterfly, zhou2023butterfly, sheshbolouki2022sgrapp, sanei2019fleet}, which are a special case of bicliques (i.e., a complete $2 \times 2$ biclique).

\begin{table}[t]
    \centering
    \resizebox{1\textwidth}{!}{%
    \begin{tabular}{|c|c|c|c|}
        \hline
        \multicolumn{1}{|c|}{\cellcolor{gray!25}\textbf{Pattern}} & 
        \multicolumn{1}{c|}{\cellcolor{gray!25}\textbf{Category}} & 
        \multicolumn{1}{c|}{\cellcolor{gray!25}\textbf{Type}} & 
        \multicolumn{1}{c|}{\cellcolor{gray!25}\textbf{Algorithms}}  \\
        \hline
        \multirow{5}{*}{Butterfly} & \multirow{3}{*}{In-memory}  & \multirow{2}{*}{Exact algorithms} & Wang-Exact \cite{wang2014rectangle}, ExactBFC \cite{sanei2018butterfly}, BFC-VP \cite{wang2019vertex, wang2023accelerated}\\
        & & & BFC-VP++ \cite{wang2019vertex, wang2023accelerated} \\
         \cline{3-4}
        & & \multirow{1}{*}{Approximate algorithms} & ESpar \cite{sanei2018butterfly}, ClrSpar \cite{sanei2018butterfly}, Zhang \cite{zhang2023scalable} \\
        \cline{2-4}
         & \multirow{2}{*}{Heterogeneous and parallel} & & PAR-rect \cite{wang2014rectangle}, MR-rect \cite{wang2014rectangle}, ParButterfly \cite{shi2022parallel}\\
         & & & G-BFC \cite{xu2022efficient, xia2024gpu}, IOBufs \cite{wang2024parallelization, wang2023efficient2}\\
        \hline
        \multirow{2}{*}{Bitriangle} & \multirow{2}{*}{In-memory} & Exact algorithms & WJ-Count \cite{yang2021efficient}, SWJ-Count \cite{yang2021efficient}, RSWJ-Count \cite{yang2021efficient}\\
        \cline{3-4}
         & & Approximate algorithms & Zhang \cite{zhang2023scalable}\\
        \hline
        \multirow{3}{*}{Biclique} & \multirow{2}{*}{In-memory}  & Exact algorithms & BCList \cite{yang2021p}, BCList++\cite{yang2021p}, EPivoter \cite{ye2023efficient}\\
         \cline{3-4}
        & & \multirow{1}{*}{Approximate algorithms} & ZigZag \cite{ye2023efficient}, ZigZag++ \cite{ye2023efficient} \\
        \cline{2-4}
         & \multirow{1}{*}{Heterogeneous and parallel} & & GBC \cite{qiu2024accelerating}\\
        \hline
    \end{tabular}%
    }
    \vspace{1mm}
    \caption{Summary of all the motif counting algorithms in bipartite graphs}
    \label{DatasetB} 
    \vspace{-8mm}
\end{table}

\subsection{\textbf{Butterfly Counting}}
 
Butterfly counting in bipartite graphs has become a key focus in network analysis due to its importance in understanding the structural properties of complex systems and its wide range of applications. Despite its simplicity, this motif proves to be a powerful tool for capturing intricate interaction patterns, making it indispensable in areas such as host-parasite network analysis \cite{walker2017uncertain}, viral marketing \cite{fain2006sponsored, liu2019efficient, wang2018efficient}, and other fields \cite{aksoy2017measuring, lind2005cycles, opsahl2013triadic, fang2020survey, peng2018efficient, zhang2018discovering, zou2016bitruss}. In this section, we review existing in-memory algorithms as well as heterogeneous and parallel algorithms for butterfly counting.


\begin{definition} 
[{\bf Butterfly}] Given a bipartite graph $G=(V=(U, L), E)$ and the four vertices $u,v,w,x \in V$ where $u, w \in U$ and $v, x \in L$, $u,v,w$ and $x$ form a butterfly if and only if $u$ and $w$ are all connected to $v$ and $x$, respectively, by edges.
\end{definition}

\subsubsection{\textbf{In-memory butterfly counting algorithms}}
In-memory algorithms for butterfly counting are designed to efficiently detect the butterfly in bipartite graphs. These techniques can be broadly classified into exact methods, which provide precise counts, and approximate methods, which employ sampling or sketching techniques to estimate butterfly counts with lower computational overhead. In this section, we review these in-memory approaches, examining their computational characteristics, efficiency trade-offs, and practical applications.

\noindent
\textbf{\underline{Exact algorithms for butterfly counting.}} In \cite{wang2014rectangle}, Wang et al. introduce the in-memory and the I/O-efficient algorithm for butterfly counting. The in-memory algorithm is designed for small to medium-sized bipartite graphs that can fit into main memory. It processes each vertex sequentially, using adjacency lists to compute the butterflies that involve the vertex. The algorithm identifies two-hop neighbors of each vertex and uses them to count butterflies efficiently. The I/O-efficient algorithm, in contrast, addresses the limitations of the in-memory approach when dealing with large graphs that cannot fit into memory. It partitions the vertex set into smaller subsets, constructs neighborhood subgraphs for each partition, and processes them sequentially. By minimizing random disk access and utilizing sequential scans, the algorithm effectively reduces I/O costs.

Although Wang et al.'s algorithm can accurately count the number of butterflies, it always starts computations from a fixed vertex partition, leading to longer runtimes. To address this limitation, Sanei-Mehri et al. present the exact butterfly counting algorithm called {\em ExactBFC} \cite{sanei2018butterfly}, which dynamically determines the optimal starting partition based on the sum of squared vertex degrees. Starting with the partition with the smaller sum, \texttt{ExactBFC} significantly reduces the total number of computations of two-hop neighbors, the most computationally expensive step. Furthermore, \texttt{ExactBFC} avoids redundant set intersections by using a hash map to efficiently count and store two-hop paths, thereby reducing both computational overhead and memory usage compared to Wang et al.'s approach, which relies on repeated intersection operations.

The existing algorithms \cite{wang2014rectangle, sanei2018butterfly} rely on layer-based approaches that prioritize processing entire vertex layers. However, these algorithms often suffer from inefficiencies as they fail to minimize unnecessary wedge enumerations. Therefore, Wang et al. propose a novel {\em vertex priority-based butterfly counting (BFC-VP)} algorithm, along with its cache-optimized extension {\em BFC-VP++}, to address these limitations \cite{wang2019vertex, wang2023accelerated}. The \texttt{BFC-VP} algorithm assigns priorities to individual vertices based on their degrees. Vertices with higher degrees are processed first as start vertices, ensuring that wedge enumeration focuses only on necessary subsets of the graph. Specifically, the algorithm processes each edge by selecting the vertex with a higher priority as the start vertex and restricting enumeration to wedges where both middle and end vertices have lower priorities. This design avoids redundant computations and reduces the number of processed wedges. To further enhance performance, the paper introduces \texttt{BFC-VP++}, a cache-aware optimization of \texttt{BFC-VP} that improves CPU memory access efficiency. \texttt{BFC-VP++} employs two key strategies: cache-aware wedge processing and cache-aware graph projection. The wedge processing strategy ensures that high-priority vertices are accessed more frequently as end vertices, enhancing memory locality and reducing cache misses. The graph projection strategy reorders vertex IDs based on their priorities, grouping high-priority vertices closer together, which further improves cache performance.

\begin{figure}[t]
	\begin{center}
            \subfigure[A bipartite graph]{
			\label{fig5.1}	
			\includegraphics[scale=0.33]{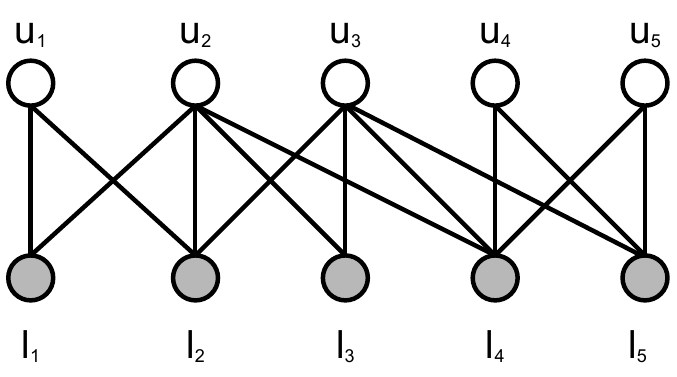}      
		}
		\subfigure[A butterfly]{\hspace{2mm}
			\label{fig5.2}
			\includegraphics[scale=0.33]{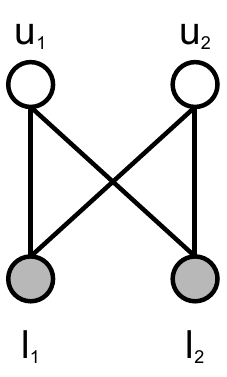}   
		}\hspace{2mm}
            \subfigure[A bitriangle]{
			\label{fig5.3}
			\includegraphics[scale=0.33]{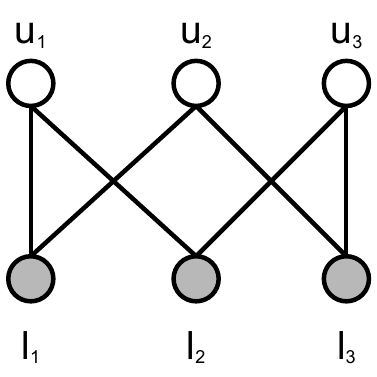}
		}\hspace{2mm}
            \subfigure[A $(4,2)$-biclique]{
			\label{fig5.4}
			\includegraphics[scale=0.33]{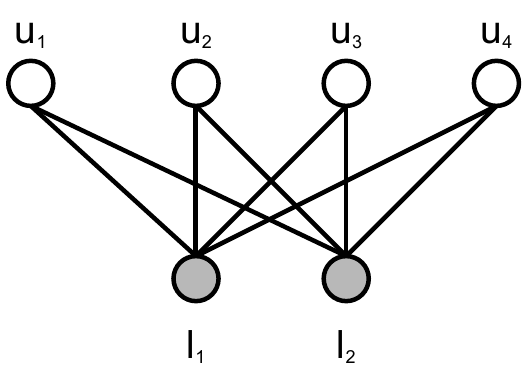}
		}
	\end{center}
        \vspace{-4.5mm}
	\caption{Examples of a bipartite, butterfly, bitriangle, and biclique.}
    
    \label{BipartiteExample}
    \vspace{-6mm}
\end{figure}

\noindent
\textbf{\underline{Approximate algorithms for butterfly counting.}} Seyed-Vahid et al. propose a suite of randomized algorithms \cite{sanei2018butterfly} to efficiently estimate the number of butterflies in large-scale bipartite graphs. They introduce two main approaches: local sampling and one-shot sparsification. Local sampling methods include vertex sampling, edge sampling, and wedge sampling, which rely on sampling subgraphs centered around specific graph elements—such as vertices, edges, or wedges—to estimate butterfly counts. In contrast, one-shot sparsification methods reduce the graph globally. The {\em edge sparsification algorithm (ESpar)} retains each edge with a certain probability and then scales the butterfly count based on the sparsified graph. A variant, {\em colorful sparsification (ClrSpar)}, incorporates dependencies between sampled edges by using vertex coloring, which enhances the likelihood of preserving dense structures like butterflies. While {\tt ClrSpar} demonstrates potential in maintaining dense motifs, its variance is generally higher than that of {\tt ESpar} due to the complex dependencies between butterfly pairs.

In \cite{zhang2023scalable}, Zhang et al. propose a novel one-sided weighted sampling algorithm, which leverages the unique structure of bipartite networks to achieve higher efficiency and accuracy compared to Seyed-Vahid et al.’s algorithm \cite{sanei2018butterfly}. The algorithm focuses on sampling pairs of vertices from one side of the bipartite graph (e.g., the left vertex set $L$) and estimates the butterfly count by analyzing their shared neighbors. Specifically, given two sampled vertices $u$ and $v$ from $L$, the algorithm computes the intersection of their neighbor sets on the opposite side (i.e., $R$), and the butterfly count for the pair is derived from the size of this intersection. To enhance efficiency, the algorithm employs a weighted sampling strategy, where vertices with higher degrees are selected with greater probability, as they are more likely to form butterflies. This prioritizes regions of the graph with higher butterfly densities. The proposed method calculates an unbiased estimate of the total butterfly count by scaling the local counts with a normalization factor that accounts for the sampling probabilities. Furthermore, to reduce variance, multiple independent sampling runs are performed, and their results are averaged.

\subsubsection{\textbf{Heterogeneous and parallel algorithms for butterfly counting}}

The in-memory butterfly counting algorithm is highly efficient for small- to medium-sized bipartite graphs, but it faces significant limitations when handling large datasets. To address these challenges, Wang et al. propose two distributed approaches: a partition-based parallel algorithm called {\em PAR-rect} and a MapReduce-based algorithm called {\em MR-rect} \cite{wang2014rectangle}. The partition-based algorithm divides the left vertex set $L_G$ of the bipartite graph into $p$ disjoint partitions, each assigned to a separate machine. For each partition $L_i$, a subgraph $NG(R_i)$ containing all edges incident to vertices in $R_i$ is constructed and processed using an in-memory algorithm to count butterflies involving vertices in $L_i$. In contrast, the MapReduce algorithm employs two rounds of Map and Reduce operations. The first round maps vertex pairs at two-hop distances via shared neighbors and aggregates adjacency information to compute intermediate results. The second round combines these results to calculate the butterfly counts.

Although Wang et al. propose a simple parallel algorithm \cite{wang2014rectangle}, it is not work-efficient. Therefore, Shi and Shun introduce the {\em ParButterfly} framework \cite{shi2022parallel}, a collection of parallel algorithms designed to efficiently process butterflies. The framework addresses tasks such as global counting, per-vertex counting, and per-edge counting, as well as vertex peeling and edge peeling. The key idea lies in aggregating wedges incident on subsets of vertices, with multiple methods for wedge aggregation, including sorting, hashing, histogramming, and batching. To optimize performance, the framework employs various vertex ranking strategies, such as side ordering, degree ordering, and complement coreness ordering, which reduce the work required for counting by focusing on specific vertex orders. Additionally, the framework supports both exact and approximate butterfly counting, using graph sparsification to handle large-scale graphs efficiently. For peeling tasks, it iteratively removes vertices or edges with the minimum butterfly count, updating counts in parallel. These algorithms match the work of the best-known sequential methods while achieving high parallel scalability.

Existing algorithms, while supporting parallelism on CPUs, face significant inefficiencies when applied to GPUs due to challenges such as synchronization overhead, workload imbalance, and the high computational cost of traversing two-hop paths. To address these issues, Xu et al. propose a GPU-based algorithm called {\em G-BFC} \cite{xu2022efficient, xia2024gpu}. \texttt{G-BFC} introduces three core innovations: a lock-free design to minimize synchronization costs, an adaptive workload partitioning strategy to balance computational loads across GPU threads, and a preprocessing phase that reduces the number of traversed wedges by employing an edge direction optimization. The lock-free design strategically reorganizes operations to eliminate serial dependencies, using GPU shared memory to significantly accelerate computations. The adaptive workload balancing dynamically adjusts thread and kernel allocations based on adjacency list lengths, effectively mitigating imbalances caused by irregular graph structures. Meanwhile, the preprocessing step reduces graph traversal overhead by implementing a novel edge direction strategy that prioritizes high-degree vertices and eliminates redundant computations.

Wang et al. propose a novel algorithmic framework, {\em IOBufs}, for butterfly counting \cite{wang2024parallelization, wang2023efficient2}, addressing the challenges of super-linear time complexity and large memory consumption through hierarchical memory configurations. This framework combines fast but limited main memory with slower, larger secondary storage. Unlike existing algorithms, which fail to achieve optimal I/O complexity, \texttt{IOBufs} introduces a semi-witnessing algorithm that minimizes I/O costs by focusing on essential substructures rather than observing the entire motif. This innovation enables IOBufs to dynamically adapt to graph density with two specialized variants: \texttt{IOBufs-edge}, tailored for sparse graphs by maintaining edges in memory, and \texttt{IOBufs-wedge}, optimized for dense graphs by prioritizing wedges. To further improve scalability and computational efficiency, the authors present \texttt{PIOBufs}, a parallelized extension of \texttt{IOBufs} designed for modern multi-core CPUs and GPUs. By employing fine-grained task division, \texttt{PIOBufs} ensures effective workload distribution while preserving I/O efficiency. On GPUs, the framework leverages warp-level primitives and hierarchical parallelism to optimize computational throughput and memory access.








\subsection{\textbf{Bitriangle Counting}}

In general graphs, triangles are among the most fundamental structures, formed by closing a wedge with an additional edge. However, this concept does not apply to bipartite graphs, as triangles cannot exist in such structures. The concept of the bitriangle is first introduced by Opsahl et al. in \cite{opsahl2013triadic}, where they propose a structure known as a 6-cycle. This structure consists of a cycle in which three vertices belong to one vertex set and three to the other in a two-mode network. Subsequent researchers refer to their proposed model as the bitriangle and develop counting algorithms based on this concept.

\begin{definition} 
[{\bf Bitriangle}] Given a bipartite graph $G=(V=(U \cup L), E)$, a bi-triangle is 6-cycle, or a cycle with a length of 6 having three vertices in $U$ and three vertices in $L$.
\end{definition}  

In \cite{yang2021efficient}, Yang et al. propose three bitriangle counting algorithms: {\em WJ-Count}, {\em SWJ-Count}, and {\em RSWJ-Count}, each of which is progressively optimized to improve counting speed and efficiency. The baseline algorithm \texttt{WJ-Count} represents each bitriangle as a combination of three wedges. It operates by iterating over each vertex and enumerating its 2-hop neighbors to form wedges, then counts bitriangles by identifying combinations of these wedges that form a 6-cycle. However, \texttt{WJ-Count} suffers from significant redundancy, as it recalculates neighbor intersections multiple times. To address this, they propose the \texttt{SWJ-Count} algorithm, which represents bitriangles as the join of super-wedges (3-edge paths) rather than individual wedges. By counting these super-wedges to find bitriangles, the algorithm can avoid the unnecessary enumeration of vertex triplets. Nevertheless, the algorithm must still handle complex invalid cases where certain super-wedge combinations do not form valid bitriangles. To further optimize bitriangle counting, the authors introduce the \texttt{RSWJ-Count} algorithm, which incorporates a vertex ranking system based on vertex degree. Additionally, to avoid duplicate counting for each bitriangle, the authors define ranked super-wedge units to ensure that each bitriangle is counted only once. In addition to the global bitriangle counting problem, the authors also address local bitriangle counting for individual vertices or edges, which can be used to calculate local clustering coefficients.

In \cite{zhang2023scalable}, Zhang et al. introduce a novel approach for approximate bitriangle counting, extending the sampling techniques originally developed for butterfly counting. Specifically, they propose a weighted one-sided triple sampling algorithm that samples three vertices, $u$, $v$, and $w$, from one side of the bipartite graph based on their degrees, where vertices with higher degrees have a greater chance of being selected.  The algorithm then uses the neighbor lists of these vertices to form intersection sets, including a common intersection and three pairwise intersections. By combining the sizes of these intersection sets, the algorithm estimates the total bitriangles involving the selected vertices.



\subsection{\textbf{Biclique Counting}}

While the butterfly, as a special case of the $(p,q)$-biclique, has received considerable attention, many graph-based tasks require more general forms of $(p,q)$-bicliques where $p$ and $q$ are not fixed to 2. For example, in the context of the aggregation of information from the {\em graph geural network (GNN)}, the enumeration of $(p,q)$-bicliques can significantly optimize \texttt{GNN} operations. This is because $(p,q)$-bicliques inherently represent a dense connection pattern between sets of vertices, thereby minimizing the need for repetitive information exchange. In this section, we introduce existing algorithms for counting $(p,q)$-bicliques.

\begin{definition} 
[{\bf $(p,q)$-biclique}] Given a bipartite graph $G=(V=(U \cup L), E)$, a biclique $B(N,R)$ is a complete subgraph of $G$, where $N \subseteq U$, $R \subseteq L$, that is $\forall (u, v) \in N \times R$, $(u, v) \in E$. $B(N, R)$ is called a $(p,q)$-biclique if $|N|=p$ and $|R|=q$.
\end{definition}

The first work to systematically study the problem of $(p,q)$-biclique counting and enumeration is proposed by Yang et al. \cite{yang2021p}. They introduce two methods: a baseline algorithm called {\em BCList} and an advanced version, {\em BCList++}. \texttt{BCList} utilizes a branch-and-bound framework with depth-first exploration and incorporates pruning techniques, such as 2-hop neighbors and vertex ordering, to reduce computational overhead. However, its complexity is exponential in $p+q$, making it inefficient for large values of these parameters. \texttt{BCList++} improves upon this by adopting a layer-based exploration strategy that anchors the search to only one partition, reducing complexity to exponential in $p$ or $q$ alone. It also introduces a cost model to dynamically select the optimal partition for anchoring, while using pre-allocated arrays and vertex labeling to minimize the overhead of creating intermediate subgraphs. Additionally, \texttt{BCList++} is optimized through parallelization to enable efficient processing of large datasets.

Although Yang et al.'s algorithm can counts $(p,q)$-bicliques \cite{yang2021p}, it requires enumerating all $(p,q)$-bicliques to obtain the count, which becomes computationally expensive for large $p$ and $q$. Additionally, it is primarily designed for counting $(p,q)$-bicliques for a single pair of $(p,q)$, making it impractical for counting bicliques across all possible pairs of $(p,q)$. To address these limitations, Ye et al. propose a series of innovative algorithms \cite{ye2023efficient}. The {\em EPivoter} algorithm employs an edge-pivoting technique to avoid exhaustive enumeration by representing bicliques combinatorially within larger structures. Furthermore, they introduce two sampling-based algorithms, {\em ZigZag} and {\em ZigZag++}, which use a novel $h$-zigzag dynamic programming method for efficient and accurate biclique estimation, particularly in dense graph regions. To further enhance efficiency, a hybrid framework combines \texttt{EPivoter} for sparse graph regions with the sampling-based algorithms for dense regions, optimizing performance across different graph densities. These advancements result in significant improvements in both speed and scalability.

Existing algorithms face significant challenges, including inefficiencies in set intersection computations, which dominate runtime due to redundant comparisons and high memory access overhead. Moreover, the backtracking-based exploration strategies commonly used result in low thread utilization on GPUs, as the diminishing size of intermediate results at deeper search levels leaves many threads idle. To address these issues, Qiu et al. propose the {\em GBC} algorithm \cite{qiu2024accelerating}, which employs a {\em hierarchical truncated bitmap (HTB)} to optimize set intersection computations by using bitwise operations to reduce redundant comparisons and memory transactions. It introduces a hybrid {\em DFS-BFS} exploration strategy that combines the global efficiency of depth-first search with the local parallelism of breadth-first search to improve thread utilization and memory bandwidth efficiency. Additionally, the algorithm incorporates a composite load-balancing mechanism that combines pre-runtime task allocation with runtime task stealing, ensuring an equitable distribution of workloads across GPU threads.





%% file: Content/5_heterogeneous_graph.tex
\section{\textbf{OTHER HETEROGENEOUS GRAPHS}}
\label{heterogeneous graph}

\begin{figure}[t]
	\begin{center}
            \subfigure[A heterogeneous graph $G$]{
			\label{fig4.1}	
			\includegraphics[scale=0.36]{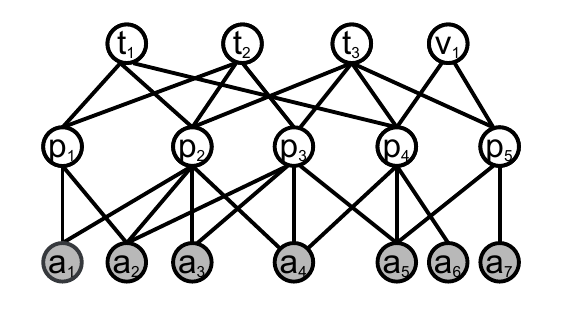}      
		}\hspace{0mm}
		\subfigure[A b-triangle and a c-triangle]{
			\label{fig4.2}
			\includegraphics[scale=0.36]{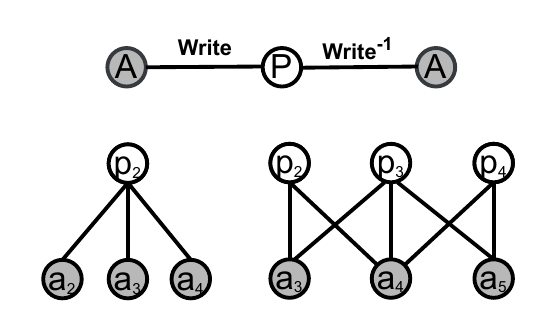}   
		}
            \subfigure[A induced graph]{\hspace{3mm}
			\label{fig4.3}
			\includegraphics[scale=0.4]{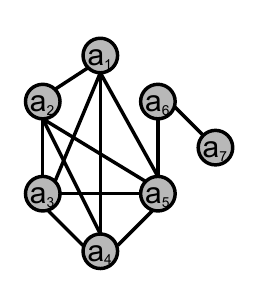}
		}\hspace{-1mm}
            \subfigure[Typed triangles ($|\mathcal{T}_{V}|=3$)]{
			\label{fig4.4}
			\includegraphics[scale=0.30]{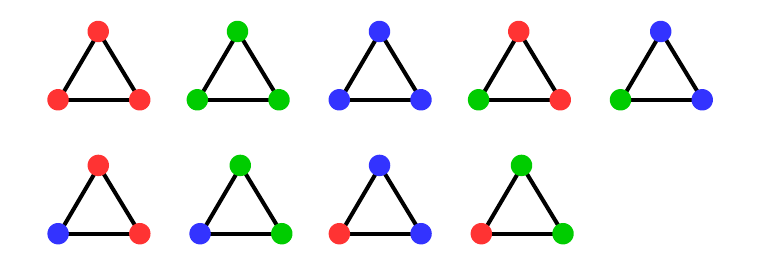}
		}
	\end{center}
        \vspace{-4.5mm}
	\caption{A heterogeneous graph $G$ and its induced graph as well as examples of all the triangles.}
    
    \label{HeterogeneousgraphExample}
    \vspace{-2mm}
\end{figure}

Heterogeneous graphs contain rich information by representing structural relationships (edges) among diverse types of vertices while incorporating unstructured content linked to each vertex. 
This characteristic makes them an important tool for capturing complex interactions across various real-world domains, such as human interactions \cite{kong2013inferring}, neural networks \cite{bassett2006small, bullmore2009complex}, computer networks \cite{rossi2013multi}, infrastructure systems \cite{wang2009cascade}, and economic systems \cite{schweitzer2009economic}. 
Currently, motifs in heterogeneous graphs can be classified into meta-path-based motifs and typed motifs. 
Meta-path-based motifs are constructed based on a predefined meta-path, capturing connectivity patterns along specific sequences of vertex types. On the other hand, typed motifs explicitly consider vertex types in addition to structural patterns, ensuring that vertex type constraints are preserved when identifying recurring subgraph structures. 
Due to their structural differences, the counting algorithms for these two types of motifs also vary. 
In this section, we explore the counting algorithms for meta-path-based and typed motifs.



\subsection{\textbf{Meta-path-based motifs}}
In heterogeneous graphs, vertices of the same type may not be directly connected. However, in certain scenarios, such as community search \cite{fang2020effective} or community detection \cite{luo2021detecting}, researchers often focus on the relationships among vertices of a specific type. To link these vertices, the concept of a meta-path is proposed, which helps establish connections between vertices of the same type.

\begin{definition}
[{\bf Meta-path}] A meta-path $P$ is a path defined on a heterogeneous graph schema $S_G=(\mathcal{T}_{V}, \mathcal{T}_{E})$, and is denoted in the form $\mathcal{T}_{V}^1 \xrightarrow{\mathcal{T}_{E}^1} \mathcal{T}_{V}^2 \xrightarrow{\mathcal{T}_{E}^2} \cdots \xrightarrow{\mathcal{T}_{E}^1} \mathcal{T}_{V}^{l+1}$ where $l$ is the length of $P$, $\mathcal{T}_{V}^i \in \mathcal{T}_{V}$, and $\mathcal{T}_{E}^i \in \mathcal{T}_E$ $(1 \leq i \leq l)$.
\end{definition}

\begin{definition}
[{\bf B-triangle}] Given a heterogeneous graph $G=(V,E)$ and a meta-path $P$, three vertices $(u,v,w)$ form a b-triangle if any two of them are connected by a path instance of $P$.
\end{definition}

\begin{definition}
[{\bf C-triangle}] Given a heterogeneous graph $G=(V,E)$ and a meta-path $P$, three vertices $(u,v,w)$ form a c-triangle, if there exist three instances of $P$, such that each connects a pair of vertices and they do not share any vertex except $u$, $v$, and $w$.
\end{definition}

In \cite{yang2020effective}, Yang et al. propose the concepts of {\em basic-triangle (b-triangle)} and {\em circle-triangle (c-triangle)} to enable community search in heterogeneous graphs (as shown in \autoref{HeterogeneousgraphExample}(b), which illustrates examples of these two triangles). 
They also introduce corresponding decomposition algorithms, which can be extended to count \texttt{b-triangles} and \texttt{c-triangles}. To identify all \texttt{b-triangles}, they first convert the heterogeneous graph into a homogeneous graph. Specifically, given a meta-path $P$, all target vertices are extracted. A homogeneous graph is then constructed by connecting any two target vertices that are linked by a path instance of $P$ in the original graph. For example, \autoref{HeterogeneousgraphExample}(c) shows an induced graph derived from \autoref{HeterogeneousgraphExample}(a) based on the meta-path author-paper-author. Finally, existing algorithms \cite{itai1977finding, burkhardt2017graphing, schank2005finding} are applied to detect \texttt{b-triangles} in the induced homogeneous graph.

While a \texttt{b-triangle} captures the relationships among three target vertices, its formulation may lead to weak cohesiveness. For example, if all path instances pass through a common vertex, removing this vertex would disconnect the triangle. To address this limitation, Yang et al. introduce \texttt{c-triangles} to offer stronger connectivity among target vertices than \texttt{b-triangles}. To identify all \texttt{c-triangles}, Yang et al. propose a verification method that first detects all \texttt{b-triangles}. The algorithm then employs either a depth-first search or a more efficient bidirectional search to verify whether three independent paths, corresponding to the given meta-path, form a circular connection. These paths must not share any intermediate vertices, ensuring that each edge remains distinct and uniquely contributes to the triangle’s formation.

\subsection{\textbf{Typed motifs}}

The existing meta-path-based motifs are primarily used for community search in heterogeneous graphs, where the focus is on vertices of the same type. As a result, these motifs are typically built on symmetric meta-paths. However, this design limits the corresponding algorithms' ability to capture information from multiple vertex types. To address this limitation, researchers introduce the concept of typed motifs, which not only define motifs but also incorporate vertex types and positions into their structure. \autoref{HeterogeneousgraphExample}(d) presents all the typed triangles when there are three vertex types in the heterogeneous graph.

\begin{definition} 
[{\bf Typed motifs}] Given a heterogeneous graph $G=(V, E, \phi, \xi)$, a typed motif is a connected induced heterogeneous subgraph $H=(V', E', \phi', \xi')$ of $G$ such that (1) $(V', E')$ is a motif of $(V, E)$, (2) $\phi'=\phi | V'$, that is, $\phi'$ is the restriction of $\phi$ to $V'$, (3) $\xi'=\xi | E'$, that is, $\xi'$ is the restriction of $\xi$ to $E'$.
\end{definition}

In \cite{rossi2019heterogeneous, rossi2020heterogeneous}, Rossi et al. propose a framework to count typed motifs in heterogeneous graphs. The framework begins by tackling the local counting problem, where for each edge $(i, j)$ in the graph, it computes the counts of typed motifs involving vertices $i$ and $j$, using typed neighbor sets, typed triangles, and typed stars to capture various connectivity patterns around the edge. To derive higher-order typed motifs from lower-order ones, the algorithm uses non-trivial combinatorial relationships, breaking down complex motifs into simpler components, which improves efficiency. To enable efficient look-ups and updates, a hash function is introduced for uniquely identifying and storing typed motifs, while a sparse storage format records only nonzero motif counts. By aggregating local counts and applying constraints to prevent redundant counting, they also produce an algorithm for accurate global counts of typed motifs. To enhance scalability, the framework incorporates a parallelization strategy that distributes the counting tasks among workers using a task queue and global broadcast channel. Each worker independently processes edges, utilizing the same hash function to maintain consistency without requiring locks. This edge-centric approach ensures better load balancing and allows efficient counting in large networks.

    

In \cite{hu2019discovering}, Hu et al. introduce the concept of a motif-clique, which generalizes traditional cliques by defining completeness not merely by edge adjacency but through user-defined typed motifs that specify the type and structure of vertex relationships. Extending this concept, they formulate the {\em maximal motif-clique enumeration (MMCE)} problem, which aims to identify all maximal motif-cliques that are not contained within any larger motif-clique. To address this complex problem, they develop the {\em maximal motif-clique enumeraTion algorithm (META)}. In detail, the \texttt{META} algorithm builds on the foundational {\em Bron-Kerbosch (BK)} framework \cite{bron1973algorithm}, which is traditionally used for maximal clique enumeration in general graphs, and extends it to handle the more complex setting of heterogeneous graphs with motif-based cliques. \texttt{META} first identifies initial motif embeddings through subgraph isomorphism techniques, using these as starting points for motif-clique discovery. For efficient vertex expansion, the algorithm incorporates dominance relationships, selectively expanding only promising candidates. Additionally, \texttt{META} introduces a custom early stop pruning strategy specific to motif-cliques, allowing it to halt expansions when further growth is unlikely to yield new maximal motif-cliques. Finally, \texttt{META} addresses the challenge of duplicate motif-cliques by using a set-trie structure to prevent redundant computation, ensuring that each maximal motif-clique is derived and reported only once. 

%% file: Content/6_hypergraph.tex
\section{\textbf{HYPERGRAPHS}}
\label{hypergraph}

Hypergraphs provide an ideal framework for modeling systems characterized by complex group or many-to-many interactions among components. Unlike general graphs, where edges connect only pairs of nodes, hypergraphs use hyperedges to link multiple nodes simultaneously, making them exceptionally suitable for representing complex relational dynamics in which multiple elements interact concurrently \cite{bretto2013hypergraph}. Most existing motif counting methods in hypergraphs focus on cliques and their special form, triangles. However, the definition of the same type of motif can vary across different studies, leading researchers to propose various counting algorithms. In this section, we introduce all the definitions of triangles and cliques in hypergraphs and discuss the corresponding counting algorithms.

\subsection{\textbf{Triangle Counting}}

\begin{figure}[t]
	\begin{center}
            \subfigure[A hypergraph $G$]{
			\label{fig6.1}	
			\includegraphics[scale=0.29]{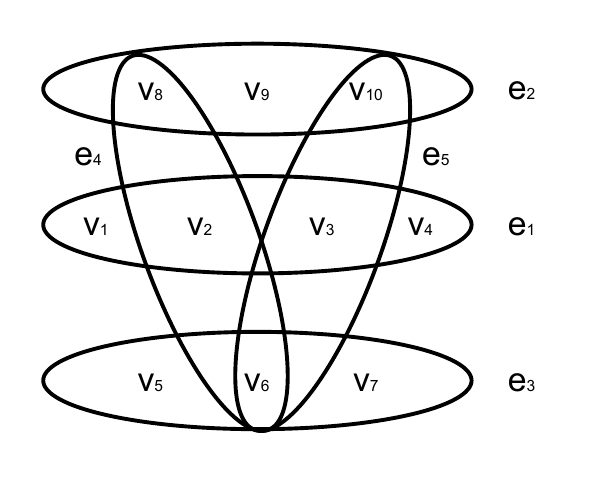}      
		}\hspace{-4mm}
		\subfigure[$G_v$ of $G$]{
			\label{fig6.2}
			\includegraphics[scale=0.55]{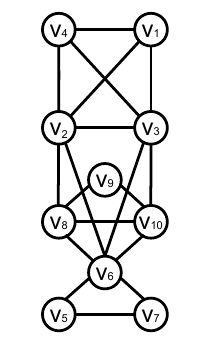}   
		}\hspace{-4mm}
            \subfigure[$G_e$ of $G$]{
			\label{fig6.3}
			\includegraphics[scale=0.55]{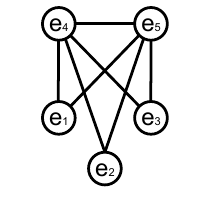}
		}\hspace{-4mm}
            \subfigure[Vertex-based triangles]{
			\label{fig6.4}
			\includegraphics[scale=0.31]{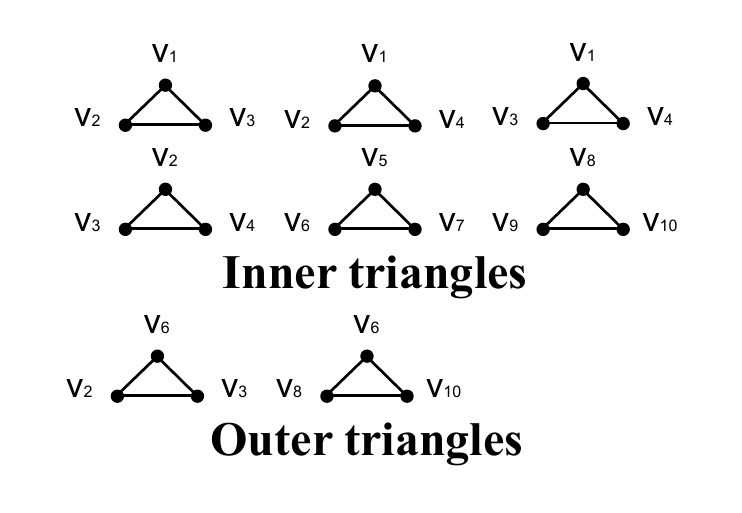}
		}\hspace{-4mm}
            \subfigure[Hyper-triangles]{
			\label{fig6.5}
			\includegraphics[scale=0.36]{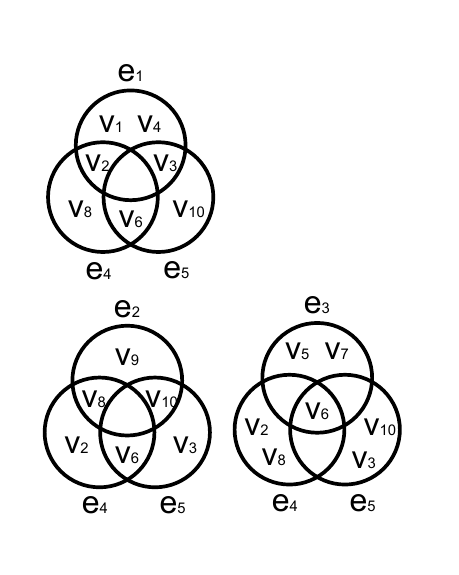}
		}
	\end{center}
        \vspace{-4.5mm}
	\caption{A hypergraph $G$ and its vertex-based/hyperedge-based pairwise graph as well as all the triangles in $G$.}
    
    \label{HypergraphExample}
    \vspace{-4mm}
\end{figure}

In hypergraphs, triangles can be categorized into two types. The first type, vertex-based triangles, is defined by the connections between vertices, capturing pairwise interactions within the structure. The second type, hyperedge-based triangles, is characterized by relationships among hyperedges, representing a more complex interaction model where connections extend across multiple vertices. This classification is essential to understand the structural and functional dynamics of hypergraphs, as each type of triangle provides unique information on the topology of the network.

\subsubsection{\textbf{Vertex-based triangle}}
The existing study \cite{zhang2023efficiently} introduces a vertex-based triangle model to describe the relationships between individual objects in a hypergraph. This model consists of two types: inner triangles and outer triangles. An inner triangle represents the connections among vertices within a single hyperedge, while an outer triangle captures the relationships among vertices spanning three distinct hyperedges. The detailed definition is as follows:

\begin{definition} 
[{\bf Inner/Outer Triangle}] Given a hypergraph \( G = (V, E) \) and three vertices $v_i$, $v_j, v_k \in V$, if $v_i$, $v_j, v_k $ are included by the same hyperedge, then they form an inner triangle. If $v_i$, $v_j, v_k $ are shared by
three hyperedge pairwise, then they form an outer triangle.
\end{definition}

\autoref{HypergraphExample}(a) illustrates a hypergraph $G$ containing six inner triangles and two outer triangles, as shown in \autoref{HypergraphExample}(d). To identify all vertex-based triangles in a hypergraph, Zhang et al. \cite{zhang2023efficiently} introduce a baseline algorithm that treats vertices within the same hyperedge as connected. Based on this approach, the hypergraph is then transformed into a vertex-based pairwise graph, explicitly capturing connections between vertex pairs. For instance, \autoref{HypergraphExample}(b) depicts the vertex-based pairwise graph derived from \autoref{HypergraphExample}(a). Using this induced graph, all vertex-based triangles can be counted or identified and further classified as inner or outer triangles based on their relationships within the original hypergraph.

In addition to the baseline algorithm, Zhang et al. propose a hyperedge-based sampling method, which includes {\em reservoir-based hyperedge sampling (HyperSV)} and {\em reservoir-based weighted sampling (HyperWSV)}. \texttt{HyperSV} processes hyperedge streams by maintaining a fixed-size reservoir, where each incoming hyperedge has a chance to replace an existing one. \texttt{HyperWSV} further optimizes this approach by assigning lower sampling probabilities to larger hyperedges, mitigating their over-representation and improving efficiency. Once sampled, these hyperedges are used to estimate triangle counts. By sampling hyperedges directly rather than converting the hypergraph into a pairwise graph, these methods avoid unnecessary computational and reduce memory consumption.


\subsubsection{\textbf{Hyperedge-based triangle}}

Unlike vertex-based triangles, which capture relationships among only three individual objects, hyperedge-based triangles consist of three pairwise connected hyperedges. For example, \autoref{HypergraphExample}(e) illustrates three hyperedge-based triangles derived from \autoref{HypergraphExample}(a). Compared to vertex-based triangles, this structure more intuitively represents relationships between groups. In \cite{zhang2023efficiently}, Zhang et al. introduce the he-triangle model, while Lee et al. \cite{lee2020hypergraph} propose a model for closed triangles in hypergraphs. Despite the different terminology, these models share the same definition. In the following parts, we refer to them collectively as hyper-triangles. The formal definition of a hyper-triangle is as follows:

\begin{figure}[t]
	\begin{center}
            \label{fig2.1}	
            \includegraphics[scale=0.21]{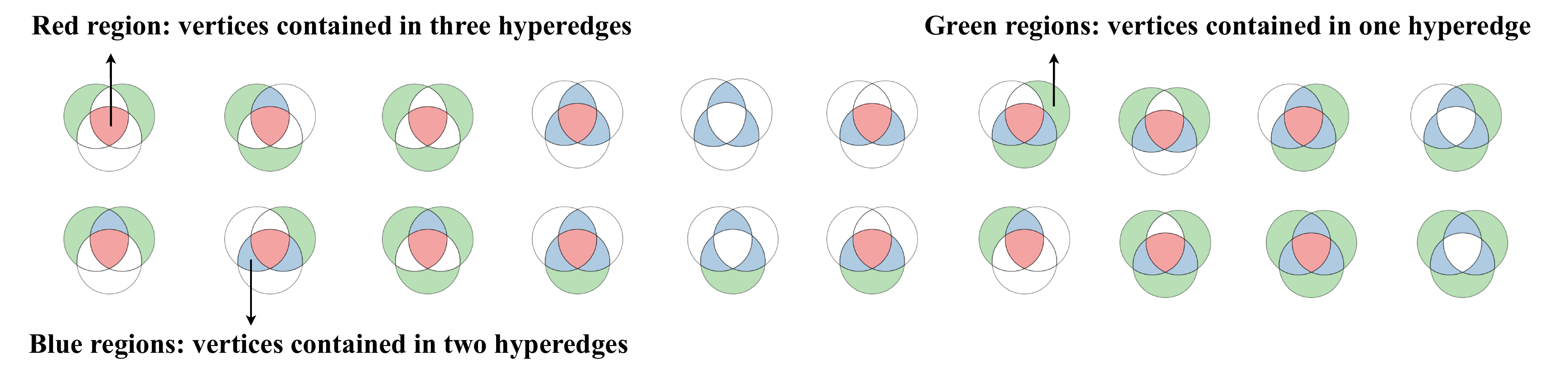}      	
	\end{center}
        \vspace{-3mm}
	\caption{All the patterns for hyperedge-based triangles.}
    
    \label{Hyper-triangle Pattern}
    \vspace{-2mm}
\end{figure}

\begin{definition} 
[{\bf Hyper-triangle}] Given a hypergraph \( G = (V, E) \) and three hyperedges $e_i$, $e_j, e_k \in E$ with $e_i \cap e_j \neq \emptyset$, $e_i \cap e_k \neq \emptyset$, and $e_j \cap e_k \neq \emptyset$, then the subgraph composed of $e_i$, $e_j$ and $e_k$ is a hyper-triangle.
\end{definition}

Existing research \cite{zhang2023efficiently, chakrabarti2021counting, lee2020hypergraph} commonly treats two hyperedges as connected if they share vertices. Based on this, a hypergraph can be transformed into a hyperedge-based pairwise graph, enabling the counting of hyper-triangles on this induced structure (\autoref{HypergraphExample}(c) illustrates the hyperedge-based pairwise graph derived from \autoref{HypergraphExample}(a)). In \cite{lee2020hypergraph}, Lee et al. refine the classification of hyper-triangles. For a hyper-triangle formed by hyperedges $e_i$, $e_j$, and $e_k$, they define seven distinct regions: (a) $e_i \setminus e_j \setminus e_k$, (b) $e_j \setminus e_k \setminus e_i$, (c) $e_k \setminus e_i \setminus e_j$, (d) $e_i \cap e_j \setminus e_k$, (e) $e_j \cap e_k \setminus e_i$, (f) $e_k \cap e_i \setminus e_j$, and (g) $e_i \cap e_j \cap e_k$. By eliminating symmetric cases and considering the presence or absence of each region, they categorize hyper-triangles into 20 distinct patterns, as presented in \autoref{Hyper-triangle Pattern}. They analyze real-world hypergraphs from various domains, measuring the frequency of each hyper-triangle pattern and evaluating its significance by comparing the counts to those in randomized hypergraphs. Their findings show that while the prevalence of hyper-triangle patterns differs across domains, their distribution remains stable within the same domain.


The concept of hyper-triangle patterns is further extended in \cite{lee2024hypergraph}. While the initial framework in \cite{lee2020hypergraph} focuses solely on the presence or absence of intersections among hyperedges, the extended approach incorporates the cardinality of these intersections, offering a more detailed analysis of hypergraph structures. This refinement classifies patterns based on whether the cardinality of each non-empty block exceeds a predefined threshold. These enhanced hyper-triangle patterns improve performance in various applications, including hypergraph classification, node classification, and hyperedge prediction.

\subsection{\textbf{Clique Counting}}

In general graphs, a clique is a subgraph in which all vertices are connected. 
However, in hypergraphs, vertices are grouped within hyperedges, making the traditional clique definition inapplicable. To fill this gap, existing studies define a specialized form of clique in $r$-uniform hypergraphs \cite{holmsen2020large, plant2019maximum, yuster2006finding}, which is also referred to as a simplex in certain works \cite{chakrabarti2021counting, barrett2024counting}. An $r$-uniform hypergraph is a hypergraph in which every hyperedge contains exactly $r$ vertices. Based on this, the definition of a simplex is as follows:

\begin{definition} 
[{\bf Simplex}] Given a $r$-uniform hypergraph \( G = (V, E) \), a subset $S \subset V$ forms a simplex in $G$ if every $r$-subset of vertices in $S$ forms
a hyperedge in $E$ (i.e., $\binom{S}{r} \subset E$), and we call $S$ as $k$-simplex where $k$ is the number of hyperedges contained in $S$ (i.e., $k=|S|-1$).
\end{definition}


In \cite{yuster2006finding}, Yuster introduces an algorithm based on fast matrix multiplication to find simplices in $r$-uniform hypergraphs. The key idea is to partition the vertices of the hypergraph into subsets and then use matrix multiplication to check whether there exists a set of vertices in the hypergraph that matches the structure of the target simplex. The algorithm also addresses finding monochromatic simplices in colored hypergraphs and provides a probabilistic method to approximate the number of independent simplices in random hypergraphs. This approach is significantly faster than the naive algorithm, which relies on exhaustive search to perform the task, especially for certain classes of labeled hypergraphs.

In \cite{chakrabarti2021counting}, Chakrabarti and Haris develop space-efficient algorithms for estimating the number of simplices in hypergraph streams. Based on the triangle-counting problem in graph streams, they extend the concept to $r$-uniform hypergraphs by focusing on $r$-simplices. Their methods employ both targeted and oblivious sampling strategies, each tailored to specific hypergraph characteristics. The targeted sampling approach selects specific hyperedges and vertices, optimizing space usage while maintaining accuracy, particularly in hypergraphs with structural properties that support efficient sampling. In contrast, the oblivious sampling strategy leverages a shadow hypergraph, detecting simplices by ensuring that all edges are monochromatic, providing robust estimates under strict space constraints. Additionally, the authors introduce a one-pass algorithm that eliminates the need for heavy/light edge partitioning while preserving accuracy in $r$-simplex counting.


%% file: Content/7_conclusion.tex
\section{\textbf{CONCLUSIONS AND FUTURE DIRECTIONS}}
\label{conclusion}
In this survey, we provided a comprehensive overview of motif counting techniques across different graph types, including general graphs, heterogeneous graphs, and hypergraphs. We classified existing algorithms based on their computational strategies, highlighting their similarities, differences, and trade-offs. Our analysis covered exact, approximate, heterogeneous, and parallelized approaches, offering insights into their efficiency and real-world applicability in various network analysis tasks.

Although motif counting has been extensively studied in various types of graphs, many challenges remain unaddressed, leaving ample opportunities for further exploration. As graphs grow in complexity and scale, existing algorithms struggle with efficiency, scalability, and adaptability to diverse network structures. Therefore, future research should focus on advancing motif counting techniques in multiple directions. One promising direction is leveraging motif-based representations to enhance {\em retrieval-augmented generation (RAG)} in {\em large language models (LLMs)}, particularly within {\em GraphRAG} frameworks. Furthermore, improving the efficiency of motif counting in large and specialized graph types, such as heterogeneous and hypergraphs, remains a critical challenge, necessitating the development of parallelized, GPU-accelerated, and distributed algorithms. Lastly, as network data continues to evolve, motif counting techniques must be extended to accommodate dynamic and higher-order graph structures, ensuring their relevance to emerging domains in complex network analysis.

\subsection{\textbf{Integrating Motif Counting into Large Language Models}}

Motif counting has long been a cornerstone of network analysis, offering insights into the higher-order structures of complex systems. For instance, motif-aware architectures could improve the interpretability of graph learning systems by explicitly linking predictions to well-defined network structures \cite{chen2023tempme}. However, its integration with \texttt{LLMs} remains largely unexplored. Given the recent advancements in \texttt{GraphRAG} frameworks, which leverage knowledge graph retrieval-augmented generation to enhance \texttt{LLM} reasoning \cite{chen2025pathrag,cai2024simgrag,li2024simple}, motif-based representations present a promising avenue for improving both retrieval efficiency and structured reasoning in \texttt{LLMs}.



In the context of \texttt{GraphRAG}, which retrieves relevant subgraphs for reasoning \cite{han2024retrieval,thakrar2024dynagrag}, current methods often lack structural selectivity. Motif structures act as query patterns, allowing \texttt{LLMs} to retrieve and reason over structurally meaningful contexts. A motif-aware framework could refine retrieval by prioritizing subgraphs enriched with significant motifs, thereby reducing irrelevant information and improving reasoning precision. By integrating motif-aware retrieval, structured pretraining, and multimodal representations, \texttt{LLMs} could substantially advance graph reasoning capabilities, positioning motif counting as a key enabler for the next generation of graph-enhanced deep learning systems.



\subsection{\textbf{Efficient Motif Counting for Large and Complex Networks}}



Motif counting in large-scale and complex networks remains a fundamental yet computationally challenging task, particularly for heterogeneous graphs and hypergraphs. While recent advances in parallel computing, approximate counting, and distributed frameworks have significantly improved motif enumeration for general graphs, scaling these techniques to more complex structures requires further innovations.

One major challenge lies in efficient motif counting for heterogeneous graphs and hypergraphs. Unlike general graphs, heterogeneous graphs contain multiple node and edge types, making it difficult to directly apply existing motif counting techniques. Similarly, hypergraphs, which capture multi-way interactions rather than simple pairwise relations, require specialized counting methods that consider higher-order connectivity patterns. To tackle these challenges, future research should explore distributed and parallel computing strategies tailored for complex graphs. GPU-accelerated methods have demonstrated significant speedups in general graph motif counting by leveraging thousands of parallel threads for edge-triangle intersection computations. Extending such GPU-based approaches to heterogeneous graphs and hypergraphs could improve scalability, but requires optimized memory management and workload balancing due to the irregular structure of these graphs. Distributed computing frameworks such as MapReduce, Apache Spark, and MPI-based systems offer another promising avenue by partitioning the graph into subgraphs and performing motif counting in parallel, reducing memory bottlenecks and enabling efficient motif analysis for networks with billions of edges.

\subsection{\textbf{Effective Motif Patterns in Complex Graphs}}

Motif counting in complex networks is a rapidly evolving field, with ongoing efforts to enhance the effectiveness of existing methodologies. As network data becomes increasingly diverse and structured, traditional motif definitions and counting techniques may fail to capture the full spectrum of meaningful subgraph patterns. Future research should focus on expanding motif definitions and improving adaptability to specialized graph settings.

One promising direction is the introduction of new motif types tailored to specific application domains. Traditional motifs such as triangles, cycles, and cliques have been extensively studied, but emerging graph structures require more expressive motifs that better capture domain-specific interactions. Moreover, beyond static graphs, motif counting in dynamic, temporal, and attributed graphs remains an open challenge. Dynamic graphs \cite{dhulipala2021parallel, bonne2019distributed}, where edges and nodes evolve over time, require incremental counting algorithms that update motif frequencies efficiently instead of recomputing them from scratch. Temporal graphs \cite{himmel2016enumerating, banerjee2019enumeration}, which encode time-stamped interactions, introduce additional complexity, as motif occurrences depend not only on structure but also on event timing and sequencing. Finally, attributed graphs \cite{inae2023motif}, where nodes and edges carry metadata, demand motif counting approaches that incorporate both structural and attribute-based constraints. To achieve more effective motif-based analysis, future work should explore novel frameworks that integrate topological, temporal, and semantic information, potentially combining exact counting with intelligent sampling or leveraging learning-based methods for motif identification. Advancing these techniques will enhance the scalability and applicability of motif counting across diverse real-world networks.

%% file: sample-acmsmall.bbl

\begin{thebibliography}{161}


\ifx \showCODEN    \undefined \def \showCODEN     #1{\unskip}     \fi
\ifx \showISBNx    \undefined \def \showISBNx     #1{\unskip}     \fi
\ifx \showISBNxiii \undefined \def \showISBNxiii  #1{\unskip}     \fi
\ifx \showISSN     \undefined \def \showISSN      #1{\unskip}     \fi
\ifx \showLCCN     \undefined \def \showLCCN      #1{\unskip}     \fi
\ifx \shownote     \undefined \def \shownote      #1{#1}          \fi
\ifx \showarticletitle \undefined \def \showarticletitle #1{#1}   \fi
\ifx \showURL      \undefined \def \showURL       {\relax}        \fi
\providecommand\bibfield[2]{#2}
\providecommand\bibinfo[2]{#2}
\providecommand\natexlab[1]{#1}
\providecommand\showeprint[2][]{arXiv:#2}

\bibitem[Abboud et~al\mbox{.}(2022)]%
        {abboud2022listing}
\bibfield{author}{\bibinfo{person}{Amir Abboud}, \bibinfo{person}{Seri Khoury}, \bibinfo{person}{Oree Leibowitz}, {and} \bibinfo{person}{Ron Safier}.} \bibinfo{year}{2022}\natexlab{}.
\newblock \showarticletitle{Listing 4-cycles}.
\newblock \bibinfo{journal}{\emph{arXiv preprint arXiv:2211.10022}} (\bibinfo{year}{2022}).
\newblock


\bibitem[Ahmed et~al\mbox{.}(2021)]%
        {ahmed2021triangle}
\bibfield{author}{\bibinfo{person}{Aly Ahmed}, \bibinfo{person}{Keanelek Enns}, {and} \bibinfo{person}{Alex Thomo}.} \bibinfo{year}{2021}\natexlab{}.
\newblock \showarticletitle{Triangle enumeration for billion-scale graphs in rdbms}. In \bibinfo{booktitle}{\emph{International Conference on Advanced Information Networking and Applications}}. Springer, \bibinfo{pages}{160--173}.
\newblock


\bibitem[Ahmed et~al\mbox{.}(2014)]%
        {ahmed2014graph}
\bibfield{author}{\bibinfo{person}{Nesreen~K Ahmed}, \bibinfo{person}{Nick Duffield}, \bibinfo{person}{Jennifer Neville}, {and} \bibinfo{person}{Ramana Kompella}.} \bibinfo{year}{2014}\natexlab{}.
\newblock \showarticletitle{Graph sample and hold: A framework for big-graph analytics}. In \bibinfo{booktitle}{\emph{Proceedings of the 20th ACM SIGKDD international conference on Knowledge discovery and data mining}}. \bibinfo{pages}{1446--1455}.
\newblock


\bibitem[Aksoy et~al\mbox{.}(2017)]%
        {aksoy2017measuring}
\bibfield{author}{\bibinfo{person}{Sinan~G Aksoy}, \bibinfo{person}{Tamara~G Kolda}, {and} \bibinfo{person}{Ali Pinar}.} \bibinfo{year}{2017}\natexlab{}.
\newblock \showarticletitle{Measuring and modeling bipartite graphs with community structure}.
\newblock \bibinfo{journal}{\emph{Journal of Complex Networks}} \bibinfo{volume}{5}, \bibinfo{number}{4} (\bibinfo{year}{2017}), \bibinfo{pages}{581--603}.
\newblock


\bibitem[Almasri et~al\mbox{.}(2021)]%
        {almasri2021k}
\bibfield{author}{\bibinfo{person}{Mohammad Almasri}, \bibinfo{person}{Izzat El~Hajj}, \bibinfo{person}{Rakesh Nagi}, \bibinfo{person}{Jinjun Xiong}, {and} \bibinfo{person}{Wen-mei Hwu}.} \bibinfo{year}{2021}\natexlab{}.
\newblock \showarticletitle{K-clique counting on gpus}.
\newblock \bibinfo{journal}{\emph{arXiv preprint arXiv:2104.13209}} (\bibinfo{year}{2021}).
\newblock


\bibitem[Almasri et~al\mbox{.}(2022)]%
        {almasri2022parallel}
\bibfield{author}{\bibinfo{person}{Mohammad Almasri}, \bibinfo{person}{Izzat~El Hajj}, \bibinfo{person}{Rakesh Nagi}, \bibinfo{person}{Jinjun Xiong}, {and} \bibinfo{person}{Wen-mei Hwu}.} \bibinfo{year}{2022}\natexlab{}.
\newblock \showarticletitle{Parallel k-clique counting on gpus}. In \bibinfo{booktitle}{\emph{Proceedings of the 36th ACM International Conference on Supercomputing}}. \bibinfo{pages}{1--14}.
\newblock


\bibitem[Alon et~al\mbox{.}(1997)]%
        {alon1997finding}
\bibfield{author}{\bibinfo{person}{Noga Alon}, \bibinfo{person}{Raphael Yuster}, {and} \bibinfo{person}{Uri Zwick}.} \bibinfo{year}{1997}\natexlab{}.
\newblock \showarticletitle{Finding and counting given length cycles}.
\newblock \bibinfo{journal}{\emph{Algorithmica}} \bibinfo{volume}{17}, \bibinfo{number}{3} (\bibinfo{year}{1997}), \bibinfo{pages}{209--223}.
\newblock


\bibitem[Angel et~al\mbox{.}(2012)]%
        {angel2012dense}
\bibfield{author}{\bibinfo{person}{Albert Angel}, \bibinfo{person}{Nikos Sarkas}, \bibinfo{person}{Nick Koudas}, {and} \bibinfo{person}{Divesh Srivastava}.} \bibinfo{year}{2012}\natexlab{}.
\newblock \showarticletitle{Dense subgraph maintenance under streaming edge weight updates for real-time story identification}.
\newblock \bibinfo{journal}{\emph{Proceedings of the VLDB Endowment}} \bibinfo{volume}{5}, \bibinfo{number}{6} (\bibinfo{year}{2012}), \bibinfo{pages}{574--585}.
\newblock


\bibitem[Arifuzzaman et~al\mbox{.}(2013)]%
        {arifuzzaman2013patric}
\bibfield{author}{\bibinfo{person}{Shaikh Arifuzzaman}, \bibinfo{person}{Maleq Khan}, {and} \bibinfo{person}{Madhav Marathe}.} \bibinfo{year}{2013}\natexlab{}.
\newblock \showarticletitle{Patric: a parallel algorithm for counting triangles in massive networks}. In \bibinfo{booktitle}{\emph{Proceedings of the 22nd ACM international conference on Information \& Knowledge Management}}. \bibinfo{pages}{529--538}.
\newblock


\bibitem[Arifuzzaman et~al\mbox{.}(2019)]%
        {arifuzzaman2019fast}
\bibfield{author}{\bibinfo{person}{Shaikh Arifuzzaman}, \bibinfo{person}{Maleq Khan}, {and} \bibinfo{person}{Madhav Marathe}.} \bibinfo{year}{2019}\natexlab{}.
\newblock \showarticletitle{Fast parallel algorithms for counting and listing triangles in big graphs}.
\newblock \bibinfo{journal}{\emph{ACM Transactions on Knowledge Discovery from Data (TKDD)}} \bibinfo{volume}{14}, \bibinfo{number}{1} (\bibinfo{year}{2019}), \bibinfo{pages}{1--34}.
\newblock


\bibitem[Azad et~al\mbox{.}(2015)]%
        {azad2015parallel}
\bibfield{author}{\bibinfo{person}{Ariful Azad}, \bibinfo{person}{Aydin Bulu{\c{c}}}, {and} \bibinfo{person}{John Gilbert}.} \bibinfo{year}{2015}\natexlab{}.
\newblock \showarticletitle{Parallel triangle counting and enumeration using matrix algebra}. In \bibinfo{booktitle}{\emph{2015 IEEE International Parallel and Distributed Processing Symposium Workshop}}. IEEE, \bibinfo{pages}{804--811}.
\newblock


\bibitem[Banerjee and Pal(2019)]%
        {banerjee2019enumeration}
\bibfield{author}{\bibinfo{person}{Suman Banerjee} {and} \bibinfo{person}{Bithika Pal}.} \bibinfo{year}{2019}\natexlab{}.
\newblock \showarticletitle{On the enumeration of maximal ($\delta$, $\gamma$)-cliques of a temporal network}. In \bibinfo{booktitle}{\emph{Proceedings of the ACM India Joint International Conference on Data Science and Management of Data}}. \bibinfo{pages}{112--120}.
\newblock


\bibitem[Barrett et~al\mbox{.}(2024)]%
        {barrett2024counting}
\bibfield{author}{\bibinfo{person}{Jordan Barrett}, \bibinfo{person}{Pawe{\l} Pra{\l}at}, \bibinfo{person}{Aaron Smith}, {and} \bibinfo{person}{Fran{\c{c}}ois Th{\'e}berge}.} \bibinfo{year}{2024}\natexlab{}.
\newblock \showarticletitle{Counting simplicial pairs in hypergraphs}.
\newblock \bibinfo{journal}{\emph{arXiv preprint arXiv:2408.11806}} (\bibinfo{year}{2024}).
\newblock


\bibitem[Bassett and Bullmore(2006)]%
        {bassett2006small}
\bibfield{author}{\bibinfo{person}{Danielle~Smith Bassett} {and} \bibinfo{person}{ED Bullmore}.} \bibinfo{year}{2006}\natexlab{}.
\newblock \showarticletitle{Small-world brain networks}.
\newblock \bibinfo{journal}{\emph{The neuroscientist}} \bibinfo{volume}{12}, \bibinfo{number}{6} (\bibinfo{year}{2006}), \bibinfo{pages}{512--523}.
\newblock


\bibitem[Batagelj and Mrvar(2001)]%
        {batagelj2001subquadratic}
\bibfield{author}{\bibinfo{person}{Vladimir Batagelj} {and} \bibinfo{person}{Andrej Mrvar}.} \bibinfo{year}{2001}\natexlab{}.
\newblock \showarticletitle{A subquadratic triad census algorithm for large sparse networks with small maximum degree}.
\newblock \bibinfo{journal}{\emph{Social networks}} \bibinfo{volume}{23}, \bibinfo{number}{3} (\bibinfo{year}{2001}), \bibinfo{pages}{237--243}.
\newblock


\bibitem[Becchetti et~al\mbox{.}(2010)]%
        {becchetti2010efficient}
\bibfield{author}{\bibinfo{person}{Luca Becchetti}, \bibinfo{person}{Paolo Boldi}, \bibinfo{person}{Carlos Castillo}, {and} \bibinfo{person}{Aristides Gionis}.} \bibinfo{year}{2010}\natexlab{}.
\newblock \showarticletitle{Efficient algorithms for large-scale local triangle counting}.
\newblock \bibinfo{journal}{\emph{ACM Transactions on Knowledge Discovery from Data (TKDD)}} \bibinfo{volume}{4}, \bibinfo{number}{3} (\bibinfo{year}{2010}), \bibinfo{pages}{1--28}.
\newblock


\bibitem[Berry et~al\mbox{.}(2011)]%
        {berry2011tolerating}
\bibfield{author}{\bibinfo{person}{Jonathan~W Berry}, \bibinfo{person}{Bruce Hendrickson}, \bibinfo{person}{Randall~A LaViolette}, {and} \bibinfo{person}{Cynthia~A Phillips}.} \bibinfo{year}{2011}\natexlab{}.
\newblock \showarticletitle{Tolerating the community detection resolution limit with edge weighting}.
\newblock \bibinfo{journal}{\emph{Physical Review E—Statistical, Nonlinear, and Soft Matter Physics}} \bibinfo{volume}{83}, \bibinfo{number}{5} (\bibinfo{year}{2011}), \bibinfo{pages}{056119}.
\newblock


\bibitem[Bisson and Fatica(2017)]%
        {bisson2017high}
\bibfield{author}{\bibinfo{person}{Mauro Bisson} {and} \bibinfo{person}{Massimiliano Fatica}.} \bibinfo{year}{2017}\natexlab{}.
\newblock \showarticletitle{High performance exact triangle counting on gpus}.
\newblock \bibinfo{journal}{\emph{IEEE Transactions on Parallel and Distributed Systems}} \bibinfo{volume}{28}, \bibinfo{number}{12} (\bibinfo{year}{2017}), \bibinfo{pages}{3501--3510}.
\newblock


\bibitem[Boekhout et~al\mbox{.}(2021)]%
        {boekhout2021investigating}
\bibfield{author}{\bibinfo{person}{Hanjo~D Boekhout}, \bibinfo{person}{Vincent~A Traag}, {and} \bibinfo{person}{Frank~W Takes}.} \bibinfo{year}{2021}\natexlab{}.
\newblock \showarticletitle{Investigating scientific mobility in co-authorship networks using multilayer temporal motifs}.
\newblock \bibinfo{journal}{\emph{Network Science}} \bibinfo{volume}{9}, \bibinfo{number}{3} (\bibinfo{year}{2021}), \bibinfo{pages}{354--386}.
\newblock


\bibitem[Bonne and Censor-Hillel(2019)]%
        {bonne2019distributed}
\bibfield{author}{\bibinfo{person}{Matthias Bonne} {and} \bibinfo{person}{Keren Censor-Hillel}.} \bibinfo{year}{2019}\natexlab{}.
\newblock \showarticletitle{Distributed detection of cliques in dynamic networks}.
\newblock \bibinfo{journal}{\emph{arXiv preprint arXiv:1904.11440}} (\bibinfo{year}{2019}).
\newblock


\bibitem[Borgatti and Everett(1997)]%
        {borgatti1997network}
\bibfield{author}{\bibinfo{person}{Stephen~P Borgatti} {and} \bibinfo{person}{Martin~G Everett}.} \bibinfo{year}{1997}\natexlab{}.
\newblock \showarticletitle{Network analysis of 2-mode data}.
\newblock \bibinfo{journal}{\emph{Social networks}} \bibinfo{volume}{19}, \bibinfo{number}{3} (\bibinfo{year}{1997}), \bibinfo{pages}{243--269}.
\newblock


\bibitem[Bretto(2013)]%
        {bretto2013hypergraph}
\bibfield{author}{\bibinfo{person}{Alain Bretto}.} \bibinfo{year}{2013}\natexlab{}.
\newblock \showarticletitle{Hypergraph theory}.
\newblock \bibinfo{journal}{\emph{An introduction. Mathematical Engineering. Cham: Springer}}  \bibinfo{volume}{1} (\bibinfo{year}{2013}).
\newblock


\bibitem[Bron and Kerbosch(1973)]%
        {bron1973algorithm}
\bibfield{author}{\bibinfo{person}{Coen Bron} {and} \bibinfo{person}{Joep Kerbosch}.} \bibinfo{year}{1973}\natexlab{}.
\newblock \showarticletitle{Algorithm 457: finding all cliques of an undirected graph}.
\newblock \bibinfo{journal}{\emph{Commun. ACM}} \bibinfo{volume}{16}, \bibinfo{number}{9} (\bibinfo{year}{1973}), \bibinfo{pages}{575--577}.
\newblock


\bibitem[Bullmore and Sporns(2009)]%
        {bullmore2009complex}
\bibfield{author}{\bibinfo{person}{Ed Bullmore} {and} \bibinfo{person}{Olaf Sporns}.} \bibinfo{year}{2009}\natexlab{}.
\newblock \showarticletitle{Complex brain networks: graph theoretical analysis of structural and functional systems}.
\newblock \bibinfo{journal}{\emph{Nature reviews neuroscience}} \bibinfo{volume}{10}, \bibinfo{number}{3} (\bibinfo{year}{2009}), \bibinfo{pages}{186--198}.
\newblock


\bibitem[Buriol et~al\mbox{.}(2006)]%
        {buriol2006counting}
\bibfield{author}{\bibinfo{person}{Luciana~S Buriol}, \bibinfo{person}{Gereon Frahling}, \bibinfo{person}{Stefano Leonardi}, \bibinfo{person}{Alberto Marchetti-Spaccamela}, {and} \bibinfo{person}{Christian Sohler}.} \bibinfo{year}{2006}\natexlab{}.
\newblock \showarticletitle{Counting triangles in data streams}. In \bibinfo{booktitle}{\emph{Proceedings of the twenty-fifth ACM SIGMOD-SIGACT-SIGART symposium on Principles of database systems}}. \bibinfo{pages}{253--262}.
\newblock


\bibitem[Burkhardt(2017)]%
        {burkhardt2017graphing}
\bibfield{author}{\bibinfo{person}{Paul Burkhardt}.} \bibinfo{year}{2017}\natexlab{}.
\newblock \showarticletitle{Graphing trillions of triangles}.
\newblock \bibinfo{journal}{\emph{Information Visualization}} \bibinfo{volume}{16}, \bibinfo{number}{3} (\bibinfo{year}{2017}), \bibinfo{pages}{157--166}.
\newblock


\bibitem[Burkhardt and Harris(2023)]%
        {burkhardt2023simple}
\bibfield{author}{\bibinfo{person}{Paul Burkhardt} {and} \bibinfo{person}{David~G Harris}.} \bibinfo{year}{2023}\natexlab{}.
\newblock \showarticletitle{Simple and efficient four-cycle counting on sparse graphs}.
\newblock \bibinfo{journal}{\emph{arXiv preprint arXiv:2303.06090}} (\bibinfo{year}{2023}).
\newblock


\bibitem[Cai et~al\mbox{.}(2023)]%
        {cai2023efficient}
\bibfield{author}{\bibinfo{person}{Xinwei Cai}, \bibinfo{person}{Xiangyu Ke}, \bibinfo{person}{Kai Wang}, \bibinfo{person}{Lu Chen}, \bibinfo{person}{Tianming Zhang}, \bibinfo{person}{Qing Liu}, {and} \bibinfo{person}{Yunjun Gao}.} \bibinfo{year}{2023}\natexlab{}.
\newblock \showarticletitle{Efficient Temporal Butterfly Counting and Enumeration on Temporal Bipartite Graphs}.
\newblock \bibinfo{journal}{\emph{arXiv preprint arXiv:2306.00893}} (\bibinfo{year}{2023}).
\newblock


\bibitem[Cai et~al\mbox{.}(2024)]%
        {cai2024simgrag}
\bibfield{author}{\bibinfo{person}{Yuzheng Cai}, \bibinfo{person}{Zhenyue Guo}, \bibinfo{person}{Yiwen Pei}, \bibinfo{person}{Wanrui Bian}, {and} \bibinfo{person}{Weiguo Zheng}.} \bibinfo{year}{2024}\natexlab{}.
\newblock \showarticletitle{SimGRAG: Leveraging Similar Subgraphs for Knowledge Graphs Driven Retrieval-Augmented Generation}.
\newblock \bibinfo{journal}{\emph{arXiv preprint arXiv:2412.15272}} (\bibinfo{year}{2024}).
\newblock


\bibitem[Chakrabarti and Haris(2021)]%
        {chakrabarti2021counting}
\bibfield{author}{\bibinfo{person}{Amit Chakrabarti} {and} \bibinfo{person}{Themistoklis Haris}.} \bibinfo{year}{2021}\natexlab{}.
\newblock \showarticletitle{Counting simplices in hypergraph streams}.
\newblock \bibinfo{journal}{\emph{arXiv preprint arXiv:2112.11016}} (\bibinfo{year}{2021}).
\newblock


\bibitem[Chen et~al\mbox{.}(2025)]%
        {chen2025pathrag}
\bibfield{author}{\bibinfo{person}{Boyu Chen}, \bibinfo{person}{Zirui Guo}, \bibinfo{person}{Zidan Yang}, \bibinfo{person}{Yuluo Chen}, \bibinfo{person}{Junze Chen}, \bibinfo{person}{Zhenghao Liu}, \bibinfo{person}{Chuan Shi}, {and} \bibinfo{person}{Cheng Yang}.} \bibinfo{year}{2025}\natexlab{}.
\newblock \showarticletitle{PathRAG: Pruning Graph-based Retrieval Augmented Generation with Relational Paths}.
\newblock \bibinfo{journal}{\emph{arXiv preprint arXiv:2502.14902}} (\bibinfo{year}{2025}).
\newblock


\bibitem[Chen and Ying(2023)]%
        {chen2023tempme}
\bibfield{author}{\bibinfo{person}{Jialin Chen} {and} \bibinfo{person}{Rex Ying}.} \bibinfo{year}{2023}\natexlab{}.
\newblock \showarticletitle{Tempme: Towards the explainability of temporal graph neural networks via motif discovery}.
\newblock \bibinfo{journal}{\emph{Advances in Neural Information Processing Systems}}  \bibinfo{volume}{36} (\bibinfo{year}{2023}), \bibinfo{pages}{29005--29028}.
\newblock


\bibitem[Chen et~al\mbox{.}(2021)]%
        {chen2021efficiently}
\bibfield{author}{\bibinfo{person}{Xiaoshuang Chen}, \bibinfo{person}{Kai Wang}, \bibinfo{person}{Xuemin Lin}, \bibinfo{person}{Wenjie Zhang}, \bibinfo{person}{Lu Qin}, {and} \bibinfo{person}{Ying Zhang}.} \bibinfo{year}{2021}\natexlab{}.
\newblock \showarticletitle{Efficiently answering reachability and path queries on temporal bipartite graphs}.
\newblock \bibinfo{journal}{\emph{Proceedings of the VLDB Endowment}} (\bibinfo{year}{2021}).
\newblock


\bibitem[Chiba and Nishizeki(1985)]%
        {chiba1985arboricity}
\bibfield{author}{\bibinfo{person}{Norishige Chiba} {and} \bibinfo{person}{Takao Nishizeki}.} \bibinfo{year}{1985}\natexlab{}.
\newblock \showarticletitle{Arboricity and subgraph listing algorithms}.
\newblock \bibinfo{journal}{\emph{SIAM Journal on computing}} \bibinfo{volume}{14}, \bibinfo{number}{1} (\bibinfo{year}{1985}), \bibinfo{pages}{210--223}.
\newblock


\bibitem[Choobdar et~al\mbox{.}(2012)]%
        {choobdar2012comparison}
\bibfield{author}{\bibinfo{person}{Sarvenaz Choobdar}, \bibinfo{person}{Pedro Ribeiro}, \bibinfo{person}{Sylwia Bugla}, {and} \bibinfo{person}{Fernando Silva}.} \bibinfo{year}{2012}\natexlab{}.
\newblock \showarticletitle{Comparison of co-authorship networks across scientific fields using motifs}. In \bibinfo{booktitle}{\emph{2012 IEEE/ACM International Conference on Advances in Social Networks Analysis and Mining}}. IEEE, \bibinfo{pages}{147--152}.
\newblock


\bibitem[Chou and Suzuki(2010)]%
        {chou2010discovering}
\bibfield{author}{\bibinfo{person}{Bin-Hui Chou} {and} \bibinfo{person}{Einoshin Suzuki}.} \bibinfo{year}{2010}\natexlab{}.
\newblock \showarticletitle{Discovering community-oriented roles of nodes in a social network}. In \bibinfo{booktitle}{\emph{International Conference on Data Warehousing and Knowledge Discovery}}. Springer, \bibinfo{pages}{52--64}.
\newblock


\bibitem[Chu and Cheng(2011)]%
        {chu2011triangle}
\bibfield{author}{\bibinfo{person}{Shumo Chu} {and} \bibinfo{person}{James Cheng}.} \bibinfo{year}{2011}\natexlab{}.
\newblock \showarticletitle{Triangle listing in massive networks and its applications}. In \bibinfo{booktitle}{\emph{Proceedings of the 17th ACM SIGKDD international conference on Knowledge discovery and data mining}}. \bibinfo{pages}{672--680}.
\newblock


\bibitem[Cohen(2009)]%
        {cohen2009graph}
\bibfield{author}{\bibinfo{person}{Jonathan Cohen}.} \bibinfo{year}{2009}\natexlab{}.
\newblock \showarticletitle{Graph twiddling in a mapreduce world}.
\newblock \bibinfo{journal}{\emph{Computing in Science \& Engineering}} \bibinfo{volume}{11}, \bibinfo{number}{4} (\bibinfo{year}{2009}), \bibinfo{pages}{29--41}.
\newblock


\bibitem[Coppersmith and Winograd(1987)]%
        {coppersmith1987matrix}
\bibfield{author}{\bibinfo{person}{Don Coppersmith} {and} \bibinfo{person}{Shmuel Winograd}.} \bibinfo{year}{1987}\natexlab{}.
\newblock \showarticletitle{Matrix multiplication via arithmetic progressions}. In \bibinfo{booktitle}{\emph{Proceedings of the nineteenth annual ACM symposium on Theory of computing}}. \bibinfo{pages}{1--6}.
\newblock


\bibitem[Coyle and Vaughn(2008)]%
        {coyle2008social}
\bibfield{author}{\bibinfo{person}{Cheryl~L Coyle} {and} \bibinfo{person}{Heather Vaughn}.} \bibinfo{year}{2008}\natexlab{}.
\newblock \showarticletitle{Social networking: Communication revolution or evolution?}
\newblock \bibinfo{journal}{\emph{Bell Labs technical journal}} \bibinfo{volume}{13}, \bibinfo{number}{2} (\bibinfo{year}{2008}), \bibinfo{pages}{13--17}.
\newblock


\bibitem[Danisch et~al\mbox{.}(2018)]%
        {danisch2018listing}
\bibfield{author}{\bibinfo{person}{Maximilien Danisch}, \bibinfo{person}{Oana Balalau}, {and} \bibinfo{person}{Mauro Sozio}.} \bibinfo{year}{2018}\natexlab{}.
\newblock \showarticletitle{Listing k-cliques in sparse real-world graphs}. In \bibinfo{booktitle}{\emph{Proceedings of the 2018 World Wide Web Conference}}. \bibinfo{pages}{589--598}.
\newblock


\bibitem[Danisch et~al\mbox{.}(2017)]%
        {danisch2017large}
\bibfield{author}{\bibinfo{person}{Maximilien Danisch}, \bibinfo{person}{T-H~Hubert Chan}, {and} \bibinfo{person}{Mauro Sozio}.} \bibinfo{year}{2017}\natexlab{}.
\newblock \showarticletitle{Large scale density-friendly graph decomposition via convex programming}. In \bibinfo{booktitle}{\emph{Proceedings of the 26th International Conference on World Wide Web}}. \bibinfo{pages}{233--242}.
\newblock


\bibitem[Demmel(1997)]%
        {demmel1997applied}
\bibfield{author}{\bibinfo{person}{James~W Demmel}.} \bibinfo{year}{1997}\natexlab{}.
\newblock \bibinfo{booktitle}{\emph{Applied numerical linear algebra}}.
\newblock \bibinfo{publisher}{SIAM}.
\newblock


\bibitem[Deveci et~al\mbox{.}(2017)]%
        {deveci2017performance}
\bibfield{author}{\bibinfo{person}{Mehmet Deveci}, \bibinfo{person}{Christian Trott}, {and} \bibinfo{person}{Sivasankaran Rajamanickam}.} \bibinfo{year}{2017}\natexlab{}.
\newblock \showarticletitle{Performance-portable sparse matrix-matrix multiplication for many-core architectures}. In \bibinfo{booktitle}{\emph{2017 IEEE International Parallel and Distributed Processing Symposium Workshops (IPDPSW)}}. IEEE, \bibinfo{pages}{693--702}.
\newblock


\bibitem[Dhulipala et~al\mbox{.}(2021)]%
        {dhulipala2021parallel}
\bibfield{author}{\bibinfo{person}{Laxman Dhulipala}, \bibinfo{person}{Quanquan~C Liu}, \bibinfo{person}{Julian Shun}, {and} \bibinfo{person}{Shangdi Yu}.} \bibinfo{year}{2021}\natexlab{}.
\newblock \showarticletitle{Parallel batch-dynamic k-clique counting}. In \bibinfo{booktitle}{\emph{Symposium on Algorithmic Principles of Computer Systems (APOCS)}}. SIAM, \bibinfo{pages}{129--143}.
\newblock


\bibitem[Dourisboure et~al\mbox{.}(2009)]%
        {dourisboure2009extraction}
\bibfield{author}{\bibinfo{person}{Yon Dourisboure}, \bibinfo{person}{Filippo Geraci}, {and} \bibinfo{person}{Marco Pellegrini}.} \bibinfo{year}{2009}\natexlab{}.
\newblock \showarticletitle{Extraction and classification of dense implicit communities in the web graph}.
\newblock \bibinfo{journal}{\emph{ACM Transactions on the Web (TWEB)}} \bibinfo{volume}{3}, \bibinfo{number}{2} (\bibinfo{year}{2009}), \bibinfo{pages}{1--36}.
\newblock


\bibitem[Eden et~al\mbox{.}(2018)]%
        {eden2018approximating}
\bibfield{author}{\bibinfo{person}{Talya Eden}, \bibinfo{person}{Dana Ron}, {and} \bibinfo{person}{C Seshadhri}.} \bibinfo{year}{2018}\natexlab{}.
\newblock \showarticletitle{On approximating the number of k-cliques in sublinear time}. In \bibinfo{booktitle}{\emph{Proceedings of the 50th annual ACM SIGACT symposium on theory of computing}}. \bibinfo{pages}{722--734}.
\newblock


\bibitem[Edwards et~al\mbox{.}(2014)]%
        {edwards2014kokkos}
\bibfield{author}{\bibinfo{person}{H~Carter Edwards}, \bibinfo{person}{Christian~R Trott}, {and} \bibinfo{person}{Daniel Sunderland}.} \bibinfo{year}{2014}\natexlab{}.
\newblock \showarticletitle{Kokkos: Enabling manycore performance portability through polymorphic memory access patterns}.
\newblock \bibinfo{journal}{\emph{Journal of parallel and distributed computing}} \bibinfo{volume}{74}, \bibinfo{number}{12} (\bibinfo{year}{2014}), \bibinfo{pages}{3202--3216}.
\newblock


\bibitem[Fain and Pedersen(2006)]%
        {fain2006sponsored}
\bibfield{author}{\bibinfo{person}{Daniel~C Fain} {and} \bibinfo{person}{Jan~O Pedersen}.} \bibinfo{year}{2006}\natexlab{}.
\newblock \showarticletitle{Sponsored search: A brief history}.
\newblock \bibinfo{journal}{\emph{Bulletin-American Society For Information Science And Technology}} \bibinfo{volume}{32}, \bibinfo{number}{2} (\bibinfo{year}{2006}), \bibinfo{pages}{12}.
\newblock


\bibitem[Fang et~al\mbox{.}(2020a)]%
        {fang2020survey}
\bibfield{author}{\bibinfo{person}{Yixiang Fang}, \bibinfo{person}{Xin Huang}, \bibinfo{person}{Lu Qin}, \bibinfo{person}{Ying Zhang}, \bibinfo{person}{Wenjie Zhang}, \bibinfo{person}{Reynold Cheng}, {and} \bibinfo{person}{Xuemin Lin}.} \bibinfo{year}{2020}\natexlab{a}.
\newblock \showarticletitle{A survey of community search over big graphs}.
\newblock \bibinfo{journal}{\emph{The VLDB Journal}}  \bibinfo{volume}{29} (\bibinfo{year}{2020}), \bibinfo{pages}{353--392}.
\newblock


\bibitem[Fang et~al\mbox{.}(2020b)]%
        {fang2020effective}
\bibfield{author}{\bibinfo{person}{Yixiang Fang}, \bibinfo{person}{Yixing Yang}, \bibinfo{person}{Wenjie Zhang}, \bibinfo{person}{Xuemin Lin}, {and} \bibinfo{person}{Xin Cao}.} \bibinfo{year}{2020}\natexlab{b}.
\newblock \showarticletitle{Effective and efficient community search over large heterogeneous information networks}.
\newblock \bibinfo{journal}{\emph{Proceedings of the VLDB Endowment}} \bibinfo{volume}{13}, \bibinfo{number}{6} (\bibinfo{year}{2020}), \bibinfo{pages}{854--867}.
\newblock


\bibitem[Finocchi et~al\mbox{.}(2015)]%
        {finocchi2015clique}
\bibfield{author}{\bibinfo{person}{Irene Finocchi}, \bibinfo{person}{Marco Finocchi}, {and} \bibinfo{person}{Emanuele~G Fusco}.} \bibinfo{year}{2015}\natexlab{}.
\newblock \showarticletitle{Clique counting in mapreduce: Algorithms and experiments}.
\newblock \bibinfo{journal}{\emph{Journal of Experimental Algorithmics (JEA)}}  \bibinfo{volume}{20} (\bibinfo{year}{2015}), \bibinfo{pages}{1--20}.
\newblock


\bibitem[Fox et~al\mbox{.}(2018)]%
        {fox2018fast}
\bibfield{author}{\bibinfo{person}{James Fox}, \bibinfo{person}{Oded Green}, \bibinfo{person}{Kasimir Gabert}, \bibinfo{person}{Xiaojing An}, {and} \bibinfo{person}{David~A Bader}.} \bibinfo{year}{2018}\natexlab{}.
\newblock \showarticletitle{Fast and adaptive list intersections on the gpu}. In \bibinfo{booktitle}{\emph{2018 IEEE High Performance extreme Computing Conference (HPEC)}}. IEEE, \bibinfo{pages}{1--7}.
\newblock


\bibitem[Gao et~al\mbox{.}(2022)]%
        {gao2022scalable}
\bibfield{author}{\bibinfo{person}{Zhongqiang Gao}, \bibinfo{person}{Chuanqi Cheng}, \bibinfo{person}{Yanwei Yu}, \bibinfo{person}{Lei Cao}, \bibinfo{person}{Chao Huang}, {and} \bibinfo{person}{Junyu Dong}.} \bibinfo{year}{2022}\natexlab{}.
\newblock \showarticletitle{Scalable motif counting for large-scale temporal graphs}. In \bibinfo{booktitle}{\emph{2022 IEEE 38th International Conference on Data Engineering (ICDE)}}. IEEE, \bibinfo{pages}{2656--2668}.
\newblock


\bibitem[Giscard et~al\mbox{.}(2017)]%
        {giscard2017evaluating}
\bibfield{author}{\bibinfo{person}{Pierre-Louis Giscard}, \bibinfo{person}{Paul Rochet}, {and} \bibinfo{person}{Richard~C Wilson}.} \bibinfo{year}{2017}\natexlab{}.
\newblock \showarticletitle{Evaluating balance on social networks from their simple cycles}.
\newblock \bibinfo{journal}{\emph{Journal of Complex Networks}} \bibinfo{volume}{5}, \bibinfo{number}{5} (\bibinfo{year}{2017}), \bibinfo{pages}{750--775}.
\newblock


\bibitem[Golub and Van~Loan(2013)]%
        {golub2013matrix}
\bibfield{author}{\bibinfo{person}{Gene~H Golub} {and} \bibinfo{person}{Charles~F Van~Loan}.} \bibinfo{year}{2013}\natexlab{}.
\newblock \bibinfo{booktitle}{\emph{Matrix computations}}.
\newblock \bibinfo{publisher}{JHU press}.
\newblock


\bibitem[Green et~al\mbox{.}(2018)]%
        {green2018logarithmic}
\bibfield{author}{\bibinfo{person}{Oded Green}, \bibinfo{person}{James Fox}, \bibinfo{person}{Alex Watkins}, \bibinfo{person}{Alok Tripathy}, \bibinfo{person}{Kasimir Gabert}, \bibinfo{person}{Euna Kim}, \bibinfo{person}{Xiaojing An}, \bibinfo{person}{Kumar Aatish}, {and} \bibinfo{person}{David~A Bader}.} \bibinfo{year}{2018}\natexlab{}.
\newblock \showarticletitle{Logarithmic radix binning and vectorized triangle counting}. In \bibinfo{booktitle}{\emph{2018 IEEE High Performance extreme Computing Conference (HPEC)}}. IEEE, \bibinfo{pages}{1--7}.
\newblock


\bibitem[Hampton et~al\mbox{.}(2011)]%
        {hampton2011social}
\bibfield{author}{\bibinfo{person}{Keith~N Hampton}, \bibinfo{person}{Lauren~Sessions Goulet}, \bibinfo{person}{Lee Rainie}, {and} \bibinfo{person}{Kristen Purcell}.} \bibinfo{year}{2011}\natexlab{}.
\newblock \bibinfo{booktitle}{\emph{Social networking sites and our lives}}. Vol.~\bibinfo{volume}{1}.
\newblock \bibinfo{publisher}{Pew Internet \& American Life Project Washington, DC}.
\newblock


\bibitem[Han et~al\mbox{.}(2024)]%
        {han2024retrieval}
\bibfield{author}{\bibinfo{person}{Haoyu Han}, \bibinfo{person}{Yu Wang}, \bibinfo{person}{Harry Shomer}, \bibinfo{person}{Kai Guo}, \bibinfo{person}{Jiayuan Ding}, \bibinfo{person}{Yongjia Lei}, \bibinfo{person}{Mahantesh Halappanavar}, \bibinfo{person}{Ryan~A Rossi}, \bibinfo{person}{Subhabrata Mukherjee}, \bibinfo{person}{Xianfeng Tang}, {et~al\mbox{.}}} \bibinfo{year}{2024}\natexlab{}.
\newblock \showarticletitle{Retrieval-augmented generation with graphs (graphrag)}.
\newblock \bibinfo{journal}{\emph{arXiv preprint arXiv:2501.00309}} (\bibinfo{year}{2024}).
\newblock


\bibitem[Hasenplaugh et~al\mbox{.}(2014)]%
        {hasenplaugh2014ordering}
\bibfield{author}{\bibinfo{person}{William Hasenplaugh}, \bibinfo{person}{Tim Kaler}, \bibinfo{person}{Tao~B Schardl}, {and} \bibinfo{person}{Charles~E Leiserson}.} \bibinfo{year}{2014}\natexlab{}.
\newblock \showarticletitle{Ordering heuristics for parallel graph coloring}. In \bibinfo{booktitle}{\emph{Proceedings of the 26th ACM symposium on Parallelism in algorithms and architectures}}. \bibinfo{pages}{166--177}.
\newblock


\bibitem[Himmel et~al\mbox{.}(2016)]%
        {himmel2016enumerating}
\bibfield{author}{\bibinfo{person}{Anne-Sophie Himmel}, \bibinfo{person}{Hendrik Molter}, \bibinfo{person}{Rolf Niedermeier}, {and} \bibinfo{person}{Manuel Sorge}.} \bibinfo{year}{2016}\natexlab{}.
\newblock \showarticletitle{Enumerating maximal cliques in temporal graphs}. In \bibinfo{booktitle}{\emph{2016 IEEE/ACM International Conference on Advances in Social Networks Analysis and Mining (ASONAM)}}. IEEE, \bibinfo{pages}{337--344}.
\newblock


\bibitem[Holmsen(2020)]%
        {holmsen2020large}
\bibfield{author}{\bibinfo{person}{Andreas~F Holmsen}.} \bibinfo{year}{2020}\natexlab{}.
\newblock \showarticletitle{Large cliques in hypergraphs with forbidden substructures}.
\newblock \bibinfo{journal}{\emph{Combinatorica}} \bibinfo{volume}{40}, \bibinfo{number}{4} (\bibinfo{year}{2020}), \bibinfo{pages}{527--537}.
\newblock


\bibitem[Hu et~al\mbox{.}(2019a)]%
        {hu2019discovering}
\bibfield{author}{\bibinfo{person}{Jiafeng Hu}, \bibinfo{person}{Reynold Cheng}, \bibinfo{person}{Kevin Chen-Chuan Chang}, \bibinfo{person}{Aravind Sankar}, \bibinfo{person}{Yixiang Fang}, {and} \bibinfo{person}{Brian~YH Lam}.} \bibinfo{year}{2019}\natexlab{a}.
\newblock \showarticletitle{Discovering maximal motif cliques in large heterogeneous information networks}. In \bibinfo{booktitle}{\emph{2019 IEEE 35th International Conference on Data Engineering (ICDE)}}. IEEE, \bibinfo{pages}{746--757}.
\newblock


\bibitem[Hu et~al\mbox{.}(2019b)]%
        {hu2019triangle}
\bibfield{author}{\bibinfo{person}{Lin Hu}, \bibinfo{person}{Naiqing Guan}, {and} \bibinfo{person}{Lei Zou}.} \bibinfo{year}{2019}\natexlab{b}.
\newblock \showarticletitle{Triangle counting on GPU using fine-grained task distribution}. In \bibinfo{booktitle}{\emph{2019 IEEE 35th International Conference on Data Engineering Workshops (ICDEW)}}. IEEE, \bibinfo{pages}{225--232}.
\newblock


\bibitem[Hu et~al\mbox{.}(2021)]%
        {hu2021accelerating}
\bibfield{author}{\bibinfo{person}{Lin Hu}, \bibinfo{person}{Lei Zou}, {and} \bibinfo{person}{Yu Liu}.} \bibinfo{year}{2021}\natexlab{}.
\newblock \showarticletitle{Accelerating triangle counting on GPU}. In \bibinfo{booktitle}{\emph{Proceedings of the 2021 International Conference on Management of Data}}. \bibinfo{pages}{736--748}.
\newblock


\bibitem[Inae et~al\mbox{.}(2023)]%
        {inae2023motif}
\bibfield{author}{\bibinfo{person}{Eric Inae}, \bibinfo{person}{Gang Liu}, {and} \bibinfo{person}{Meng Jiang}.} \bibinfo{year}{2023}\natexlab{}.
\newblock \showarticletitle{Motif-aware attribute masking for molecular graph pre-training}.
\newblock \bibinfo{journal}{\emph{arXiv preprint arXiv:2309.04589}} (\bibinfo{year}{2023}).
\newblock


\bibitem[Itai and Rodeh(1977)]%
        {itai1977finding}
\bibfield{author}{\bibinfo{person}{Alon Itai} {and} \bibinfo{person}{Michael Rodeh}.} \bibinfo{year}{1977}\natexlab{}.
\newblock \showarticletitle{Finding a minimum circuit in a graph}. In \bibinfo{booktitle}{\emph{Proceedings of the ninth annual ACM symposium on Theory of computing}}. \bibinfo{pages}{1--10}.
\newblock


\bibitem[Jain and Seshadhri(2017)]%
        {jain2017fast}
\bibfield{author}{\bibinfo{person}{Shweta Jain} {and} \bibinfo{person}{C Seshadhri}.} \bibinfo{year}{2017}\natexlab{}.
\newblock \showarticletitle{A fast and provable method for estimating clique counts using tur{\'a}n's theorem}. In \bibinfo{booktitle}{\emph{Proceedings of the 26th international conference on world wide web}}. \bibinfo{pages}{441--449}.
\newblock


\bibitem[Jain and Seshadhri(2020)]%
        {jain2020power}
\bibfield{author}{\bibinfo{person}{Shweta Jain} {and} \bibinfo{person}{C Seshadhri}.} \bibinfo{year}{2020}\natexlab{}.
\newblock \showarticletitle{The power of pivoting for exact clique counting}. In \bibinfo{booktitle}{\emph{Proceedings of the 13th International Conference on Web Search and Data Mining}}. \bibinfo{pages}{268--276}.
\newblock


\bibitem[Jin et~al\mbox{.}(2024)]%
        {jin2024listing}
\bibfield{author}{\bibinfo{person}{Ce Jin}, \bibinfo{person}{Virginia~Vassilevska Williams}, {and} \bibinfo{person}{Renfei Zhou}.} \bibinfo{year}{2024}\natexlab{}.
\newblock \showarticletitle{Listing 6-cycles}. In \bibinfo{booktitle}{\emph{2024 Symposium on Simplicity in Algorithms (SOSA)}}. SIAM, \bibinfo{pages}{19--27}.
\newblock


\bibitem[Klamt and von Kamp(2009)]%
        {klamt2009computing}
\bibfield{author}{\bibinfo{person}{Steffen Klamt} {and} \bibinfo{person}{Axel von Kamp}.} \bibinfo{year}{2009}\natexlab{}.
\newblock \showarticletitle{Computing paths and cycles in biological interaction graphs}.
\newblock \bibinfo{journal}{\emph{BMC bioinformatics}}  \bibinfo{volume}{10} (\bibinfo{year}{2009}), \bibinfo{pages}{1--11}.
\newblock


\bibitem[Kollias et~al\mbox{.}(2024)]%
        {kollias2024counting}
\bibfield{author}{\bibinfo{person}{Georgios Kollias}, \bibinfo{person}{Vassilis Kalantzis}, \bibinfo{person}{Lior Horesh}, \bibinfo{person}{Shashanka Ubaru}, {and} \bibinfo{person}{Panagiotis~A Traganitis}.} \bibinfo{year}{2024}\natexlab{}.
\newblock \showarticletitle{Counting Triangles of Graphs via Matrix Partitioning}. In \bibinfo{booktitle}{\emph{2024 IEEE 34th International Workshop on Machine Learning for Signal Processing (MLSP)}}. IEEE, \bibinfo{pages}{1--6}.
\newblock


\bibitem[Kong et~al\mbox{.}(2013)]%
        {kong2013inferring}
\bibfield{author}{\bibinfo{person}{Xiangnan Kong}, \bibinfo{person}{Jiawei Zhang}, {and} \bibinfo{person}{Philip~S Yu}.} \bibinfo{year}{2013}\natexlab{}.
\newblock \showarticletitle{Inferring anchor links across multiple heterogeneous social networks}. In \bibinfo{booktitle}{\emph{Proceedings of the 22nd ACM international conference on Information \& Knowledge Management}}. \bibinfo{pages}{179--188}.
\newblock


\bibitem[Kovanen et~al\mbox{.}(2011)]%
        {kovanen2011temporal}
\bibfield{author}{\bibinfo{person}{Lauri Kovanen}, \bibinfo{person}{M{\'a}rton Karsai}, \bibinfo{person}{Kimmo Kaski}, \bibinfo{person}{J{\'a}nos Kert{\'e}sz}, {and} \bibinfo{person}{Jari Saram{\"a}ki}.} \bibinfo{year}{2011}\natexlab{}.
\newblock \showarticletitle{Temporal motifs in time-dependent networks}.
\newblock \bibinfo{journal}{\emph{Journal of Statistical Mechanics: Theory and Experiment}} \bibinfo{volume}{2011}, \bibinfo{number}{11} (\bibinfo{year}{2011}), \bibinfo{pages}{P11005}.
\newblock


\bibitem[Latapy(2006)]%
        {latapy2006theory}
\bibfield{author}{\bibinfo{person}{Matthieu Latapy}.} \bibinfo{year}{2006}\natexlab{}.
\newblock \showarticletitle{Theory and practice of triangle problems in very large (sparse (power-law)) graphs}.
\newblock \bibinfo{journal}{\emph{arXiv preprint cs/0609116}} (\bibinfo{year}{2006}).
\newblock


\bibitem[Latapy(2008)]%
        {latapy2008main}
\bibfield{author}{\bibinfo{person}{Matthieu Latapy}.} \bibinfo{year}{2008}\natexlab{}.
\newblock \showarticletitle{Main-memory triangle computations for very large (sparse (power-law)) graphs}.
\newblock \bibinfo{journal}{\emph{Theoretical computer science}} \bibinfo{volume}{407}, \bibinfo{number}{1-3} (\bibinfo{year}{2008}), \bibinfo{pages}{458--473}.
\newblock


\bibitem[Latapy et~al\mbox{.}(2008)]%
        {latapy2008basic}
\bibfield{author}{\bibinfo{person}{Matthieu Latapy}, \bibinfo{person}{Cl{\'e}mence Magnien}, {and} \bibinfo{person}{Nathalie Del~Vecchio}.} \bibinfo{year}{2008}\natexlab{}.
\newblock \showarticletitle{Basic notions for the analysis of large two-mode networks}.
\newblock \bibinfo{journal}{\emph{Social networks}} \bibinfo{volume}{30}, \bibinfo{number}{1} (\bibinfo{year}{2008}), \bibinfo{pages}{31--48}.
\newblock


\bibitem[L{\'e}cuyer et~al\mbox{.}(2023)]%
        {lecuyer2023tailored}
\bibfield{author}{\bibinfo{person}{Fabrice L{\'e}cuyer}, \bibinfo{person}{Louis Jachiet}, \bibinfo{person}{Cl{\'e}mence Magnien}, {and} \bibinfo{person}{Lionel Tabourier}.} \bibinfo{year}{2023}\natexlab{}.
\newblock \showarticletitle{Tailored vertex ordering for faster triangle listing in large graphs}. In \bibinfo{booktitle}{\emph{2023 Proceedings of the Symposium on Algorithm Engineering and Experiments (ALENEX)}}. SIAM, \bibinfo{pages}{77--85}.
\newblock


\bibitem[Lee et~al\mbox{.}(2020)]%
        {lee2020hypergraph}
\bibfield{author}{\bibinfo{person}{Geon Lee}, \bibinfo{person}{Jihoon Ko}, {and} \bibinfo{person}{Kijung Shin}.} \bibinfo{year}{2020}\natexlab{}.
\newblock \showarticletitle{Hypergraph motifs: concepts, algorithms, and discoveries}.
\newblock \bibinfo{journal}{\emph{arXiv preprint arXiv:2003.01853}} (\bibinfo{year}{2020}).
\newblock


\bibitem[Lee et~al\mbox{.}(2024)]%
        {lee2024hypergraph}
\bibfield{author}{\bibinfo{person}{Geon Lee}, \bibinfo{person}{Seokbum Yoon}, \bibinfo{person}{Jihoon Ko}, \bibinfo{person}{Hyunju Kim}, {and} \bibinfo{person}{Kijung Shin}.} \bibinfo{year}{2024}\natexlab{}.
\newblock \showarticletitle{Hypergraph motifs and their extensions beyond binary}.
\newblock \bibinfo{journal}{\emph{The VLDB Journal}} \bibinfo{volume}{33}, \bibinfo{number}{3} (\bibinfo{year}{2024}), \bibinfo{pages}{625--665}.
\newblock


\bibitem[Lee et~al\mbox{.}(2010)]%
        {lee2010survey}
\bibfield{author}{\bibinfo{person}{Victor~E Lee}, \bibinfo{person}{Ning Ruan}, \bibinfo{person}{Ruoming Jin}, {and} \bibinfo{person}{Charu Aggarwal}.} \bibinfo{year}{2010}\natexlab{}.
\newblock \showarticletitle{A survey of algorithms for dense subgraph discovery}.
\newblock \bibinfo{journal}{\emph{Managing and mining graph data}} (\bibinfo{year}{2010}), \bibinfo{pages}{303--336}.
\newblock


\bibitem[Letsios et~al\mbox{.}(2016)]%
        {letsios2016finding}
\bibfield{author}{\bibinfo{person}{Matthaios Letsios}, \bibinfo{person}{Oana~Denisa Balalau}, \bibinfo{person}{Maximilien Danisch}, \bibinfo{person}{Emmanuel Orsini}, {and} \bibinfo{person}{Mauro Sozio}.} \bibinfo{year}{2016}\natexlab{}.
\newblock \showarticletitle{Finding heaviest k-subgraphs and events in social media}. In \bibinfo{booktitle}{\emph{2016 IEEE 16th International Conference on Data Mining Workshops (ICDMW)}}. IEEE, \bibinfo{pages}{113--120}.
\newblock


\bibitem[Li et~al\mbox{.}(2024)]%
        {li2024simple}
\bibfield{author}{\bibinfo{person}{Mufei Li}, \bibinfo{person}{Siqi Miao}, {and} \bibinfo{person}{Pan Li}.} \bibinfo{year}{2024}\natexlab{}.
\newblock \showarticletitle{Simple is effective: The roles of graphs and large language models in knowledge-graph-based retrieval-augmented generation}.
\newblock \bibinfo{journal}{\emph{arXiv preprint arXiv:2410.20724}} (\bibinfo{year}{2024}).
\newblock


\bibitem[Li et~al\mbox{.}(2020a)]%
        {li2020ordering}
\bibfield{author}{\bibinfo{person}{Ronghua Li}, \bibinfo{person}{Sen Gao}, \bibinfo{person}{Lu Qin}, \bibinfo{person}{Guoren Wang}, \bibinfo{person}{Weihua Yang}, {and} \bibinfo{person}{Jeffrey~Xu Yu}.} \bibinfo{year}{2020}\natexlab{a}.
\newblock \showarticletitle{Ordering Heuristics for k-clique Listing.}
\newblock \bibinfo{journal}{\emph{Proc. VLDB Endow.}} (\bibinfo{year}{2020}).
\newblock


\bibitem[Li et~al\mbox{.}(2020b)]%
        {li2020hierarchical}
\bibfield{author}{\bibinfo{person}{Zhao Li}, \bibinfo{person}{Xin Shen}, \bibinfo{person}{Yuhang Jiao}, \bibinfo{person}{Xuming Pan}, \bibinfo{person}{Pengcheng Zou}, \bibinfo{person}{Xianling Meng}, \bibinfo{person}{Chengwei Yao}, {and} \bibinfo{person}{Jiajun Bu}.} \bibinfo{year}{2020}\natexlab{b}.
\newblock \showarticletitle{Hierarchical bipartite graph neural networks: Towards large-scale e-commerce applications}. In \bibinfo{booktitle}{\emph{2020 IEEE 36th International Conference on Data Engineering (ICDE)}}. IEEE, \bibinfo{pages}{1677--1688}.
\newblock


\bibitem[Lind et~al\mbox{.}(2005)]%
        {lind2005cycles}
\bibfield{author}{\bibinfo{person}{Pedro~G Lind}, \bibinfo{person}{Marta~C Gonzalez}, {and} \bibinfo{person}{Hans~J Herrmann}.} \bibinfo{year}{2005}\natexlab{}.
\newblock \showarticletitle{Cycles and clustering in bipartite networks}.
\newblock \bibinfo{journal}{\emph{Physical Review E—Statistical, Nonlinear, and Soft Matter Physics}} \bibinfo{volume}{72}, \bibinfo{number}{5} (\bibinfo{year}{2005}), \bibinfo{pages}{056127}.
\newblock


\bibitem[Liu et~al\mbox{.}(2019)]%
        {liu2019efficient}
\bibfield{author}{\bibinfo{person}{Boge Liu}, \bibinfo{person}{Long Yuan}, \bibinfo{person}{Xuemin Lin}, \bibinfo{person}{Lu Qin}, \bibinfo{person}{Wenjie Zhang}, {and} \bibinfo{person}{Jingren Zhou}.} \bibinfo{year}{2019}\natexlab{}.
\newblock \showarticletitle{Efficient ($\alpha$, $\beta$)-core computation: An index-based approach}. In \bibinfo{booktitle}{\emph{The World Wide Web Conference}}. \bibinfo{pages}{1130--1141}.
\newblock


\bibitem[Liu et~al\mbox{.}(2021)]%
        {liu2021temporal}
\bibfield{author}{\bibinfo{person}{Penghang Liu}, \bibinfo{person}{Valerio Guarrasi}, {and} \bibinfo{person}{Ahmet~Erdem Sar{\i}y{\"u}ce}.} \bibinfo{year}{2021}\natexlab{}.
\newblock \showarticletitle{Temporal network motifs: Models, limitations, evaluation}.
\newblock \bibinfo{journal}{\emph{IEEE Transactions on Knowledge and Data Engineering}} \bibinfo{volume}{35}, \bibinfo{number}{1} (\bibinfo{year}{2021}), \bibinfo{pages}{945--957}.
\newblock


\bibitem[Liu et~al\mbox{.}(2020)]%
        {liu2020truss}
\bibfield{author}{\bibinfo{person}{Qing Liu}, \bibinfo{person}{Minjun Zhao}, \bibinfo{person}{Xin Huang}, \bibinfo{person}{Jianliang Xu}, {and} \bibinfo{person}{Yunjun Gao}.} \bibinfo{year}{2020}\natexlab{}.
\newblock \showarticletitle{Truss-based community search over large directed graphs}. In \bibinfo{booktitle}{\emph{Proceedings of the 2020 ACM SIGMOD International Conference on Management of Data}}. \bibinfo{pages}{2183--2197}.
\newblock


\bibitem[Low et~al\mbox{.}(2017)]%
        {low2017first}
\bibfield{author}{\bibinfo{person}{Tze~Meng Low}, \bibinfo{person}{Varun~Nagaraj Rao}, \bibinfo{person}{Matthew Lee}, \bibinfo{person}{Doru Popovici}, \bibinfo{person}{Franz Franchetti}, {and} \bibinfo{person}{Scott McMillan}.} \bibinfo{year}{2017}\natexlab{}.
\newblock \showarticletitle{First look: Linear algebra-based triangle counting without matrix multiplication}. In \bibinfo{booktitle}{\emph{2017 IEEE High Performance Extreme Computing Conference (HPEC)}}. IEEE, \bibinfo{pages}{1--6}.
\newblock


\bibitem[Luo et~al\mbox{.}(2021)]%
        {luo2021detecting}
\bibfield{author}{\bibinfo{person}{Linhao Luo}, \bibinfo{person}{Yixiang Fang}, \bibinfo{person}{Xin Cao}, \bibinfo{person}{Xiaofeng Zhang}, {and} \bibinfo{person}{Wenjie Zhang}.} \bibinfo{year}{2021}\natexlab{}.
\newblock \showarticletitle{Detecting communities from heterogeneous graphs: A context path-based graph neural network model}. In \bibinfo{booktitle}{\emph{Proceedings of the 30th ACM international conference on information \& knowledge management}}. \bibinfo{pages}{1170--1180}.
\newblock


\bibitem[Ma et~al\mbox{.}(2019)]%
        {ma2019linc}
\bibfield{author}{\bibinfo{person}{Chenhao Ma}, \bibinfo{person}{Reynold Cheng}, \bibinfo{person}{Laks~VS Lakshmanan}, \bibinfo{person}{Tobias Grubenmann}, \bibinfo{person}{Yixiang Fang}, {and} \bibinfo{person}{Xiaodong Li}.} \bibinfo{year}{2019}\natexlab{}.
\newblock \showarticletitle{Linc: a motif counting algorithm for uncertain graphs}.
\newblock \bibinfo{journal}{\emph{Proceedings of the VLDB Endowment}} \bibinfo{volume}{13}, \bibinfo{number}{2} (\bibinfo{year}{2019}), \bibinfo{pages}{155--168}.
\newblock


\bibitem[Makino and Uno(2004)]%
        {makino2004new}
\bibfield{author}{\bibinfo{person}{Kazuhisa Makino} {and} \bibinfo{person}{Takeaki Uno}.} \bibinfo{year}{2004}\natexlab{}.
\newblock \showarticletitle{New algorithms for enumerating all maximal cliques}. In \bibinfo{booktitle}{\emph{Algorithm Theory-SWAT 2004: 9th Scandinavian Workshop on Algorithm Theory, Humleb{\ae}k, Denmark, July 8-10, 2004. Proceedings 9}}. Springer, \bibinfo{pages}{260--272}.
\newblock


\bibitem[Marcus and Shavitt(2010)]%
        {marcus2010efficient}
\bibfield{author}{\bibinfo{person}{Dror Marcus} {and} \bibinfo{person}{Yuval Shavitt}.} \bibinfo{year}{2010}\natexlab{}.
\newblock \showarticletitle{Efficient counting of network motifs}. In \bibinfo{booktitle}{\emph{2010 IEEE 30th International Conference on Distributed Computing Systems Workshops}}. IEEE, \bibinfo{pages}{92--98}.
\newblock


\bibitem[Masoudi-Nejad et~al\mbox{.}(2012)]%
        {masoudi2012building}
\bibfield{author}{\bibinfo{person}{Ali Masoudi-Nejad}, \bibinfo{person}{Falk Schreiber}, {and} \bibinfo{person}{Zahra Razaghi~Moghadam Kashani}.} \bibinfo{year}{2012}\natexlab{}.
\newblock \showarticletitle{Building blocks of biological networks: a review on major network motif discovery algorithms}.
\newblock \bibinfo{journal}{\emph{IET systems biology}} \bibinfo{volume}{6}, \bibinfo{number}{5} (\bibinfo{year}{2012}), \bibinfo{pages}{164--174}.
\newblock


\bibitem[Opsahl(2013)]%
        {opsahl2013triadic}
\bibfield{author}{\bibinfo{person}{Tore Opsahl}.} \bibinfo{year}{2013}\natexlab{}.
\newblock \showarticletitle{Triadic closure in two-mode networks: Redefining the global and local clustering coefficients}.
\newblock \bibinfo{journal}{\emph{Social networks}} \bibinfo{volume}{35}, \bibinfo{number}{2} (\bibinfo{year}{2013}), \bibinfo{pages}{159--167}.
\newblock


\bibitem[Pandey et~al\mbox{.}(2019)]%
        {pandey2019h}
\bibfield{author}{\bibinfo{person}{Santosh Pandey}, \bibinfo{person}{Xiaoye~Sherry Li}, \bibinfo{person}{Aydin Buluc}, \bibinfo{person}{Jiejun Xu}, {and} \bibinfo{person}{Hang Liu}.} \bibinfo{year}{2019}\natexlab{}.
\newblock \showarticletitle{H-index: Hash-indexing for parallel triangle counting on GPUs}. In \bibinfo{booktitle}{\emph{2019 IEEE high performance extreme computing conference (HPEC)}}. IEEE, \bibinfo{pages}{1--7}.
\newblock


\bibitem[Pandey et~al\mbox{.}(2021)]%
        {pandey2021trust}
\bibfield{author}{\bibinfo{person}{Santosh Pandey}, \bibinfo{person}{Zhibin Wang}, \bibinfo{person}{Sheng Zhong}, \bibinfo{person}{Chen Tian}, \bibinfo{person}{Bolong Zheng}, \bibinfo{person}{Xiaoye Li}, \bibinfo{person}{Lingda Li}, \bibinfo{person}{Adolfy Hoisie}, \bibinfo{person}{Caiwen Ding}, \bibinfo{person}{Dong Li}, {et~al\mbox{.}}} \bibinfo{year}{2021}\natexlab{}.
\newblock \showarticletitle{Trust: Triangle counting reloaded on GPUs}.
\newblock \bibinfo{journal}{\emph{IEEE Transactions on Parallel and Distributed Systems}} \bibinfo{volume}{32}, \bibinfo{number}{11} (\bibinfo{year}{2021}), \bibinfo{pages}{2646--2660}.
\newblock


\bibitem[Park et~al\mbox{.}(2016)]%
        {park2016pte}
\bibfield{author}{\bibinfo{person}{Ha-Myung Park}, \bibinfo{person}{Sung-Hyon Myaeng}, {and} \bibinfo{person}{U Kang}.} \bibinfo{year}{2016}\natexlab{}.
\newblock \showarticletitle{Pte: Enumerating trillion triangles on distributed systems}. In \bibinfo{booktitle}{\emph{Proceedings of the 22nd ACM SIGKDD International Conference on Knowledge Discovery and Data Mining}}. \bibinfo{pages}{1115--1124}.
\newblock


\bibitem[Pashanasangi and Seshadhri(2021)]%
        {pashanasangi2021faster}
\bibfield{author}{\bibinfo{person}{Noujan Pashanasangi} {and} \bibinfo{person}{C Seshadhri}.} \bibinfo{year}{2021}\natexlab{}.
\newblock \showarticletitle{Faster and generalized temporal triangle counting, via degeneracy ordering}. In \bibinfo{booktitle}{\emph{Proceedings of the 27th ACM SIGKDD Conference on Knowledge Discovery \& Data Mining}}. \bibinfo{pages}{1319--1328}.
\newblock


\bibitem[Peng et~al\mbox{.}(2018)]%
        {peng2018efficient}
\bibfield{author}{\bibinfo{person}{You Peng}, \bibinfo{person}{Ying Zhang}, \bibinfo{person}{Wenjie Zhang}, \bibinfo{person}{Xuemin Lin}, {and} \bibinfo{person}{Lu Qin}.} \bibinfo{year}{2018}\natexlab{}.
\newblock \showarticletitle{Efficient probabilistic k-core computation on uncertain graphs}. In \bibinfo{booktitle}{\emph{2018 IEEE 34th International Conference on Data Engineering (ICDE)}}. IEEE, \bibinfo{pages}{1192--1203}.
\newblock


\bibitem[Plant and Moura(2019)]%
        {plant2019maximum}
\bibfield{author}{\bibinfo{person}{Lachlan Plant} {and} \bibinfo{person}{Lucia Moura}.} \bibinfo{year}{2019}\natexlab{}.
\newblock \showarticletitle{Maximum Clique Exhaustive Search in Circulant k-Hypergraphs}. In \bibinfo{booktitle}{\emph{Combinatorial Algorithms: 30th International Workshop, IWOCA 2019, Pisa, Italy, July 23--25, 2019, Proceedings 30}}. Springer, \bibinfo{pages}{378--392}.
\newblock


\bibitem[Polak(2016)]%
        {polak2016counting}
\bibfield{author}{\bibinfo{person}{Adam Polak}.} \bibinfo{year}{2016}\natexlab{}.
\newblock \showarticletitle{Counting triangles in large graphs on GPU}. In \bibinfo{booktitle}{\emph{2016 IEEE International Parallel and Distributed Processing Symposium Workshops (IPDPSW)}}. IEEE, \bibinfo{pages}{740--746}.
\newblock


\bibitem[Qiu et~al\mbox{.}(2024)]%
        {qiu2024accelerating}
\bibfield{author}{\bibinfo{person}{Linshan Qiu}, \bibinfo{person}{Zhonggen Li}, \bibinfo{person}{Xiangyu Ke}, \bibinfo{person}{Lu Chen}, {and} \bibinfo{person}{Yunjun Gao}.} \bibinfo{year}{2024}\natexlab{}.
\newblock \showarticletitle{Accelerating Biclique Counting on GPU}.
\newblock \bibinfo{journal}{\emph{arXiv preprint arXiv:2403.07858}} (\bibinfo{year}{2024}).
\newblock


\bibitem[Radicchi et~al\mbox{.}(2004)]%
        {radicchi2004defining}
\bibfield{author}{\bibinfo{person}{Filippo Radicchi}, \bibinfo{person}{Claudio Castellano}, \bibinfo{person}{Federico Cecconi}, \bibinfo{person}{Vittorio Loreto}, {and} \bibinfo{person}{Domenico Parisi}.} \bibinfo{year}{2004}\natexlab{}.
\newblock \showarticletitle{Defining and identifying communities in networks}.
\newblock \bibinfo{journal}{\emph{Proceedings of the national academy of sciences}} \bibinfo{volume}{101}, \bibinfo{number}{9} (\bibinfo{year}{2004}), \bibinfo{pages}{2658--2663}.
\newblock


\bibitem[Rossi et~al\mbox{.}(2013)]%
        {rossi2013multi}
\bibfield{author}{\bibinfo{person}{Ryan Rossi}, \bibinfo{person}{Sonia Fahmy}, {and} \bibinfo{person}{Nilothpal Talukder}.} \bibinfo{year}{2013}\natexlab{}.
\newblock \showarticletitle{A multi-level approach for evaluating internet topology generators}. In \bibinfo{booktitle}{\emph{2013 IFIP Networking Conference}}. IEEE, \bibinfo{pages}{1--9}.
\newblock


\bibitem[Rossi et~al\mbox{.}(2019)]%
        {rossi2019heterogeneous}
\bibfield{author}{\bibinfo{person}{Ryan~A Rossi}, \bibinfo{person}{Nesreen~K Ahmed}, \bibinfo{person}{Aldo Carranza}, \bibinfo{person}{David Arbour}, \bibinfo{person}{Anup Rao}, \bibinfo{person}{Sungchul Kim}, {and} \bibinfo{person}{Eunyee Koh}.} \bibinfo{year}{2019}\natexlab{}.
\newblock \showarticletitle{Heterogeneous network motifs}.
\newblock \bibinfo{journal}{\emph{arXiv preprint arXiv:1901.10026}} (\bibinfo{year}{2019}).
\newblock


\bibitem[Rossi et~al\mbox{.}(2020)]%
        {rossi2020heterogeneous}
\bibfield{author}{\bibinfo{person}{Ryan~A Rossi}, \bibinfo{person}{Nesreen~K Ahmed}, \bibinfo{person}{Aldo Carranza}, \bibinfo{person}{David Arbour}, \bibinfo{person}{Anup Rao}, \bibinfo{person}{Sungchul Kim}, {and} \bibinfo{person}{Eunyee Koh}.} \bibinfo{year}{2020}\natexlab{}.
\newblock \showarticletitle{Heterogeneous graphlets}.
\newblock \bibinfo{journal}{\emph{ACM Transactions on Knowledge Discovery from Data (TKDD)}} \bibinfo{volume}{15}, \bibinfo{number}{1} (\bibinfo{year}{2020}), \bibinfo{pages}{1--43}.
\newblock


\bibitem[Sanei-Mehri et~al\mbox{.}(2018)]%
        {sanei2018butterfly}
\bibfield{author}{\bibinfo{person}{Seyed-Vahid Sanei-Mehri}, \bibinfo{person}{Ahmet~Erdem Sariyuce}, {and} \bibinfo{person}{Srikanta Tirthapura}.} \bibinfo{year}{2018}\natexlab{}.
\newblock \showarticletitle{Butterfly counting in bipartite networks}. In \bibinfo{booktitle}{\emph{Proceedings of the 24th ACM SIGKDD International Conference on Knowledge Discovery \& Data Mining}}. \bibinfo{pages}{2150--2159}.
\newblock


\bibitem[Sanei-Mehri et~al\mbox{.}(2019)]%
        {sanei2019fleet}
\bibfield{author}{\bibinfo{person}{Seyed-Vahid Sanei-Mehri}, \bibinfo{person}{Yu Zhang}, \bibinfo{person}{Ahmet~Erdem Sariy{\"u}ce}, {and} \bibinfo{person}{Srikanta Tirthapura}.} \bibinfo{year}{2019}\natexlab{}.
\newblock \showarticletitle{Fleet: Butterfly estimation from a bipartite graph stream}. In \bibinfo{booktitle}{\emph{Proceedings of the 28th ACM International Conference on Information and Knowledge Management}}. \bibinfo{pages}{1201--1210}.
\newblock


\bibitem[Schank(2007)]%
        {schank2007algorithmic}
\bibfield{author}{\bibinfo{person}{Thomas Schank}.} \bibinfo{year}{2007}\natexlab{}.
\newblock \showarticletitle{Algorithmic aspects of triangle-based network analysis}.
\newblock  (\bibinfo{year}{2007}).
\newblock


\bibitem[Schank and Wagner(2005)]%
        {schank2005finding}
\bibfield{author}{\bibinfo{person}{Thomas Schank} {and} \bibinfo{person}{Dorothea Wagner}.} \bibinfo{year}{2005}\natexlab{}.
\newblock \showarticletitle{Finding, counting and listing all triangles in large graphs, an experimental study}. In \bibinfo{booktitle}{\emph{International workshop on experimental and efficient algorithms}}. Springer, \bibinfo{pages}{606--609}.
\newblock


\bibitem[Schweitzer et~al\mbox{.}(2009)]%
        {schweitzer2009economic}
\bibfield{author}{\bibinfo{person}{Frank Schweitzer}, \bibinfo{person}{Giorgio Fagiolo}, \bibinfo{person}{Didier Sornette}, \bibinfo{person}{Fernando Vega-Redondo}, \bibinfo{person}{Alessandro Vespignani}, {and} \bibinfo{person}{Douglas~R White}.} \bibinfo{year}{2009}\natexlab{}.
\newblock \showarticletitle{Economic networks: The new challenges}.
\newblock \bibinfo{journal}{\emph{science}} \bibinfo{volume}{325}, \bibinfo{number}{5939} (\bibinfo{year}{2009}), \bibinfo{pages}{422--425}.
\newblock


\bibitem[Sheshbolouki and {\"O}zsu(2022)]%
        {sheshbolouki2022sgrapp}
\bibfield{author}{\bibinfo{person}{Aida Sheshbolouki} {and} \bibinfo{person}{M~Tamer {\"O}zsu}.} \bibinfo{year}{2022}\natexlab{}.
\newblock \showarticletitle{sGrapp: Butterfly approximation in streaming graphs}.
\newblock \bibinfo{journal}{\emph{ACM Transactions on Knowledge Discovery from Data (TKDD)}} \bibinfo{volume}{16}, \bibinfo{number}{4} (\bibinfo{year}{2022}), \bibinfo{pages}{1--43}.
\newblock


\bibitem[Shi et~al\mbox{.}(2021)]%
        {shi2021parallel}
\bibfield{author}{\bibinfo{person}{Jessica Shi}, \bibinfo{person}{Laxman Dhulipala}, {and} \bibinfo{person}{Julian Shun}.} \bibinfo{year}{2021}\natexlab{}.
\newblock \showarticletitle{Parallel clique counting and peeling algorithms}. In \bibinfo{booktitle}{\emph{SIAM Conference on Applied and Computational Discrete Algorithms (ACDA21)}}. SIAM, \bibinfo{pages}{135--146}.
\newblock


\bibitem[Shi and Shun(2022)]%
        {shi2022parallel}
\bibfield{author}{\bibinfo{person}{Jessica Shi} {and} \bibinfo{person}{Julian Shun}.} \bibinfo{year}{2022}\natexlab{}.
\newblock \showarticletitle{Parallel algorithms for butterfly computations}.
\newblock In \bibinfo{booktitle}{\emph{Massive Graph Analytics}}. \bibinfo{publisher}{Chapman and Hall/CRC}, \bibinfo{pages}{287--330}.
\newblock


\bibitem[Stefani et~al\mbox{.}(2017)]%
        {stefani2017triest}
\bibfield{author}{\bibinfo{person}{Lorenzo~De Stefani}, \bibinfo{person}{Alessandro Epasto}, \bibinfo{person}{Matteo Riondato}, {and} \bibinfo{person}{Eli Upfal}.} \bibinfo{year}{2017}\natexlab{}.
\newblock \showarticletitle{Triest: Counting local and global triangles in fully dynamic streams with fixed memory size}.
\newblock \bibinfo{journal}{\emph{ACM Transactions on Knowledge Discovery from Data (TKDD)}} \bibinfo{volume}{11}, \bibinfo{number}{4} (\bibinfo{year}{2017}), \bibinfo{pages}{1--50}.
\newblock


\bibitem[Strassen(1986)]%
        {strassen1986asymptotic}
\bibfield{author}{\bibinfo{person}{Volker Strassen}.} \bibinfo{year}{1986}\natexlab{}.
\newblock \showarticletitle{The asymptotic spectrum of tensors and the exponent of matrix multiplication}. In \bibinfo{booktitle}{\emph{27th Annual Symposium on Foundations of Computer Science (sfcs 1986)}}. IEEE, \bibinfo{pages}{49--54}.
\newblock


\bibitem[Suri and Vassilvitskii(2011)]%
        {suri2011counting}
\bibfield{author}{\bibinfo{person}{Siddharth Suri} {and} \bibinfo{person}{Sergei Vassilvitskii}.} \bibinfo{year}{2011}\natexlab{}.
\newblock \showarticletitle{Counting triangles and the curse of the last reducer}. In \bibinfo{booktitle}{\emph{Proceedings of the 20th international conference on World wide web}}. \bibinfo{pages}{607--614}.
\newblock


\bibitem[Takeaki(2012)]%
        {takeaki2012implementation}
\bibfield{author}{\bibinfo{person}{U Takeaki}.} \bibinfo{year}{2012}\natexlab{}.
\newblock \showarticletitle{Implementation issues of clique enumeration algorithm}.
\newblock \bibinfo{journal}{\emph{Special issue: theoretical computer science and discrete mathematics, Progress in Informatics}} (\bibinfo{year}{2012}), \bibinfo{pages}{25}.
\newblock


\bibitem[Thakrar(2024)]%
        {thakrar2024dynagrag}
\bibfield{author}{\bibinfo{person}{Karishma Thakrar}.} \bibinfo{year}{2024}\natexlab{}.
\newblock \showarticletitle{DynaGRAG: Improving Language Understanding and Generation through Dynamic Subgraph Representation in Graph Retrieval-Augmented Generation}.
\newblock \bibinfo{journal}{\emph{arXiv preprint arXiv:2412.18644}} (\bibinfo{year}{2024}).
\newblock


\bibitem[Tsourakakis(2008)]%
        {tsourakakis2008fast}
\bibfield{author}{\bibinfo{person}{Charalampos~E Tsourakakis}.} \bibinfo{year}{2008}\natexlab{}.
\newblock \showarticletitle{Fast counting of triangles in large real networks without counting: Algorithms and laws}. In \bibinfo{booktitle}{\emph{2008 Eighth IEEE International Conference on Data Mining}}. IEEE, \bibinfo{pages}{608--617}.
\newblock


\bibitem[Tsourakakis et~al\mbox{.}(2009)]%
        {tsourakakis2009doulion}
\bibfield{author}{\bibinfo{person}{Charalampos~E Tsourakakis}, \bibinfo{person}{U Kang}, \bibinfo{person}{Gary~L Miller}, {and} \bibinfo{person}{Christos Faloutsos}.} \bibinfo{year}{2009}\natexlab{}.
\newblock \showarticletitle{Doulion: counting triangles in massive graphs with a coin}. In \bibinfo{booktitle}{\emph{Proceedings of the 15th ACM SIGKDD international conference on Knowledge discovery and data mining}}. \bibinfo{pages}{837--846}.
\newblock


\bibitem[Tsukiyama et~al\mbox{.}(1977)]%
        {tsukiyama1977new}
\bibfield{author}{\bibinfo{person}{Shuji Tsukiyama}, \bibinfo{person}{Mikio Ide}, \bibinfo{person}{Hiromu Ariyoshi}, {and} \bibinfo{person}{Isao Shirakawa}.} \bibinfo{year}{1977}\natexlab{}.
\newblock \showarticletitle{A new algorithm for generating all the maximal independent sets}.
\newblock \bibinfo{journal}{\emph{SIAM J. Comput.}} \bibinfo{volume}{6}, \bibinfo{number}{3} (\bibinfo{year}{1977}), \bibinfo{pages}{505--517}.
\newblock


\bibitem[Turn(1941)]%
        {turn1941extremal}
\bibfield{author}{\bibinfo{person}{Paul Turn}.} \bibinfo{year}{1941}\natexlab{}.
\newblock \showarticletitle{On an extremal problem in graph theory}.
\newblock \bibinfo{journal}{\emph{Mat. Fiz. Lapok}}  \bibinfo{volume}{48} (\bibinfo{year}{1941}), \bibinfo{pages}{436452}.
\newblock


\bibitem[Walker et~al\mbox{.}(2017)]%
        {walker2017uncertain}
\bibfield{author}{\bibinfo{person}{Josephine~G Walker}, \bibinfo{person}{Michaela Plein}, \bibinfo{person}{Eric~R Morgan}, {and} \bibinfo{person}{Peter~A Vesk}.} \bibinfo{year}{2017}\natexlab{}.
\newblock \showarticletitle{Uncertain links in host--parasite networks: lessons for parasite transmission in a multi-host system}.
\newblock \bibinfo{journal}{\emph{Philosophical Transactions of the Royal Society B: Biological Sciences}} \bibinfo{volume}{372}, \bibinfo{number}{1719} (\bibinfo{year}{2017}), \bibinfo{pages}{20160095}.
\newblock


\bibitem[Wang and Cheng(2012)]%
        {wang2012truss}
\bibfield{author}{\bibinfo{person}{Jia Wang} {and} \bibinfo{person}{James Cheng}.} \bibinfo{year}{2012}\natexlab{}.
\newblock \showarticletitle{Truss decomposition in massive networks}.
\newblock \bibinfo{journal}{\emph{arXiv preprint arXiv:1205.6693}} (\bibinfo{year}{2012}).
\newblock


\bibitem[Wang et~al\mbox{.}(2014)]%
        {wang2014rectangle}
\bibfield{author}{\bibinfo{person}{Jia Wang}, \bibinfo{person}{Ada Wai-Chee Fu}, {and} \bibinfo{person}{James Cheng}.} \bibinfo{year}{2014}\natexlab{}.
\newblock \showarticletitle{Rectangle counting in large bipartite graphs}. In \bibinfo{booktitle}{\emph{2014 IEEE International Congress on Big Data}}. IEEE, \bibinfo{pages}{17--24}.
\newblock


\bibitem[Wang et~al\mbox{.}(2020)]%
        {wang2020efficient}
\bibfield{author}{\bibinfo{person}{Jingjing Wang}, \bibinfo{person}{Yanhao Wang}, \bibinfo{person}{Wenjun Jiang}, \bibinfo{person}{Yuchen Li}, {and} \bibinfo{person}{Kian-Lee Tan}.} \bibinfo{year}{2020}\natexlab{}.
\newblock \showarticletitle{Efficient sampling algorithms for approximate temporal motif counting}. In \bibinfo{booktitle}{\emph{Proceedings of the 29th ACM international conference on information \& knowledge management}}. \bibinfo{pages}{1505--1514}.
\newblock


\bibitem[Wang et~al\mbox{.}(2023c)]%
        {wang2023efficient}
\bibfield{author}{\bibinfo{person}{Jianhua Wang}, \bibinfo{person}{Jianye Yang}, \bibinfo{person}{Ziyi Ma}, \bibinfo{person}{Chengyuan Zhang}, \bibinfo{person}{Shiyu Yang}, {and} \bibinfo{person}{Wenjie Zhang}.} \bibinfo{year}{2023}\natexlab{c}.
\newblock \showarticletitle{Efficient maximal biclique enumeration on large uncertain bipartite graphs}.
\newblock \bibinfo{journal}{\emph{IEEE Transactions on Knowledge and Data Engineering}} \bibinfo{volume}{35}, \bibinfo{number}{12} (\bibinfo{year}{2023}), \bibinfo{pages}{12634--12648}.
\newblock


\bibitem[Wang et~al\mbox{.}(2022)]%
        {wang2022efficient}
\bibfield{author}{\bibinfo{person}{Jianhua Wang}, \bibinfo{person}{Jianye Yang}, \bibinfo{person}{Chengyuan Zhang}, {and} \bibinfo{person}{Xuemin Lin}.} \bibinfo{year}{2022}\natexlab{}.
\newblock \showarticletitle{Efficient maximum edge-weighted biclique search on large bipartite graphs}.
\newblock \bibinfo{journal}{\emph{IEEE Transactions on Knowledge and Data Engineering}} \bibinfo{volume}{35}, \bibinfo{number}{8} (\bibinfo{year}{2022}), \bibinfo{pages}{7921--7934}.
\newblock


\bibitem[Wang and Rong(2009)]%
        {wang2009cascade}
\bibfield{author}{\bibinfo{person}{Jian-Wei Wang} {and} \bibinfo{person}{Li-Li Rong}.} \bibinfo{year}{2009}\natexlab{}.
\newblock \showarticletitle{Cascade-based attack vulnerability on the US power grid}.
\newblock \bibinfo{journal}{\emph{Safety science}} \bibinfo{volume}{47}, \bibinfo{number}{10} (\bibinfo{year}{2009}), \bibinfo{pages}{1332--1336}.
\newblock


\bibitem[Wang et~al\mbox{.}(2018)]%
        {wang2018efficient}
\bibfield{author}{\bibinfo{person}{Kai Wang}, \bibinfo{person}{Xin Cao}, \bibinfo{person}{Xuemin Lin}, \bibinfo{person}{Wenjie Zhang}, {and} \bibinfo{person}{Lu Qin}.} \bibinfo{year}{2018}\natexlab{}.
\newblock \showarticletitle{Efficient computing of radius-bounded k-cores}. In \bibinfo{booktitle}{\emph{2018 IEEE 34th international conference on data engineering (ICDE)}}. IEEE, \bibinfo{pages}{233--244}.
\newblock


\bibitem[Wang et~al\mbox{.}(2019)]%
        {wang2019vertex}
\bibfield{author}{\bibinfo{person}{Kai Wang}, \bibinfo{person}{Xuemin Lin}, \bibinfo{person}{Lu Qin}, \bibinfo{person}{Wenjie Zhang}, {and} \bibinfo{person}{Ying Zhang}.} \bibinfo{year}{2019}\natexlab{}.
\newblock \showarticletitle{Vertex Priority Based Butterfly Counting for Large-scale Bipartite Networks.}
\newblock \bibinfo{journal}{\emph{PVLDB}} (\bibinfo{year}{2019}).
\newblock


\bibitem[Wang et~al\mbox{.}(2023b)]%
        {wang2023accelerated}
\bibfield{author}{\bibinfo{person}{Kai Wang}, \bibinfo{person}{Xuemin Lin}, \bibinfo{person}{Lu Qin}, \bibinfo{person}{Wenjie Zhang}, {and} \bibinfo{person}{Ying Zhang}.} \bibinfo{year}{2023}\natexlab{b}.
\newblock \showarticletitle{Accelerated butterfly counting with vertex priority on bipartite graphs}.
\newblock \bibinfo{journal}{\emph{The VLDB Journal}} \bibinfo{volume}{32}, \bibinfo{number}{2} (\bibinfo{year}{2023}), \bibinfo{pages}{257--281}.
\newblock


\bibitem[Wang et~al\mbox{.}(2024b)]%
        {wang2024efficient}
\bibfield{author}{\bibinfo{person}{Kaixin Wang}, \bibinfo{person}{Kaiqiang Yu}, {and} \bibinfo{person}{Cheng Long}.} \bibinfo{year}{2024}\natexlab{b}.
\newblock \showarticletitle{Efficient k-Clique Listing: An Edge-Oriented Branching Strategy}.
\newblock \bibinfo{journal}{\emph{Proceedings of the ACM on Management of Data}} \bibinfo{volume}{2}, \bibinfo{number}{1} (\bibinfo{year}{2024}), \bibinfo{pages}{1--26}.
\newblock


\bibitem[Wang et~al\mbox{.}(2023a)]%
        {wang2023efficient2}
\bibfield{author}{\bibinfo{person}{Zhibin Wang}, \bibinfo{person}{Longbin Lai}, \bibinfo{person}{Yixue Liu}, \bibinfo{person}{Bing Shui}, \bibinfo{person}{Chen Tian}, {and} \bibinfo{person}{Sheng Zhong}.} \bibinfo{year}{2023}\natexlab{a}.
\newblock \showarticletitle{I/o-efficient butterfly counting at scale}.
\newblock \bibinfo{journal}{\emph{Proceedings of the ACM on Management of Data}} \bibinfo{volume}{1}, \bibinfo{number}{1} (\bibinfo{year}{2023}), \bibinfo{pages}{1--27}.
\newblock


\bibitem[Wang et~al\mbox{.}(2024a)]%
        {wang2024parallelization}
\bibfield{author}{\bibinfo{person}{Zhibin Wang}, \bibinfo{person}{Longbin Lai}, \bibinfo{person}{Yixue Liu}, \bibinfo{person}{Bing Shui}, \bibinfo{person}{Chen Tian}, {and} \bibinfo{person}{Sheng Zhong}.} \bibinfo{year}{2024}\natexlab{a}.
\newblock \showarticletitle{Parallelization of butterfly counting on hierarchical memory}.
\newblock \bibinfo{journal}{\emph{The VLDB Journal}} (\bibinfo{year}{2024}), \bibinfo{pages}{1--32}.
\newblock


\bibitem[Watts and Strogatz(1998)]%
        {watts1998collective}
\bibfield{author}{\bibinfo{person}{Duncan~J Watts} {and} \bibinfo{person}{Steven~H Strogatz}.} \bibinfo{year}{1998}\natexlab{}.
\newblock \showarticletitle{Collective dynamics of ‘small-world’networks}.
\newblock \bibinfo{journal}{\emph{nature}} \bibinfo{volume}{393}, \bibinfo{number}{6684} (\bibinfo{year}{1998}), \bibinfo{pages}{440--442}.
\newblock


\bibitem[Wolf et~al\mbox{.}(2017)]%
        {wolf2017fast}
\bibfield{author}{\bibinfo{person}{Michael~M Wolf}, \bibinfo{person}{Mehmet Deveci}, \bibinfo{person}{Jonathan~W Berry}, \bibinfo{person}{Simon~D Hammond}, {and} \bibinfo{person}{Sivasankaran Rajamanickam}.} \bibinfo{year}{2017}\natexlab{}.
\newblock \showarticletitle{Fast linear algebra-based triangle counting with kokkoskernels}. In \bibinfo{booktitle}{\emph{2017 IEEE High Performance Extreme Computing Conference (HPEC)}}. IEEE, \bibinfo{pages}{1--7}.
\newblock


\bibitem[Wong et~al\mbox{.}(2012)]%
        {wong2012biological}
\bibfield{author}{\bibinfo{person}{Elisabeth Wong}, \bibinfo{person}{Brittany Baur}, \bibinfo{person}{Saad Quader}, {and} \bibinfo{person}{Chun-Hsi Huang}.} \bibinfo{year}{2012}\natexlab{}.
\newblock \showarticletitle{Biological network motif detection: principles and practice}.
\newblock \bibinfo{journal}{\emph{Briefings in bioinformatics}} \bibinfo{volume}{13}, \bibinfo{number}{2} (\bibinfo{year}{2012}), \bibinfo{pages}{202--215}.
\newblock


\bibitem[Xia et~al\mbox{.}(2024)]%
        {xia2024gpu}
\bibfield{author}{\bibinfo{person}{Yifei Xia}, \bibinfo{person}{Feng Zhang}, \bibinfo{person}{Qingyu Xu}, \bibinfo{person}{Mingde Zhang}, \bibinfo{person}{Zhiming Yao}, \bibinfo{person}{Lv Lu}, \bibinfo{person}{Xiaoyong Du}, \bibinfo{person}{Dong Deng}, \bibinfo{person}{Bingsheng He}, {and} \bibinfo{person}{Siqi Ma}.} \bibinfo{year}{2024}\natexlab{}.
\newblock \showarticletitle{GPU-based butterfly counting}.
\newblock \bibinfo{journal}{\emph{The VLDB Journal}} (\bibinfo{year}{2024}), \bibinfo{pages}{1--25}.
\newblock


\bibitem[Xu et~al\mbox{.}(2022)]%
        {xu2022efficient}
\bibfield{author}{\bibinfo{person}{Qingyu Xu}, \bibinfo{person}{Feng Zhang}, \bibinfo{person}{Zhiming Yao}, \bibinfo{person}{Lv Lu}, \bibinfo{person}{Xiaoyong Du}, \bibinfo{person}{Dong Deng}, {and} \bibinfo{person}{Bingsheng He}.} \bibinfo{year}{2022}\natexlab{}.
\newblock \showarticletitle{Efficient load-balanced butterfly counting on GPU}.
\newblock \bibinfo{journal}{\emph{Proceedings of the VLDB Endowment}} \bibinfo{volume}{15}, \bibinfo{number}{11} (\bibinfo{year}{2022}), \bibinfo{pages}{2450--2462}.
\newblock


\bibitem[Xu et~al\mbox{.}(2007)]%
        {xu2007scan}
\bibfield{author}{\bibinfo{person}{Xiaowei Xu}, \bibinfo{person}{Nurcan Yuruk}, \bibinfo{person}{Zhidan Feng}, {and} \bibinfo{person}{Thomas~AJ Schweiger}.} \bibinfo{year}{2007}\natexlab{}.
\newblock \showarticletitle{Scan: a structural clustering algorithm for networks}. In \bibinfo{booktitle}{\emph{Proceedings of the 13th ACM SIGKDD international conference on Knowledge discovery and data mining}}. \bibinfo{pages}{824--833}.
\newblock


\bibitem[Yang et~al\mbox{.}(2021b)]%
        {yang2021p}
\bibfield{author}{\bibinfo{person}{Jianye Yang}, \bibinfo{person}{Yun Peng}, {and} \bibinfo{person}{Wenjie Zhang}.} \bibinfo{year}{2021}\natexlab{b}.
\newblock \showarticletitle{(p, q)-biclique counting and enumeration for large sparse bipartite graphs}.
\newblock \bibinfo{journal}{\emph{Proceedings of the VLDB Endowment}} \bibinfo{volume}{15}, \bibinfo{number}{2} (\bibinfo{year}{2021}), \bibinfo{pages}{141--153}.
\newblock


\bibitem[Yang et~al\mbox{.}(2020)]%
        {yang2020effective}
\bibfield{author}{\bibinfo{person}{Yixing Yang}, \bibinfo{person}{Yixiang Fang}, \bibinfo{person}{Xuemin Lin}, {and} \bibinfo{person}{Wenjie Zhang}.} \bibinfo{year}{2020}\natexlab{}.
\newblock \showarticletitle{Effective and efficient truss computation over large heterogeneous information networks}. In \bibinfo{booktitle}{\emph{2020 IEEE 36th international conference on data engineering (ICDE)}}. IEEE, \bibinfo{pages}{901--912}.
\newblock


\bibitem[Yang et~al\mbox{.}(2021a)]%
        {yang2021efficient}
\bibfield{author}{\bibinfo{person}{Yixing Yang}, \bibinfo{person}{Yixiang Fang}, \bibinfo{person}{Maria~E Orlowska}, \bibinfo{person}{Wenjie Zhang}, {and} \bibinfo{person}{Xuemin Lin}.} \bibinfo{year}{2021}\natexlab{a}.
\newblock \showarticletitle{Efficient bi-triangle counting for large bipartite networks}.
\newblock \bibinfo{journal}{\emph{Proceedings of the VLDB Endowment}} \bibinfo{volume}{14}, \bibinfo{number}{6} (\bibinfo{year}{2021}), \bibinfo{pages}{984--996}.
\newblock


\bibitem[Ye et~al\mbox{.}(2022)]%
        {ye2022lightning}
\bibfield{author}{\bibinfo{person}{Xiaowei Ye}, \bibinfo{person}{Rong-Hua Li}, \bibinfo{person}{Qiangqiang Dai}, \bibinfo{person}{Hongzhi Chen}, {and} \bibinfo{person}{Guoren Wang}.} \bibinfo{year}{2022}\natexlab{}.
\newblock \showarticletitle{Lightning fast and space efficient k-clique counting}. In \bibinfo{booktitle}{\emph{Proceedings of the ACM Web Conference 2022}}. \bibinfo{pages}{1191--1202}.
\newblock


\bibitem[Ye et~al\mbox{.}(2023a)]%
        {ye2023efficient2}
\bibfield{author}{\bibinfo{person}{Xiaowei Ye}, \bibinfo{person}{Rong-Hua Li}, \bibinfo{person}{Qiangqiang Dai}, \bibinfo{person}{Hongzhi Chen}, {and} \bibinfo{person}{Guoren Wang}.} \bibinfo{year}{2023}\natexlab{a}.
\newblock \showarticletitle{Efficient k-Clique Counting on Large Graphs: The Power of Color-Based Sampling Approaches}.
\newblock \bibinfo{journal}{\emph{IEEE Transactions on Knowledge and Data Engineering}} (\bibinfo{year}{2023}).
\newblock


\bibitem[Ye et~al\mbox{.}(2023b)]%
        {ye2023efficient}
\bibfield{author}{\bibinfo{person}{Xiaowei Ye}, \bibinfo{person}{Rong-Hua Li}, \bibinfo{person}{Qiangqiang Dai}, \bibinfo{person}{Hongchao Qin}, {and} \bibinfo{person}{Guoren Wang}.} \bibinfo{year}{2023}\natexlab{b}.
\newblock \showarticletitle{Efficient biclique counting in large bipartite graphs}.
\newblock \bibinfo{journal}{\emph{Proceedings of the ACM on Management of Data}} \bibinfo{volume}{1}, \bibinfo{number}{1} (\bibinfo{year}{2023}), \bibinfo{pages}{1--26}.
\newblock


\bibitem[Yin et~al\mbox{.}(2017)]%
        {yin2017local}
\bibfield{author}{\bibinfo{person}{Hao Yin}, \bibinfo{person}{Austin~R Benson}, \bibinfo{person}{Jure Leskovec}, {and} \bibinfo{person}{David~F Gleich}.} \bibinfo{year}{2017}\natexlab{}.
\newblock \showarticletitle{Local higher-order graph clustering}. In \bibinfo{booktitle}{\emph{Proceedings of the 23rd ACM SIGKDD international conference on knowledge discovery and data mining}}. \bibinfo{pages}{555--564}.
\newblock


\bibitem[Yu et~al\mbox{.}(2020)]%
        {yu2020aot}
\bibfield{author}{\bibinfo{person}{Michael Yu}, \bibinfo{person}{Lu Qin}, \bibinfo{person}{Ying Zhang}, \bibinfo{person}{Wenjie Zhang}, {and} \bibinfo{person}{Xuemin Lin}.} \bibinfo{year}{2020}\natexlab{}.
\newblock \showarticletitle{Aot: Pushing the efficiency boundary of main-memory triangle listing}. In \bibinfo{booktitle}{\emph{Database Systems for Advanced Applications: 25th International Conference, DASFAA 2020, Jeju, South Korea, September 24--27, 2020, Proceedings, Part II 25}}. Springer, \bibinfo{pages}{516--533}.
\newblock


\bibitem[Yuan et~al\mbox{.}(2017)]%
        {yuan2017effective}
\bibfield{author}{\bibinfo{person}{Long Yuan}, \bibinfo{person}{Lu Qin}, \bibinfo{person}{Xuemin Lin}, \bibinfo{person}{Lijun Chang}, {and} \bibinfo{person}{Wenjie Zhang}.} \bibinfo{year}{2017}\natexlab{}.
\newblock \showarticletitle{Effective and efficient dynamic graph coloring}.
\newblock \bibinfo{journal}{\emph{Proceedings of the VLDB Endowment}} \bibinfo{volume}{11}, \bibinfo{number}{3} (\bibinfo{year}{2017}), \bibinfo{pages}{338--351}.
\newblock


\bibitem[Yuan et~al\mbox{.}(2022)]%
        {yuan2022efficient}
\bibfield{author}{\bibinfo{person}{Zhirong Yuan}, \bibinfo{person}{You Peng}, \bibinfo{person}{Peng Cheng}, \bibinfo{person}{Li Han}, \bibinfo{person}{Xuemin Lin}, \bibinfo{person}{Lei Chen}, {and} \bibinfo{person}{Wenjie Zhang}.} \bibinfo{year}{2022}\natexlab{}.
\newblock \bibinfo{title}{Efficient k-clique listing with set intersection speedup. ICDE}.
\newblock


\bibitem[Yuster(2006)]%
        {yuster2006finding}
\bibfield{author}{\bibinfo{person}{Raphael Yuster}.} \bibinfo{year}{2006}\natexlab{}.
\newblock \showarticletitle{Finding and counting cliques and independent sets in r-uniform hypergraphs}.
\newblock \bibinfo{journal}{\emph{Inform. Process. Lett.}} \bibinfo{volume}{99}, \bibinfo{number}{4} (\bibinfo{year}{2006}), \bibinfo{pages}{130--134}.
\newblock


\bibitem[Zhang et~al\mbox{.}(2023a)]%
        {zhang2023scalable}
\bibfield{author}{\bibinfo{person}{Fangyuan Zhang}, \bibinfo{person}{Dechuang Chen}, \bibinfo{person}{Sibo Wang}, \bibinfo{person}{Yin Yang}, {and} \bibinfo{person}{Junhao Gan}.} \bibinfo{year}{2023}\natexlab{a}.
\newblock \showarticletitle{Scalable approximate butterfly and bi-triangle counting for large bipartite networks}.
\newblock \bibinfo{journal}{\emph{Proceedings of the ACM on Management of Data}} \bibinfo{volume}{1}, \bibinfo{number}{4} (\bibinfo{year}{2023}), \bibinfo{pages}{1--26}.
\newblock


\bibitem[Zhang et~al\mbox{.}(2018)]%
        {zhang2018discovering}
\bibfield{author}{\bibinfo{person}{Fan Zhang}, \bibinfo{person}{Long Yuan}, \bibinfo{person}{Ying Zhang}, \bibinfo{person}{Lu Qin}, \bibinfo{person}{Xuemin Lin}, {and} \bibinfo{person}{Alexander Zhou}.} \bibinfo{year}{2018}\natexlab{}.
\newblock \showarticletitle{Discovering strong communities with user engagement and tie strength}. In \bibinfo{booktitle}{\emph{International Conference on Database Systems for Advanced Applications}}. Springer, \bibinfo{pages}{425--441}.
\newblock


\bibitem[Zhang et~al\mbox{.}(2023b)]%
        {zhang2023efficiently}
\bibfield{author}{\bibinfo{person}{Lingling Zhang}, \bibinfo{person}{Zhiwei Zhang}, \bibinfo{person}{Guoren Wang}, \bibinfo{person}{Ye Yuan}, {and} \bibinfo{person}{Kangfei Zhao}.} \bibinfo{year}{2023}\natexlab{b}.
\newblock \showarticletitle{Efficiently Counting Triangles for Hypergraph Streams by Reservoir-Based Sampling}.
\newblock \bibinfo{journal}{\emph{IEEE Transactions on Knowledge and Data Engineering}} \bibinfo{volume}{35}, \bibinfo{number}{11} (\bibinfo{year}{2023}), \bibinfo{pages}{11328--11341}.
\newblock


\bibitem[Zhou et~al\mbox{.}(2021)]%
        {zhou2021butterfly}
\bibfield{author}{\bibinfo{person}{Alexander Zhou}, \bibinfo{person}{Yue Wang}, {and} \bibinfo{person}{Lei Chen}.} \bibinfo{year}{2021}\natexlab{}.
\newblock \showarticletitle{Butterfly counting on uncertain bipartite graphs}.
\newblock \bibinfo{journal}{\emph{Proceedings of the VLDB Endowment}} \bibinfo{volume}{15}, \bibinfo{number}{2} (\bibinfo{year}{2021}), \bibinfo{pages}{211--223}.
\newblock


\bibitem[Zhou et~al\mbox{.}(2023)]%
        {zhou2023butterfly}
\bibfield{author}{\bibinfo{person}{Alexander Zhou}, \bibinfo{person}{Yue Wang}, {and} \bibinfo{person}{Lei Chen}.} \bibinfo{year}{2023}\natexlab{}.
\newblock \showarticletitle{Butterfly counting and bitruss decomposition on uncertain bipartite graphs}.
\newblock \bibinfo{journal}{\emph{The VLDB Journal}} \bibinfo{volume}{32}, \bibinfo{number}{5} (\bibinfo{year}{2023}), \bibinfo{pages}{1013--1036}.
\newblock


\bibitem[Zou(2016)]%
        {zou2016bitruss}
\bibfield{author}{\bibinfo{person}{Zhaonian Zou}.} \bibinfo{year}{2016}\natexlab{}.
\newblock \showarticletitle{Bitruss decomposition of bipartite graphs}. In \bibinfo{booktitle}{\emph{International conference on database systems for advanced applications}}. Springer, \bibinfo{pages}{218--233}.
\newblock


\end{thebibliography}
